\documentstyle[prl,aps,multicol,epsf,graphicx,amssymb]{revtex}

\def\be{\begin{eqnarray}}
\def\ee{\end{eqnarray}}

\def\bml{\begin{mathletters}}
\def\eml{\end{mathletters}}

\def\nn{\nonumber}
\def\eq{&=&}

\def\cC{{\cal C}}
\def\cD{{\cal D}}

\def\cG{{\cal G}}
\def\cH{{\cal H}}
\def\cN{{\cal N}}

\def\cP{{\cal P}}
\def\cT{{\cal T}}

\def\cX{{\cal X}}
\def\cZ{{\cal Z}}

\def\bA{{\bf A}}
\def\bB{{\bf B}}

\def\bQ{{\bf Q}}

\def\bX{{\bf X}}

\def\bk{{\bf k}}
\def\bn{{\bf n}}
\def\bp{{\bf p}}
\def\bq{{\bf q}}
\def\br{{\bf r}}

\def\bmat{\begin{pmatrix}}
\def\emat{\end{pmatrix}}

\def\sar{\sigma^{\sc{ar}}}

\def\str{\sigma^{\sc{tr}}}

\def\Tr{{\,\rm Str\,}}
\def\tr{{\,\rm tr\,}}

\begin{document}

\title{Quantum in-plane magnetoresistance in 2D electron systems}

\author{Julia S. Meyer$^1$, Vladimir I. Fal'ko$^2$, and B.L. Altshuler$^3$}

\address{$^1$ Institut f\"ur Theoretische Physik, Universit\"at zu K\"oln, 
  50937 K\"oln,  Germany\\
$^2$ Physics Department, Lancaster
  University, LA1 4YB, Lancaster, UK\\
$^3$ Physics Department, Princeton University, Princeton,
  NJ 08544, USA;\\ NEC Research Institute, 4 Independence Way,
  Princeton, NJ 08540, USA}
\date{\today}
\maketitle
\begin{abstract}
We review various aspects of magnetoresistance in
(quasi-)twodimensional systems subject to an in-plane magnetic
field. Concentrating on single-particle effects, three mechanisms
leading to magnetoresistance are discussed: the orbital effect of the magnetic
field -- due to inter-subband mixing -- and the sensitivity of
this effect to the geometrical symmetry of the system,
the interplay between spin-orbit coupling and Zeeman splitting,
and the influence of the field on spin scattering at magnetic
impurities.


\end{abstract}


\section{Introduction}
\label{sec-Introduction}

Studies of low-dimensional electron systems subject to an in-plane magnetic field
have come to the focus of intense attention recently. In several
experiments the influence of an in-plane magnetic field on two-
and one-dimensional electrons in semiconductor
heterostructures~\cite{2D,PepperWires} as well as in lateral quantum
dot devices~\cite{MarcusSpin} has been investigated. The use of the
fairly unconventional in-plane field geometry aimed at achieving a stronger
magnetic field influence on the 2D electron {spin}, thus, compensating the
dominance of orbital effects in most of the known
semiconductor materials. Then, by manipulating the field-induced spin
polarization, one may extract information about the ground state properties of interacting
electrons. At low temperatures -- i.e., in the regime where electron-electron
interactions open the possibility of a metal-insulator transition or
the ferromagnetic Stoner instability -- interaction effects coexist with
single-particle interference effects. The aim of this article is to provide
a theoretical overview of possible influences of an in-plane magnetic
field on single-particle quantum transport phenomena in
semiconductor heterostructures, quantum wells, and lateral dots.

Quantum transport phenomena have been the subject of extensive experimental and
theoretical investigations during recent decades~\cite{b-AlLe,b-leshouches95,b-meso,ib-SiAl}. There exists a broad variety of
electronic systems, where spectacular effects produced by the
interference of electron waves have been observed. These include weak localization
(WL)~\cite{GoLa79}, leading to the low-temperature magnetoresistance in
low-dimensional electron structures, such as two-dimensional electron 
and hole gases in
semiconductor heterostructures and quantum wells, as well as universal
conductance fluctuations
(UCF)~\cite{AltshulerKhm1,AltshulerKhm2,AltshulerKhm3,Washburn} in
quantum wires and
dots.

\begin{figure}[h]
\begin{center}
\includegraphics[scale=0.3]{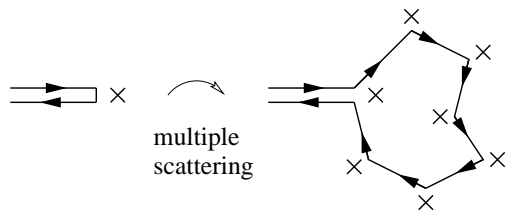}
\caption{Back-scattering from an impurity:
    Contributions due to multiple scattering events.}
\label{fig-Intro1} 
\end{center}
\end{figure}

\begin{figure}[h]
\begin{center}
\includegraphics[scale=0.35]{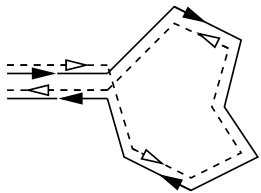}
\caption{Enhanced back-scattering caused by the
    interference of electrons traveling along the same geometrical
    path in opposite directions.}
\label{fig-Intro2} 
\end{center}
\end{figure}

Weak localization of electron waves is the result of an enhancement of
back-scattering  from a scattering center in a disordered
environment. This enhancement of back-scattering suppresses electrical
conductivity of a quantum conductor, as compared to the value expected on
the basis of a classical Drude formula. The Drude result is
determined by the mean free path corresponding to a single-impurity
scattering cross section whereas multiple scattering events as depicted in
Fig.~\ref{fig-Intro1} are responsible for the enhanced back-scattering. 
It is generated by
the constructive interference between pairs of electron waves that
scatter from the same impurities, i.e., two electron waves arrive
together at a particular impurity, scatter from it to follow a random-walk
path (in the form of a closed loop) visiting the same set of surrounding
impurities -- either in clockwise or anti-clockwise direction -- and finally
scatter into the exactly backward direction. An illustration of 
electron paths involved in such
a non-local scattering process is shown in
Fig.~\ref{fig-Intro2}. 
Obviously, the
observation of such an interference effect requires phase-coherent 
propagation along the closed-loop part of the geometrical
trajectory. Furthermore, the magnitude of the observable contribution to the
conductivity is sensitive to the fundamental symmetries of the system, in
particular, to the presence or absence of time-reversal (${\cal T}$) symmetry.

The effect of enhanced back-scattering off an impurity inside a disordered
metal can be discussed in the same terms as the enhancement of back-scattering
of waves from a diffusive medium. The probability $w_{{\bf k},-{\bf k}}$  of
a wave to be reflected back from a medium after multiple scattering inside
is determined as the modulus of the sum of amplitudes $A_j$
corresponding to particular paths $j$, 
\begin{eqnarray}
w_{{\bf k},-{\bf k}}=\Bigg| \sum_{j}\left[ A_{j}^{\circlearrowright }({\bf k}
,-{\bf k})+A_{j}^{\circlearrowleft }({\bf k},-{\bf k})\right] \Bigg| ^{2}.
\label{Intro-1}
\end{eqnarray}
Here, $\circlearrowright$ ($\circlearrowleft$) stands for clockwise
(anti-clockwise) propagation along the path.

In an ideally time-reversible system, the two amplitudes  related to the electron propagation along the same
geometrical loop `$j$' in opposite directions, $A_{j}^{
\circlearrowright }({\bf k},-{\bf k})$ and $A_{j}^{\circlearrowleft }({\bf k}
,-{\bf k})$, are
identical. Thus, $A_{j}^{\circlearrowright }A_{j}^{\circlearrowleft
\ast }=|A_{j}^{\circlearrowright }|^{2}$, though for each particular loop $
A_{j}$ contains a possibly large random semiclassical phase of
propagation. The cancellation of phase factors happens because for an electron described by a ${\cal T}$-symmetric Hamiltonian, (a) the phases acquired along the same `free'
segment of the path are equal for waves propagating with the wave number ${\bf k}$
and $-{\bf k}$, and (b) the amplitudes of intermediate scattering processes, $
{\bf k}_{a}\to {\bf k}_{b}$ for ${\circlearrowright }$ and $-
{\bf k}_{b}\to -{\bf k}_{a}$ for ${\circlearrowleft }$ also
equal each other at each node of the loop $j$. Therefore, after averaging
over different loops which is equivalent to the averaging over disorder, one
can identify two phase-insensitive contributions to the back-scattering probability,
\begin{eqnarray}
w_{{\bf k},-{\bf k}} &=&\sum_{j}\left\langle \big| A_{j}^{\circlearrowright
}\big| ^{2}+\big| A_{j}^{\circlearrowleft }\big| ^{2}\right\rangle +2
\mathop{\rm Re}
\sum_{j}\left\langle A_{j}^{\circlearrowright }A_{j}^{\circlearrowleft \ast
}\right\rangle   \nonumber \\
&=&w^{{\rm class}}+\delta w^{{\rm WL}},  \label{Intro-2}
\end{eqnarray}
where $w^{{\rm class}}$ is the back-scattering probability of a
classical particle whereas $\delta w^{{\rm WL}}$ is an additional
quantum contribution due to
constructive interference between time-reversed paths. This second
term in Eq.~(\ref{Intro-2}) increases the return probability, i.e.,
the probability for an electron to
revisit the same scatterer again -- which explains the name `weak localization'
given to this effect.

The length of closed paths contributing to the interference between
back-scattered waves is limited by the phase coherence time
$\tau_\phi$, yielding a maximal length ${\cal L}_{\rm max}=v_{\rm
  F}\tau_\phi$. In practice, this time scale is determined either by
phase relaxation within the system or the
electron escape from a fully phase-coherent conductor to bulk
reservoirs. Estimates for the weak localization corrections to the
conductance read
\begin{eqnarray}
\delta g^{{\rm WL}}\sim -\,\frac{e^{2}}{h}\times \left\{ 
\begin{array}{cl}
\ln 
{\displaystyle \frac{\tau _\phi}{\tau}}
 & \qquad {\rm 2D,} \\ 
{\displaystyle \frac1{\sqrt{D\tau _\phi}}}
& \qquad {\rm Q1D \enspace wire,} \\ 
1 & \qquad {\rm 0D \enspace dot.}
\end{array}
\right.   \label{Intro-4a}
\end{eqnarray}
For the case of a dot, by $g$ we understand its full
conductance. Furthermore, $\tau $ is
the elastic mean free path, and $D=v_{\rm F}^2\tau/d$ ($v_{\rm F}$
Fermi velocity, $d$ dimensionality of the system) is the diffusion
constant. 

The violation of ${\cal T}$-invariance lifts the equivalence between
the amplitudes $A_{j}^{\circlearrowright }$ and
$A_{j}^{\circlearrowleft }$ and, thus, 
suppresses  contributions to the enhanced back-scattering probability from longer
paths. An obvious reason for $
t\to -t$ symmetry breaking is an external magnetic field:
thus, provided the orbital
electron motion is coupled to the field, it affects the interference
effects in the phase-coherent electron transport. For two-dimensional electrons in a heterostructure
subject to a magnetic field {perpendicular} to the 2D plane, the difference
between the amplitudes $A_{j}^{\circlearrowright }$ and $A_{j}^{\circlearrowleft }$ arises
due to the Aharonov-Bohm phases accumulated along the path. For
clockwise and anti-clockwise propagation around the
oriented area $S_{j}$, the Aharonov-Bohm phases,
$\delta\varphi_{\rm AB}=B_{z}S_{j}/\phi _{0}$ ($\phi_0$ flux quantum),
have opposite signs. As for random-walk trajectories the encircled area $
S_{j}$ is random, contributions from paths encircling a magnetic 
flux larger than typically the flux quantum ($\left| B_{z}S_{j}\right|
\gtrsim \phi _{0}$) cancel out after disorder averaging: $\langle A_{j}^{\circlearrowright }A_{j}^{\circlearrowleft \ast
}\rangle= \langle \,|A_{j}^{\circlearrowright
}|^2\times\exp[2i\delta \varphi^{\circlearrowright
}_{\rm AB}]\,\rangle\to 0$.

When the orientation of the applied magnetic field lies within the
plane of the heterostructure where an effectively two-dimensional
electron moves (i.e., $B_{z}=0$), an analysis of its influence on the
weak localization properties requires taking
into account more subtle arguments than a straightforward consideration of
the Aharonov-Bohm effect. In fact, a strictly 2D system is insensitive 
to the orbital $\cT$-breaking effect of an in-plane magnetic field. A possibility to couple a 2D orbital motion of
the electrons to the in-plane magnetic field and, therefore, to break the
time-reversal symmetry in a 2D system appears only after taking into account~\cite{Falko89,FalkoRMF,MAA01,Thesis,FaJu01} the finite extent of electronic
wave functions in the confinement direction of a quantum well as well
as subband mixing by
the magnetic
field~\cite{AnisMass1,AnisMass2,Spectroscopy1,Spectroscopy2} and
possibly by
disorder. 

In {Chapter~\ref{sec-Orbital}}, 
we discuss this purely orbital effect of an in-plane magnetic
field on quantum interference in 2D semiconductor structures. 
As for the confinement of
the transverse motion, we also
describe the crossover from the quantum limit 
(where the width of the system is of order of
the Fermi wavelength)
to the semiclassical regime corresponding to thin films of
disordered and pure metals
~\cite{AlAr81,DuKh84,HoutenBeenakker,Falko89,numFal,MAA01,FaJu01}. 
The calculation aims at identifying the time scale $\tau _{B}$ after which the
time-reversal symmetry breaking sufficiently affects the phase of an
electron to destroy constructive interference. Correspondingly, only
paths with lengths shorter than ${\cal L}(B)\sim v_{{\rm
    F}}\tau _{B}$ contribute to weak localization. Thus, $\tau _{B}$ is the
time scale that determines the estimates for the weak localization correction to the
conductivity, i.e., it replaces the phase coherence time in
Eq.~(\ref{Intro-4a}). Therefore the conductance is field-dependent, and
the so-called magnetoconductance $\Delta g_B=\delta
g^{{\rm WL}}(B)-\delta g^{{\rm WL}}(0)$ obtains as
\begin{eqnarray}
\Delta g_B\sim \,\frac{e^{2}}{h}\times \left\{ 
\begin{array}{cl}
\ln 
{\displaystyle \frac{\tau _\phi}{\tau _{B} }}
 & \qquad {\rm 2D,} \\ 
{\displaystyle \frac1{\sqrt{D\tau _\phi}}}-
{\displaystyle \frac1{\sqrt{D\tau _{B}}}}
& \qquad {\rm Q1D \enspace wire.}
\end{array}
\right.   \label{Intro-4b}
\end{eqnarray}
More details about the multi-subband case can be found in chapter~\ref{sec-SUSY}.

Another possibility for the in-plane magnetic field to affect the
interference of 2D electron waves arises due to the coupling between orbital
motion and electron spin, which is a pronounced feature of the electronic
band structure of most III-V semiconductors with the zinc-blend type
crystalline lattice and a unit cell without inversion symmetry. In a
heterostructure or quantum well, spin-orbit coupling for a spin-$\frac12
$ carrier is linear, both in the planar momentum ${\bf k}$ and the spin operator 
$\underline{\sigma }$. For a given initial spin-polarized state, it produces an
electron spin precession, where the precession axis as well as the
precession frequency depend on the direction and value of the 
Fermi momentum of the propagating
electron. For a diffusing electron, such a precession randomly changes
direction, thus, providing spin relaxation for an
initially spin-polarized particle -- the mechanism known as Dyakonov-Perel
relaxation~\cite{DyakonovPerel}. However, spin-orbit coupling alone does not
violate time-reversal symmetry, though it does modify the
above-described interference effects. In particular, strong spin-orbit 
coupling changes the
constructive interference between electrons encircling the same closed
geometrical path in opposite directions into destructive
interference~\cite{HiLa80,LyandaGeller1,LyandaGeller2}. To trace the
origin of this change, let us consider 
the extreme limit of dominant spin-orbit coupling, where an
electron always remain in the same chiral state along its orbital
motion (for instance, $\xi =\frac\bk{|\bk|}
\cdot \underline{\sigma }=+1$). Thus, an electron approaching the scattering
medium in the initial state $|{\bf k},\xi \rangle $ (which can be viewed as $
|{\bf k},\rightarrow \rangle $) ends up in the final backward moving state $|\!-\!
{\bf k},\xi \rangle $ (which can be viewed as $|\!-\!{\bf
  k},\leftarrow \rangle $). This implies a rotation of its spin within
the $x-y$ plane by the angle $|\theta_j|=(2n+1)\pi $ (where $n\in{\mathbb{N}}$)
while moving along a diffusive path connecting these two states. For two
spin-$\frac{1}{2}$ electrons encircling the same closed loop in
opposite directions, $\theta _{j}^{\circlearrowleft }=-
\theta _{j}^{\circlearrowright }=(2n+1)\pi $. As a result, their
interference produces a negative contribution to the back-scattering,
\begin{eqnarray}
A_{j}^{\circlearrowright }A_{j}^{\circlearrowleft \ast
}=\big|A_{j}^{\circlearrowright }\big|^{2}\exp \left[\frac{i}{2}(\theta
_{j}^{\circlearrowright }-\theta _{j}^{\circlearrowleft
})\right]=-\big|A_{j}^{\circlearrowright }\big|^{2},  \label{Intro-6}
\end{eqnarray}
which increases the quantum conductance as compared to its
classical Drude value -- the effect known as weak {anti}-localization.  

\begin{figure}[h]
\begin{center}
\includegraphics[scale=0.5]{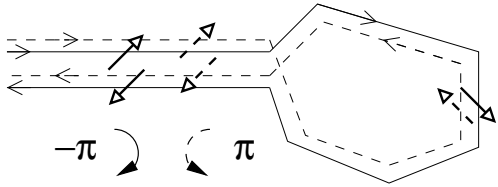}
\caption{Suppressed back-scattering in the
    presence of strong spin-orbit coupling.}
\label{fig-Intro3} 
\end{center}
\end{figure}

Now, the application of an in-plane magnetic field causes a Zeeman splitting of
spin-$\uparrow$ and spin-$\downarrow$ electron states which -- in
combination with spin-orbit coupling --
affects the interference of back-scattered waves. {Chapter~\ref{sec-Spin-orbit}} is devoted to the
description of the interplay between Zeeman splitting and spin-orbit
coupling in determining quantum transport characteristics (in particular,
of lateral dots).

Interference between different paths not only affects the
mean value of the conductance (i.e., $\langle g\rangle=g_{\rm class}+\delta
g^{\rm WL}$) but also characteristic deviations from that mean value,
$g=\langle g\rangle+\delta g$. Universal conductance fluctuations (UCF) represent another interference
phenomenon observed in mesoscopic conductors and quantum dots at low
temperatures. To discuss this effect, let us consider the transmission 
probability through a disordered piece of metal. For a  diffusing
electron wave, there are many scenarios leading it from the same initial
state $|{\bf k}\rangle _{{\rm left}}$ on the left to the same final state $|
{\bf k}^{\prime }\rangle _{{\rm right}}$ on the right, thus, creating the
possibility for interference. Depending on the individual distribution of
scatterers in the sample, different paths $j$ have different weights
-- and, thus, the waves following them acquire different phases -- which makes the transmission
probability, 
\begin{eqnarray}
P_{{\bf k},{\bf k}^{\prime }}=\Bigg| \sum_{j}A_{j}({\bf k},{\bf k}^{\prime
})\Bigg| ^{2}=\sum_{j}\big| A_{j}\big| ^{2}+\sum_{ji}A_{j}A_{i}^{\ast
}\;,  \label{Intro-8}
\end{eqnarray}
sample-specific as well as dependent on specific interference conditions,
namely the energy
(wavelength) of the incident electron or the presence of an external magnetic field. In measurements,
this manifests itself in a  random but reproducible dependence of the sample
conductance on the Fermi energy $E_{{\rm F}}$ and  the applied field
$B_{z}$. Furthermore, gate voltages (varying the shape of 
a semiconductor dot) represent an additional control parameter.

In a phase-coherent system, the variance of conductance fluctuations has
a value,
\begin{eqnarray}
\langle \delta g^{2}\rangle =c_{{\rm geom}}\,\alpha \times
\,\left( \frac{e^{2}}{h}\right) ^{2},  \label{Intro-10}
\end{eqnarray}
which is
universal~\cite{AltshulerKhm1,AltshulerKhm2,AltshulerKhm3,Beenakker,AleinerFalko}
up to a geometry-dependent
numerical factor $c_{{\rm geom}}\sim 1$. Furthermore, $\alpha$
reflects the fundamental symmetries of the system, expressed by three coefficients, 
\begin{eqnarray}
\alpha =\frac{s}{\beta \Sigma }.  \label{Intro-12}
\end{eqnarray}
The symmetry plays such an important role here as it determines the number of
independently fluctuating components of the generically random amplitudes $A_{j}$
related to each particular geometrical path. The higher the symmetry,
the more correlations are implicit between real and imaginary parts as 
well as spin
components of $A_{j}$. In particular, $s=1$ or $2$ depending on whether
Kramers' degeneracy is lifted or not. When the orbital electron
motion is independent of the electron spin state, i.e., when the
system possesses spin-rotation invariance, the $\cT$-symmetric system
is described by $\beta =1$ whereas $\beta =2$ if
time-reversal symmetry is broken. The value $\beta =4$ (with $s=1$) represents the case
of efficient spin-relaxation due to spin-orbit scattering. Finally, an
additional symmetry class number $\Sigma $~\cite{AleinerFalko} characterizes
whether a violation of time-reversal symmetry in the spin-sector
affects ($\Sigma =2$ -- or not, $\Sigma =1$) the
interference in the electron orbital motion via spin-orbit
coupling. The crossover between the latter two
symmetry classes may be achieved by varying the Zeeman splitting. Its
features will be described in Chapter~\ref{sec-Spin-orbit}, together with weak localization
effects.

As phase coherence is crucial for the occurrence of interference
effects, the influence of symmetry breaking on conductance
measurements is  not observable when the corresponding time scale is
larger than the decoherence time, $\tau _{B}>\tau _{\varphi}$. As for 
finite systems, when decoherence of electrons in a
dot or a short wire happens faster than the escape into bulk reservoirs, $
\tau _{\varphi }<\tau _{{\rm esc}}$, both the enhanced back-scattering
effect and UCFs get suppressed~\cite{MarcusDephasing1,MarcusDephasing2} as compared to the
conductance quantum, $e^{2}/h$. In these situations, no influence of an in-plane
magnetic field on quantum transport characteristics of the system would be
expected -- unless the field affects the decoherence time.

This is the case when the latter is
caused by spin-flip scattering at paramagnetic impurities~\cite{WeakLoc}. In
all foreseeable regimes~\cite{WeakLoc,FalkoJETP}, degenerate paramagnetic
impurities destroy weak localization via spin-flip
scattering, the relevant time scale being the spin-flip scattering
time $\tau_{\rm s}$. Furthermore, they
suppress conductance fluctuations~\cite
{BobkovFalkoKhm,Chandrasekhar,FalkoJP,Benoit,Geim} as the flipping of
impurity spins effectively changes the realization of disorder in the sample
and, thus, leads to a self-averaging of the conductance. However, a
Zeeman splitting of the
magnetic impurity levels by a magnetic field causes an energy threshold for
electron spin-flip processes which slows down the electron spin relaxation.
At a high field, the electron spin relaxation is possible
only due to very few thermally activated impurity spins, so its
rate is reduced by the exponential factor,  
$\tau _{s}^{-1}(B)\propto \tau _{s}^{-1}\times \exp [-g^{{\rm
    (imp)}}\mu_{\rm B}B/T]$, where $\mu_{\rm B}$ is the Bohr magneton
and 
$g^{{\rm (imp)}}$ the $g$-factor of the impurities.
Thus, a sufficiently strong magnetic field restores 
UCFs~\cite{BobkovFalkoKhm,FalkoJP,Benoit,Geim} as
well as weak
localization (if the orbital effect of the magnetic field
is excluded). The in-plane magnetoresistance of magnetically
contaminated 2D semiconductors and other features related to the dynamics of
paramagnetic impurities are discussed in
{Chapter~\ref{sec-Spin}}.

For all practical purposes, the study of symmetry breaking rates,
determining the
parallel field effect on the weak localization corrections to the
conductivity of a 2D electron gas or a semiconductor wire, can be completed
using the diagrammatic perturbation theory for disordered systems. Although
one can find more details about this technique in several textbooks and
review articles (see, e.g.,~\cite{Efetov}), for the sake of completeness, we
describe its necessary elements in {
  Chapter~\ref{sec-Techniques}}. As an alternative, field theoretic
methods may be used. In chapter~\ref{sec-SUSY}, we 
illustrate the supersymmetric non-linear sigma-model in
application to one of the problems listed above, i.e., the orbital
effect of an in-plane magnetic field (via subband mixing) on the transport
properties of 2D electrons.


\section{The diagrammatic technique: diffusons and Cooperons}
\label{sec-Techniques}

Diagrammatics is a perturbative method which allows one to
a) classify different contributions to the perturbation series and
b) sum up the relevant terms. 
To be specific, here -- for simplicity -- we choose the random
disorder potential $V$ to be drawn from a Gaussian white noise distribution:
\begin{eqnarray}
\langle V(\br) \rangle = 0, \qquad
\langle V(\br)V(\br') \rangle = \frac1{2\pi\nu\tau}\delta(\br-\br'),
\label{wn}
\end{eqnarray}
where $\nu$ is the density of states (DoS) at the Fermi level and $\tau$ is
the mean free scattering time. Note that throughout this review we will use 
units
where $\hbar=c=e=1$.

\begin{figure}[h]
\begin{center}
\includegraphics[scale=0.4]{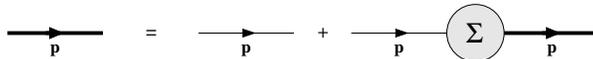}
\caption{Diagrammatic representation of the Dyson
    equation.}
\label{fig-dys}
\end{center}
\end{figure}

The starting point of the perturbative approach is the representation of the 
Green function as a series in
powers of $V$, i.e., $G=G_0+G_0\sum_n(VG_0)^n$. Under averaging,
diagrams with single impurity lines vanish (due to $\langle
V\rangle=0$). Thus, the diagrammatic expansion involves only
diagrams with paired impurity lines. 
The average Green function is then given by a
Dyson equation, depicted schematically in Fig.~\ref{fig-dys}:
\begin{eqnarray}
\langle G \rangle= G_0 + G_0\sum_{n=1}^\infty (\Sigma G_0)^n = G_0 +
G_0 \Sigma \langle G \rangle \quad \Longleftrightarrow \quad \langle G \rangle= 
\frac{G_0}{1-\Sigma G_0},\nn
\end{eqnarray}
where the self-energy $\Sigma$ contains all {irreducible}
diagrams, i.e., diagrams that cannot be split
into two by cutting one $G_0$-line.

The dominant contribution to the self-energy reads
\begin{eqnarray} 
\Sigma_0^\pm(\bp)=\left\langle\int \! d\bq \; 
V(\bq)G_0^\pm(\bp-\bq;\epsilon)V(-\bq)\right\rangle , 
\end{eqnarray}
where the shorthand notation $d\bq=d^dq/(2\pi)^d$\index{$d\bq=d^dq/(2\pi)^d$} 
has 
been introduced. 
Since the
potential is $\delta$-correlated in real space, its Fourier transform
does not depend on momentum. Hence, $\Sigma_0^\pm = \int \! d\bp \, 
G_0^\pm(\bp;\epsilon)/(2\pi\nu\tau)$.

The real part of $\Sigma$, leading to a shift in energy, can be
absorbed in the ground state energy. Using that the unperturbed Green
function reads
$G_0^\pm(\bp;\epsilon)=(\epsilon\pm i0+
\xi_{\bp})^{-1}$, where $\xi_{\bp}=\epsilon_{\rm F}-p^2/(2m)$, and the 
definition of
the density of states, $\nu(\epsilon)=L^{-d}\sum_\bp\delta (\epsilon +
\xi_{\bp})$, the 
imaginary part of $\Sigma$ obtains
\begin{eqnarray}
\Im \,\Sigma_0^\pm = \mp \frac1{2\nu\tau} \int \! d\bp \,
\delta (\epsilon + \xi_{\bp})=\mp \frac{1}{2\tau}.
\label{imsig}
\end{eqnarray}
The associated length scale
$\ell$ is the decay length of the average Green function as we will see 
shortly. In the case of weak disorder, $\tau^{-1} \ll \epsilon_{\rm
  F}$, all other contributions are small in $1/(k_{\rm F}\ell)$,  and 
Eq.~(\ref{imsig})
determines the self-energy $\Sigma$ in the {self-consistent Born
approximation} (SCBA).

Inserting the above result into the Dyson equation yields
\begin{eqnarray}
\langle G^\pm\rangle(\bp;\epsilon)\eq\frac{1}{\epsilon+\xi_{\bp} \pm 
\frac{i}{2\tau}}.
\end{eqnarray}
In real space representation, this leads to a decay of the average Green
function on the scale of the mean free path,
\begin{eqnarray}
\langle G\rangle(\br,\br')=G_0(\br,\br')\, e^{-\frac{|\br-\br'|}{2\ell}}.
\end{eqnarray}

\begin{figure}[h]
\begin{center}
\includegraphics[scale=0.35]{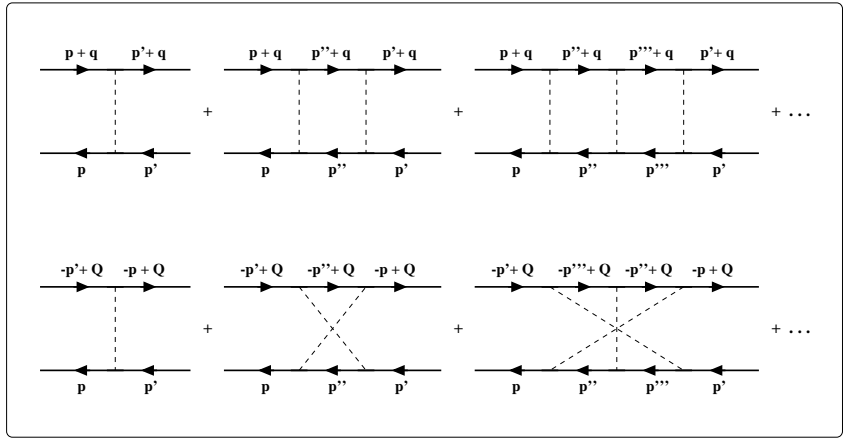}
\caption{Ladder and maximally crossed diagrams.}
\label{fig-dico}
\end{center}
\end{figure}

Having found the average Green function, the next step is to calculate 
correlation functions of the form\footnote{The connected part of averages of the form $\langle
G^+G^+\rangle$ or $\langle G^-G^-\rangle$ vanishes, i.e., no
long-ranged correlations exist.}
\begin{eqnarray} 
F(\bp_1,\bp_1',\bp_2,\bp_2';\omega)=\langle G^- (\bp_1,\bp_1';
\epsilon) G^+(\bp_2,\bp_2'; \epsilon+\omega) \rangle.
\end{eqnarray}
Again -- as for the averaged Green function -- the diagrammatic perturbation 
series involves summing up
diagrams with paired impurity lines. The dominant contributions are
series of {ladder diagrams} and {maximally crossed diagrams},
see Fig.~\ref{fig-dico}. With the notation $\langle AB\rangle_c = \langle
AB\rangle-\langle A\rangle\langle B\rangle$, i.e., subtracting the
disconnected part of the correlator, the contribution of connected diagrams can 
be 
written as
\begin{eqnarray}
&& \langle G^- (\bp_1,\bp_1'; \epsilon) G^+ (\bp_2,\bp_2'; \epsilon+\omega) 
\rangle_c \label{vert}\\
\eq \langle G^-\rangle(\bp_1,\epsilon) \, \langle G^-\rangle(\bp_1',\epsilon) 
\,  \langle G^+\rangle(\bp_2,\epsilon+\omega) \,  \langle 
G^+\rangle(\bp_2',\epsilon+\omega)\rangle\Gamma(\bp_1,\bp_
2,\bp_1',\bp_2';
\omega) \, \delta(\bp_1-\bp_2-\bp_1'+\bp_2'),\nn
\end{eqnarray}
thus defining the reducible\footnote{In the case of two-particle
  functions, a diagram is called reducible if  it can be split into
  two separate diagrams by cutting {two}
  $\langle G\rangle$-lines.} vertex function $\Gamma(\bp_1,\bp_2,\bp_1',\bp_2';
\omega)$.
 
Consider first the sum of ladder diagrams depicted in the upper part of 
Fig.~\ref{fig-dico}. 
Due to
momentum conservation at each vertex, the momentum difference 
$\bq\equiv\bp_1\!-\!\bp_2 (= \bp_1'\!-\!\bp_2')$ is constant, and 
$\Gamma$ depends on this difference only. Then $\Gamma(\bq,\omega)$ is given by 
a Bethe-Salpeter equation, the two-particle
analogue of the Dyson equation:
\begin{eqnarray} 
\Gamma(\bq,\omega) = \gamma + \gamma \Big[\underbrace{
  \int d\bp''\, \langle G^-\rangle(\bp''+\bq,\epsilon+\omega) \langle 
G^+\rangle(\bp'',\epsilon+\omega)}_{\textstyle \equiv\Pi(\bq,\omega)}\Big] 
 \Gamma(\bq,\omega),\nn
\end{eqnarray}
where $\gamma=1/(2\pi\nu\tau)$.
Anticipating the result that $\Gamma$ diverges for $\bq,\omega
\rightarrow 0$, we can approximate the irreducible vertex function $\Pi$ for 
small $|\bq|,\omega$:
\begin{eqnarray}
\Pi(\bq,\omega) \simeq 2\pi \nu\tau(1+i\omega\tau-D\bq^2\tau)
, \nn
\end{eqnarray}
where $D=v_{\rm F}^2\tau/d$\index{$D=\frac1d v_{\rm F}^2\tau$ diffusion 
constant} is the diffusion constant ($d$ dimensionality of the system).
Thus, the
sum of ladder diagrams finally yields
\begin{eqnarray}
\Gamma(\bq,\omega) = \frac{\gamma}{1-\gamma\,
  \Pi(\bq,\omega)}= \frac{1}{2\pi
  \nu\tau^2}\frac{1}{D\bq^2-i\omega}\equiv\cD(\bq,\omega).
\label{di}
\end{eqnarray}
This is a diffusion pole or so-called {diffuson}.
It can easily be seen that the diffuson is not affected by  a weak
magnetic field: Minimal coupling implies that the magnetic field
shifts all momenta by the corresponding vector potential,
$\bp\to\bp-\bA$; however, $(\bp_1-\bA)-(\bp_2-\bA)=\bp_1-\bp_2=\bq$. 

A maximally crossed diagram~\cite{LaNe66} (for a detailed discussion see 
e.g.~\cite{Ber84}) can be converted to a ladder
by reversing one $G$-line. Since now all
the arrows point in the same direction, momentum conservation at the
vertices requires the momentum
sum $\bQ =\bp_1\!+\!\bp_2'$ to be
constant. 
Apart from that, the structure of all equations is the same as for the
ladder diagrams. Thus,
\begin{eqnarray} 
\Gamma(\bQ,\omega) = \frac{1}{2\pi
  \nu\tau^2}\frac{1}{D\bQ^2-i\omega}\equiv\cC(\bQ,\omega).
\label{co}
\end{eqnarray}
This is called a {Cooperon} -- in analogy to superconductivity, as
it corresponds to a correlator in the particle-particle channel
(as opposed to the particle-hole channel for the diffuson).
Now the presence of a magnetic field requires replacing
$\bQ\to\bQ-2\bA$ which leads to a decaying of the Cooperon.

Taking into account the spin of the electrons, 
diffusons and Cooperons split into singlet and triplet modes. 
This will become important when considering spin-orbit coupling 
or spin scattering. Details will be given in the corresponding 
chapters \ref{sec-Spin-orbit} and \ref{sec-Spin}.

In coordinate representation, the expression obtained in 
Eq.~(\ref{co}) has the form of a diffusion equation, 
\begin{eqnarray}
[D(\nabla +i2{\bf A})^2 -i\omega ] 
{\cal C} =\delta ({\bf x}-{\bf x'}).
\end{eqnarray}
(A similar equation holds for the diffuson, Eq.~(\ref{di}).)
In a finite size geometry, this equation should be complemented by 
boundary conditions. For an `open' surface, as, e.g., the interface with 
a bulk reservoir, the boundary condition reads
\begin{eqnarray}
{\cal C} =0.
\end{eqnarray}
For an insulating surface, for example, the edge of a metallic dot, 
the corresponding boundary condition 
expresses the lack of a phase-coherent current density 
across the sample surface,
\begin{eqnarray}
{\bf n}_{\parallel }\cdot(\nabla +i2{\bf A}){\cal C} =0,
\end{eqnarray}
where ${\bf n}_{\parallel }$ is a unit vector normal to the sample surface.
In the case of  zero magnetic field, this boundary condition 
indicates that the lowest Cooperon and diffuson modes correspond to 
the constant ($Q=0$) solution and are gapless. This choice of the
 lowest mode, being constant in space, is called the zero-dimensional
(0D) approximation. In the presence of a magnetic field, 
one has to  give the analysis of the lowest modes in 
closed systems more attention, applying a gauge transformation to the gauge where
${\bf n}_{\parallel }\cdot{\bf A}=0$, and only then making
the 0D approximation. Subsequently, the discrete spectrum of relaxational modes
describing the Cooperon and diffuson can be found
using the standard Hamiltonian perturbation theory methods. 
This is a useful trick which enables us find
the Cooperon and diffuson functions and, in particular, their
lowest eigenmodes in finite size geometries.


\section{In-plane magnetoresistance effect due to intersubband mixing}
\label{sec-Orbital}

As pointed out in the introduction, the influence of an in-plane magnetic field
on the quantum transport characteristics of a 2D or quasi-2D system can be
divided into  orbital and spin-related effects. In this section, we
consider the orbital effect -- treating electrons as spinless. 

There are two features that make the issue of the influence of an in-plane
magnetic field on quantum interference effects in inversion layers and thin
metallic films non-trivial. The first issue is related to the fact that a finite spatial
extent of electron wave functions along the confinement direction, i.e., the $z$-direction perpendicular to the 2D plane, is needed in order that the electron can accumulate an
additional magnetic field-induced phase if the field is exactly parallel to the 2D plane. Furthermore, the possible accumulation of such an Aharonov-Bohm phase is
sensitive to the inversion symmetry properties of the system.
In particular, for a quantum well with a potential profile that is symmetric with
respect to the transformation ${\cal P}_{z}:z\rightarrow -z$, an in-plane
magnetic field cannot induce any dephazing, and the field-dependent rate $
\tau _{B}^{-1}$  is  equal to zero. This property reflects the Berry-Robnik
symmetry effect~\cite{RoBe86}: Consider a system described by a Hamiltonian ${\hat\cH}$ that at $B=0$ is invariant under a discrete symmetry transformation, such as $
{\cal P}_{z}$, and at finite fields remains invariant under the combined time
and $z$-coordinate inversion transformation, ${\cal P}_{z}{\cal T}$, i.e., $
{\cal P}_{z}{\cal T}{\hat\cH}(B)={\hat\cH}(B)$. Then there always exist pairs
of paths having exactly the same value of the semiclassical
action which, therefore, interfere constructively even in the presence of a
magnetic field. As a result, the suppression of weak localization by an
in-plane magnetic field is possible only if  either the scattering
potential in the quantum well is $z$-dependent~\cite{Falko89} (since a generic $z$-dependence does not respect $\cP_z$-symmetry), or if the form of the confinement potential in $z$-direction has
no inversion symmetry~\cite{MAA01,Thesis,FaJu01} (as usually is the case for a 
inversion layer in a heterostructure). In the first subsection, we provide a
quantitative analysis of this effect. In the second subsection, this
analysis is extended to multi-subband systems~\cite{MAA01,Thesis}, and the results are compared to the magnetoresistance in thin films with
diffusive transverse motion.

The second non-trivial issue concerns the so-called geometrical flux
cancellation characteristic for ballistic metallic films (wires), where
electrons scatter only from surface (edge) defects at two parallel surfaces.
In such a `quasi-ballistic' film, the finite resistance is provided by the
diffusive scattering of carriers from the surface. Quantum
corrections to the conductivity are related to the interference between
waves propagating along paths consisting of a sequence of ballistic flights
between the film edges. Among those paths, the trajectories containing ballistic segments that are almost parallel to the film surface and,
therefore, much longer than the film width $d$  play an
important role. These so-called L\'{e}vy flights (in the
terminology of anomalous diffusion theory~\cite{Georges}) 
produce a logarithmically large contribution to
the sample conductance~\cite{Pippard}. The oriented
area encircled by a closed loop made
of quasi-ballistic paths
is exactly equal to zero~\cite{DuKh84}, thus suppressing 
the magnetic field effect on the interference of electron waves 
propagating along them.
An electron can accumulate a magnetic flux
only due to a curving of its trajectory by the magnetic field
itself. This makes the role of L\'{e}vy flights important for the
interference-related magnetoresistance of ballistic films and wires:
curving of the longest ballistic flights -- which is a purely classical effect --
switches their scattering at the surface roughness from one film surface
(or wire edge) to the other~\cite{numFal}. This peculiar
situation and the resulting unexpected influence of a magnetic field on the
localization properties in quasi-ballistic wires are discussed at the end of this
section.


\subsection{Phase-breaking by an in-plane magnetic field in 2D inversion
layers}

\label{sub-onesub} 

The influence of an in-plane magnetic field on the orbital motion of
carriers in a heterostructure or quantum well is a result of the finite
width $d$ of any 2D layer. This effect has been discussed previously in
various contexts~\cite{AnisMass1,AnisMass2,Spectroscopy1,Spectroscopy2}. The Lorentz force on
the electrons generated
by the in-plane field ${\bf B}_{\Vert }$ mixes up the electron motion along 
and across the confinement
direction, thus resulting in a modification of the 2D dispersion, 
$E(p)\rightarrow E({\bf B}_{\Vert },{\bf p})$. In particular, a) the 2D
electron mass increases in the direction perpendicular to ${\bf
  B}_{\Vert }$. Furthermore, in heterostructures with a confining
potential that does not possess inversion symmetry, b) the in-plane field also 
lifts the ${\bf p}
\to -{\bf p}$ symmetry in the dispersion law~\cite{AnisMass1,AnisMass2}: 
$$E(
{\bf B}_{\Vert },{\bf p})-E({\bf B}_{\Vert },-{\bf p})\propto 
\left({\bf p}
\cdot [{\bf B}_{\Vert }\times {\bf e}_{z}]\right)^{3}\neq 0,$$
where ${\bf e}_{z}$ is a unit vector perpendicular to the plane.
This leads to a field-dependent suppression of the WL signal as will
be discussed below. 
By contrast, for the $z$-inversion symmetric problem, due to the
Berry-Robnik phenomenon~\cite{RoBe86}, the Cooperon mode remains
massless, and there is no magnetoresistance. 

In the presence of an in-plane magnetic field, the effective 2D Hamiltonian
for electrons in a heterostructure with the confining potential profile $W(z)$ can be
obtained from the 3D Hamiltonian,
\begin{eqnarray}
{\hat\cH}_{3D}=\frac{
(-i\nabla -{\bf A})^{2}}{2m}-\frac{\partial _{z}^{2}}{2m}+W(z)+u({\bf r},z),  \label{H3d}
\end{eqnarray}
using a plane wave representation, $\Psi _{{\bf p}}=e^{i{\bf pr}}\phi _{0}^{({\bf p})}(z)$ for the electrons in the lowest subband. Here, $\phi_{k}^{({\bf p})}(z)$ are the eigenfunctions of the $z$-dependent part of the Hamiltonian. In particular, at $B_\Vert=0$, the in-plane and perpendicular motion are separable, and the corresponding eigenfunctions are given by $|k\rangle\equiv\phi_{k}^{({0})}$. Due to the mixing between subbands $|0\rangle $ and $|k\!>\!0\rangle $ by an
in-plane magnetic field, the eigenfunction $\phi _{0}^{({\bf p})}(z)$ in the presence of the field depend on the in-plane momentum ${\bf p}$. In order to find $\phi _{0}^{({\bf p})}(z)$ and the
corresponding energy $E({\bf B}_{\Vert },{\bf p})$ for each plane wave state a
perturbation theory analysis -- as described in the following -- or a numerical
self-consistent-field technique may be used.

Furthermore, in Eq.~(\ref{H3d}), the vector potential reads ${\bf A}
=(z-z_{0}){\bf B}_{\Vert }\times {\bf e}_{z}$, where $
z_{0}=\langle 0|z|0\rangle $ is the center of mass position of the
electron wave function in the
lowest subband (at $B_{\Vert }=0$), and the potential $u({\bf r},z)$ is a combination of the
Coulomb potential of impurities and the lateral potential forming the quantum
dot. 

For weak to intermediate magnetic fields ${\bf B}_{\Vert }$, the
effective 2D Hamiltonian takes the form 
\begin{eqnarray}
{\hat\cH}_{2D}=\frac{{\bf p}^{2}}{2m}-p_{\bot }^{2}\gamma (B_{\Vert
})+p_{\bot }^{3}\beta (B_{\Vert })+u({\bf r}).  \label{H2D}
\end{eqnarray}
Here, ${\bf p}=-i\nabla -{\bf a}(\br)$
is a purely 2D momentum operator and $p_{\bot }={\bf p}\cdot \lbrack \frac{{\bf B}
_{\Vert }}{B_{\Vert }}\times {\bf e}_{z}]$ its component perpendicular to $
{\bf B}_{\Vert }$. 

Thus, the application of an in-plane magnetic fields results in a) an effective vector potential ${\bf a }(\br)$,  and b) two additional terms in the free electron dispersion due to $p_{\bot }$-dependent
inter-subband mixing. The first term,  $\delta\hat\cH^{(2)}=-p_{\bot }^{2}\gamma$, lifts the rotational symmetry by causing
an anisotropic mass enhancement~\cite{AnisMass1,AnisMass2}. It increases the 2D density
of states and, for a 2D gas with fixed sheet density, it reduces the Fermi
energy calculated from the bottom of the 2D conduction band, i.e., $E_{{\rm F}
}(B_{\Vert })=E_{{\rm F}}^{0}-{\gamma (B_{\Vert })}p_{{\rm F}}^{2}/2$.
The second term, $\delta\hat\cH^{(3)}=p_{\bot }^{3}\beta$,  is related to the time-reversal
symmetry breaking by $B_{\Vert }$. 
A perturbative analysis of the problem results in the field dependences $\gamma \sim B_{\Vert }^2$
and $\beta \sim B_{\Vert }^3$. Indeed, an expansion of the plane-wave energy up to the
third order in $p_{\bot }$  yields 
\begin{eqnarray*}
\gamma(B_\Vert) &\approx &\frac{B_{\Vert }^{2}}{m^{2}}\sum_{k\geq
1}\frac{\langle k|z|0\rangle ^{2}}{\epsilon _{k}-\epsilon _{0}}\sim 
\frac{1}{m}\left( \frac{d}{\lambda _{B}}\right) ^{4},
\\
\beta(B_\Vert) &\approx &\frac{B_{\Vert }^{3}}{m^{3}}\sum_{k,k'\geq
1}\frac{\langle 0|z|k\rangle \langle k|z|k'\rangle \langle k'|z|0\rangle }{
(\epsilon _{k}-\epsilon _{0})(\epsilon _{k'}-\epsilon _{0})}\sim 
\frac{d}{m}\left( \frac{d}{\lambda
_{B}}\right) ^{6},
\end{eqnarray*}
where $\epsilon_k$ are the subband energies and $\lambda _{B}={1/\sqrt{B_{\Vert }}}$ is the magnetic length.

In ${\hat\cH}_{2D}$ given by Eq.~(\ref{H2D}), the ``in-plane'' disorder is incorporated in the
form of a scattering potential $u({\bf r})\approx $ $\langle 0|u({\bf r}
,z)|0\rangle $. It can be characterized by the mean free
path, $\ell\gg 1/p_{{\rm F}}$ or, equivalently, the momentum relaxation time $\tau $, related to
the diffusion coefficient $D=v_{{\rm F}}^{2}\tau /2$. The modification of
the electron density of states by $B_{\Vert }$ only slightly affects the
value of the electron mean free path. However, the presence of a parallel field also
changes the symmetry of the Born scattering amplitudes between plane waves, $f_{{\bf pp}
^{\prime }}=\langle \Psi _{{\bf p}}|u({\bf r},z)|\Psi _{{\bf p}^{\prime
}}\rangle $. Due to the momentum-dependent subband mixing, $f_{{\bf pp}
^{\prime }}$ acquires an additional contribution, $f_{{\bf pp}^{\prime }}=f_{{\bf pp}
^{\prime }}^{(0)}\left\{ 1+(p_{\bot }+p_{\bot }^{\prime })B_{\Vert }\zeta
\right\} $, where 
\[
\zeta =\frac{1}{m\langle 0|u({\bf p}-{\bf p}^{\prime
},z)|0\rangle }\sum_{k\geq 1}\frac{\langle 0|u({\bf p}
-{\bf p}^{\prime },z)|k\rangle \langle k|z|0\rangle }{\epsilon _{k}-\epsilon _{0}
}.
\] 
This corresponds to the presence of a random gauge field in the
effective 2D Hamiltonian~\cite{Falko89,FalkoRMF,MathurBaranger}, 
\[
{\bf a}(\br)=2\sum_{k\geq 1}\frac{\langle 0|u(
{\bf r},z)|k\rangle \langle k|z|0\rangle }{\epsilon _{k}-\epsilon _{0}}
[{\bf B}_{\Vert }\times {\bf e}_{z}],
\]
which can be interpreted as a result of an effective `curving' of the 2D
plane by impurities with a $z$-dependent scattering potential in the system.
In the presence of an in-plane magnetic field this curving generates a random
effective perpendicular field component, $b_{\bot }=\left[\nabla\times{\bf a}
\right] _{z}$. In systems, where scattering is dominated by Coulomb centers
separated from the 2D plane by a spacer and, thus, is almost independent of $z$, a smaller effect may be
taken into account, namely $\delta {\bf a}\sim\lbrack {\bf B}_{\Vert }\times {\bf 
e}_{z}]([{\bf B}_{\Vert }\times {\bf e}_{z}]\cdot \nabla )^{2}u({\bf r})$.
However, $\delta {\bf a}$ has a negligible influence on the quantum
transport characteristics of 2D electrons as compared to the effect of
the field-dependent electron dispersion.

The perturbative calculation of two-particle correlation functions, i.e.,
diffusons ${\cal 
D}({\bf x},{\bf x}^{\prime };\omega)$ and Cooperons ${\cal C}({\bf x},{\bf x}^{\prime };\omega)$~\cite{AltshulerKhm1,AltshulerKhm2,AltshulerKhm3}, requires taking
into account all field-dependent terms in the effective 2D Hamiltonian Eq.~(\ref{H2D}). Admitting for different magnetic fields $\bB_{\Vert1}$ and $\bB_{\Vert2}$, the result for the diffuson has the form
\begin{eqnarray}
\left[ -D\nabla ^{2}-i\tilde\omega+\tau _{{\rm d}}^{-1}\right] {\cal D
}=\delta ({\bf x}-{\bf x}^{\prime }),  \label{diffusonEq}
\end{eqnarray}
where $\tilde{\omega}=\omega +\delta $ with $\delta =p_{{\rm F}
}^{2}[\gamma (B_{\Vert 1})-\gamma (B_{\Vert 2})]/2 $. The field-dependent rate $ \tau _{{\rm d}}^{-1}(B_{\Vert 1},B_{\Vert 2})$ reads
\begin{eqnarray}
\tau _{{\rm d}}^{-1}=\frac{\tau p_{{\rm F}}^{4}}{8}[\gamma
(B_{\Vert 1})-\gamma (B_{\Vert 2})]^{2}+\frac{\zeta ^{2}}{2\tau }p_{{\rm F}
}^{2}[B_{\Vert 1}-B_{\Vert 2}]^{2}  \label{TauD}
\end{eqnarray}
which vanishes for $B_{\Vert 1}=B_{\Vert 2}$.

The first term in Eq.~(\ref{TauD}) comes from a deformation of the Fermi
circle by the magnetic field $B_{\Vert }$. The second term takes into account the field effect on
the scattering of plane waves. Furthermore, Eq.~(\ref{diffusonEq}) contains the
difference between the electron kinetic energies in two measurements of
conductance, $\tilde\omega =E_{{\rm F}}(B_{\Vert 1})-E_{{\rm F}}(B_{\Vert 2})$,
each of them shifted by the magnetic field with respect to the Fermi energy $E_{{\rm F
}}^{0}$ at $B_{\Vert }=0$ of the electron gas with the same sheet density.
The latter fact is important, since, for lateral dots where the electron density
is fixed, one should substitute $\tilde{\omega}=0$.  

The Cooperon equation can be represented in the form
\begin{eqnarray}
\left[ D(-i\nabla -{\bf A}_0)^{2}-i\tilde{\omega}+\tau _{{\rm d}}^{-1}+\tau
_{\rm c}^{-1}\right] {\cal C}=\delta ({\bf x}-{\bf x}^{\prime }).
\label{CooperonEq}
\end{eqnarray}
It contains an additional decay rate $\tau _{{\rm c}}^{-1}(B_{\Vert
1},B_{\Vert 2})$ given as
\begin{eqnarray}
\tau _{{\rm c}}^{-1}=\frac{\tau p_{{\rm F}}^{6}}{8}\left[ \frac{
\beta (B_{\Vert 1})+\beta (B_{\Vert 2})}{2}\right] ^{2}+\frac{\zeta ^{2}p_{
{\rm F}}^{2}}{2\tau }[B_{\Vert 1}+B_{\Vert 2}]^{2}
\end{eqnarray}
which accounts for the dephazing of electrons encircling the same chaotic
trajectory in opposite directions. Thus, the result of lifting time-reversal symmetry by an in-plane magnetic field can be described by
the Cooperon phase-breaking rate   
\begin{eqnarray}
\tau _{B\Vert }^{-1}\equiv \tau _{{\rm c}}^{-1}(B_{\Vert },B_{\Vert })=\frac{
\tau p_{{\rm F}}^{6}}{8}\beta^2 (B_{\Vert })+
\frac{2\zeta ^{2}p_{{\rm F}}^{2}}{\tau }B_{\Vert}^{2}\sim aB_\Vert^6+bB_\Vert^2.  \label{TauB}
\end{eqnarray}
Note the gauge shift in the Cooperon equation~(\ref{CooperonEq}). The `vector potential' ${\bf A}_0=
\frac{3}{2}p_{{\rm F}}^{2}m\beta (B_{\Vert })[\frac{{\bf B}_{\Vert }}{B_{\Vert }}\times 
{\bf e}_{z}]$ results from the following artifact: the cubic
term in the effective electron dispersion not only lifts the $\bp\to-\bp$ inversion
symmetry of the line $E(B_{\Vert },{\bf p})=E_{{\rm F}}(B_{\Vert })$, but
also shifts its geometrical center with respect to the bottom of the 2D
conduction band. However, since for the conductance only electrons with energy $\epsilon=E_{
{\rm F}}$ matter, such a shift can be eliminated by choosing a slightly
modified initial gauge. This can be corrected for now by applying a gauge
transformation ${\cal C}\to e^{i{\bf qx}}{\cal C}$ directly to the
Cooperon. In other words, the phase-coherent transport is only
affected by the $B_{\Vert }$-induced ${\bf p}\rightarrow -{\bf p}$
 asymmetric distortion of the Fermi circle into an oval, but not by a
shift of its center in the momentum space. That is, only the $B_{\Vert }$-dependent anisotropy of the electron wavelength along the Fermi line
affects the interference pattern of current carriers. Note that both terms
in Eq.~(\ref{TauB}) are non-vanishing only in the absence of $z\to-z$ inversion symmetry.\footnote{In fact, the first term requires an asymmetric confining potential whereas for the second term a generic $z$-dependence of the impurity potential is sufficient.}

The phase-breaking rate in Eq.~(\ref{TauB}) can also be used to describe magnetoresistance and universal conductance fluctuations
(UCF) in lateral semiconductor dots. For quantum dots, weak
localization corrections and the variance of UCF
can be represented as 
\[
g_{{\rm WL}}(B_{\Vert })\propto \int (d{x})\;W({\bf x})\,{\cal C}({\bf x},
{\bf x};0)\equiv \langle g(B_{\Vert }) \rangle -\langle
g\rangle _{u},
\]
where the subscript `$u$' stands for unitary, and 
\[
\langle \delta g(B_{\Vert 1})\delta g(B_{\Vert 2}) \rangle
\propto  \int (d{x})(d{x}^{\prime })\;W({\bf x})W({\bf x}^{\prime })\sum_{
{\cal P=D},{\cal C}}|{\cal P}({\bf x},{\bf x}^{\prime };\omega)|^{2},
\]
where $\omega =E_{{\rm F}}(B_{\Vert 1})-E_{{\rm F}}(B_{\Vert 2})$, and $E_{
{\rm F}}(B_{\Vert })$ is the Fermi energy of the 2D electron gas calculated from the
bottom of the conduction band.
The dispersionless weight factors $W({\bf x})$ take care of the particle number
conservation upon diffusion inside the dot~\cite{UCF-ends} and incorporate
the coupling parameters to the leads. In the zero-dimensional limit, both, $g_{
{\rm WL}}$ and $\langle \delta g(B_{\Vert 1})\delta g(B_{\Vert
2})\rangle$, are dominated by the lowest (spatially homogeneous) Cooperon (diffuson)
relaxation mode $\lambda_0$, which is 
determined by the rate  of escape to the reservoirs, $\tau _{{\rm esc}}^{-1}$
$\ll D/L^{2}$ (Thouless energy), and the Cooperon suppression by time-reversal symmetry
breaking described by $\tau _{B\Vert }^{-1}$. The latter expresses the efficient reduction of the fundamental symmetry of
the system from orthogonal ($o$) to unitary ($u$). In the absence of spin-orbit
effects, the parameters $\tau _{{\rm esc}}^{-1}$ and $\tau _{B\Vert }^{-1}$ describe the
value of WL corrections as well as  the variance of UCF, $ \langle
\delta g^{2}(B_{\Vert })\rangle $, as compared to their nominal
values, $g_{{\rm WL}}(0)\equiv \langle g \rangle _{o}-
\langle g\rangle_{u}$ and $\langle\delta g^{2}\rangle_{u}$: 
\begin{eqnarray}
g_{{\rm WL}}(B_{\Vert }) &=&g_{{\rm WL}}(0)\,\frac1{1+\tau _{\rm 
esc}/\tau_B},\\
\langle\delta g^{2}(B_{\Vert })\rangle  &=&\langle
\delta g^{2}\rangle_{u}\left(1+\frac1{(1+\tau
_{\rm esc}/\tau _B)^2}\right).
\end{eqnarray}
Thus, these two parameters can be studied from the UCF fingerprints measured by
changing the shape of a dot in multi-gate devices or by
varying the Fermi energy in back-gated dots.

\begin{figure}[h]
\begin{center}
\includegraphics[scale=0.35]{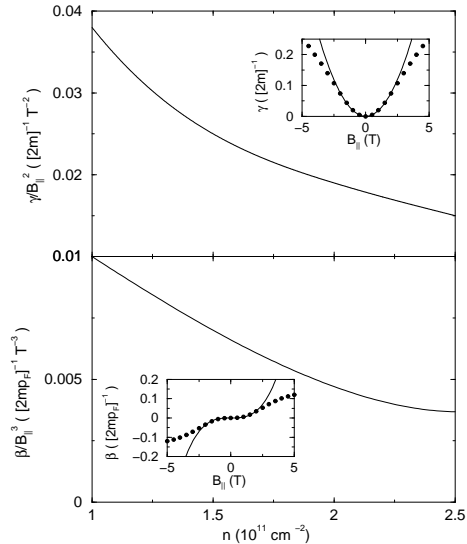}
\caption{Calculated dependence of the parameters
    $\gamma$ and $\beta$ on the sheet density of 2D electrons. The
    insets show the effect of $B_\Vert$ on the symmetric, 
 $[E(\bp)+E(-\bp)]/2$, and anti-symmetric, $[E(\bp)-E(-\bp)]/2$, parts of the 2D 
 electron dispersion in a
 broader range of $B_\Vert$, where a perturbative expansion is not 
 applicable.}
\label{fig-jungwirth}
 \end{center}
 \end{figure}

In ballistic billiards or
heterostructures with a $z$-independent impurity scattering potential, the in-plane magnetic field effects are governed by the unusual $B_{\Vert }^{6}$-dependence of $\tau_{B\Vert}^{-1}$~\cite{MAA01,Thesis,FaJu01} that we attribute to the effect of a cubic term in the 2D electron dispersion
generated by the field. This should give
rise to a relatively sharp crossover between `flat' regions related to
regimes with orthogonal and unitary symmetry, respectively. To get an idea about the relevance
of this effect for the large-area ($\sim$8$\mu $m$^{2}$) quantum dots studied
by Folk {\it et al}~\cite{MarcusSpin}, quantitative estimates for the parameters $\gamma (B_\Vert)$ and $\beta(B_\Vert)$  can be 
obtained by evaluating the electron dispersion within a fully
self-consistent numerical method~\cite{FaJu01}. Using the nominal growth 
parameters
of the Al$_{.34}$Ga$_{.66}$As/GaAs sample studied in~\cite{MarcusSpin}, the 
dependences of $\gamma $ and $\beta $ on the in-plane magnetic field for an
electron sheet density of 2$\times $10$^{11}$~cm$^{-2}$ shown in
the insets
of Fig.~\ref{fig-jungwirth} a) and b) were found. At low fields, $\gamma \sim 
B_{\parallel }^{2}$ and $\beta \sim B_{\parallel }^{3}$, as anticipated in the 
perturbation
theory treatment. The proportionality coefficients are plotted in Fig.~\ref
{fig-jungwirth} a) and b) versus the electron sheet density. Both, the 
effective
mass renormalization in the quadratic term of the energy dispersion and the
time-reversal symmetry breaking cubic term are larger at lower 2D electron
gas densities, due to the weaker confining electric field, i.e., the
increased width $d$ of the potential well. From this analysis, one estimates the field necessary to suppress the weak localization effect completely as $B_{\Vert
}=0.6\div 0.8$T, which is in agreement with the observed complete suppression of WL corrections beyond $B_{\Vert }\approx0.6$T in experiment~\cite{Marcus-private}. 


\subsection{In-plane magnetoresistance in inversion layers with several
filled subbands and in diffusive metallic films. }

\label{sub-multisub} 

\begin{figure}[h]
\begin{center}
\includegraphics[scale=0.25]{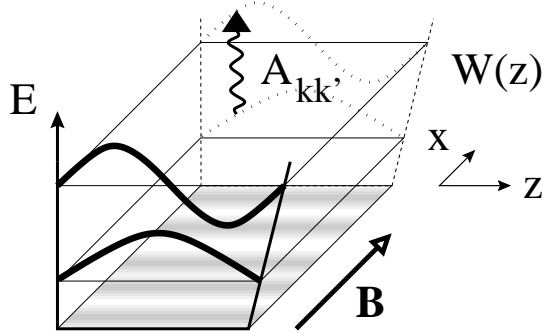}
\caption{Schematic picture of the quantum
    well. Two exemplary subband wavefunctions are shown. The spatial
    profile of the impurity potential is sketched on the bottom of the 
    well.}
\label{fig-sub}
\end{center}
\end{figure}

If the width of the quantum well exceeds the Fermi wavelength of the electrons, several subbands become occupied -- as shown in  Fig.~\ref
{fig-sub} -- and, thus,
contribute to the lateral transport. As for the one-subband case discussed above, the effect of an in-plane magnetic field
depends sensitively on the microscopic structure of the wavefunctions in the $z$-direction
perpendicular to the plane, that determines possible
couplings between the subbands. In particular, in the absence of $z$-dependent impurity scattering, the
influence of $B_{\Vert }$ is different for quantum wells with a symmetric, $
W(z)=W(-z)$, and asymmetric, $W(z)\neq W(-z)$, confinement potential due to the Berry-Robnik symmetry effect. 

In the absence of an external magnetic field ($B_\Vert=0$), the subbands are uncoupled and, thus, each subband contributes separately to the conductivity. The weak localization correction for a film
with $M$ occupied subbands is, therefore, given as
\[
g_{{\rm WL}}=M\frac{e^{2}}{\pi h}\ln ({\tau /\tau _{\phi }}),
\]
where phase coherent propagation in each subband is limited by the inelastic decoherence time $\tau _{\phi }^{-1}\propto T^{p}$.

An external magnetic field now plays two
complementary roles: a) it breaks time-reversal (${\cal T}$) symmetry and b) it couples the different subbands (i.e., if the random disorder potential is $z$-independent, it represents
the only possible coupling between subbands). In fact, these
two roles are linked: if the subbands remained completely decoupled, ${\cal T}$-invariance of the 2D motion in each
of the $M$ subbands would be preserved
because the vector potential of the parallel field can be gauged out in each
particular subband. The field-induced subband mixing results in
a spectrum of effective dephazing rates $1/\tau _{B}^{(k)}$ (where $k=0,\dots,M-1$) with the following properties: at least $M-1$ dephazing rates are quadratic in the magnetic field, $1/\tau
_{B}^{(k>0)}\propto B_{\Vert }^{2}$. The remaining  $1/\tau
_{B}^{(0)}$ is exactly
equal to zero for a ${\cal P}_{z}$-symmetric system whereas it acquires a quadratic field dependence as well if the symmetry is broken (due to an asymmetric confining potential or $z$-dependent impurity scattering).  As a result, the
field and temperature dependence of the 
magnetoconductivity are different for a symmetric vs asymmetric
confinement potential,
\begin{eqnarray}
g_{\rm WL}(B_{\Vert },T)=\frac{e^{2}}{\pi h}\left\{ 
\begin{array}{cc}
p\ln T+2(M-1)\ln B_{\Vert } & \quad {\rm for}\enspace W(z)=W(-z), \\ 
2M\ln B_{\Vert } & \quad {\rm for}\enspace W(z)\neq W(-z).
\end{array}
\right.  \label{ManySubbands}
\end{eqnarray}
The derivation of the above weak localization magnetoconductivity formula is most conveniently performed within the non-linear sigma-model formulation that will be presented in chapter~\ref{sec-SUSY}.

For strong inter-subband impurity scattering, i.e., when the inter-subband scattering rate exceeds the subband splitting
at the Fermi level, the weak localization suppression by the in-plane
magnetic field happens in the same way as in thin metallic films with
diffusive transverse motion, $d\gg \ell$. In a film with diffusive transverse
motion, the electron encircles a random flux of order $\Phi\sim B_{\Vert }d^{2}$
during its diffusive transverse flight, where the time of flight is given as $\tau
_{z}\sim d^{2}/D$. While diffusing along the wire for a longer time, $t\gg
\tau _{z}$, the electron encircles a random flux with the variance 
\[
\langle \Phi ^{2}\rangle \sim \frac{t}{\tau _{z}}\times (B_{\Vert
}d^{2})^{2}=tD(B_{\Vert }d)^{2},
\]
which results in a field-induced phase breaking at the rate~\cite{AlAr81} 
\begin{eqnarray}
\tau _{B}^{-1}\sim D(B_{\Vert }d)^{2}.  \label{DiffFilmEstimate}
\end{eqnarray}
Note that the same geometrical argument is applicable to a thin diffusive
wire with cross sectional dimensions $\sim d$ in a magnetic field applied along
the wire axis

\begin{figure}[h]
\begin{center}
     \includegraphics[scale=0.25]{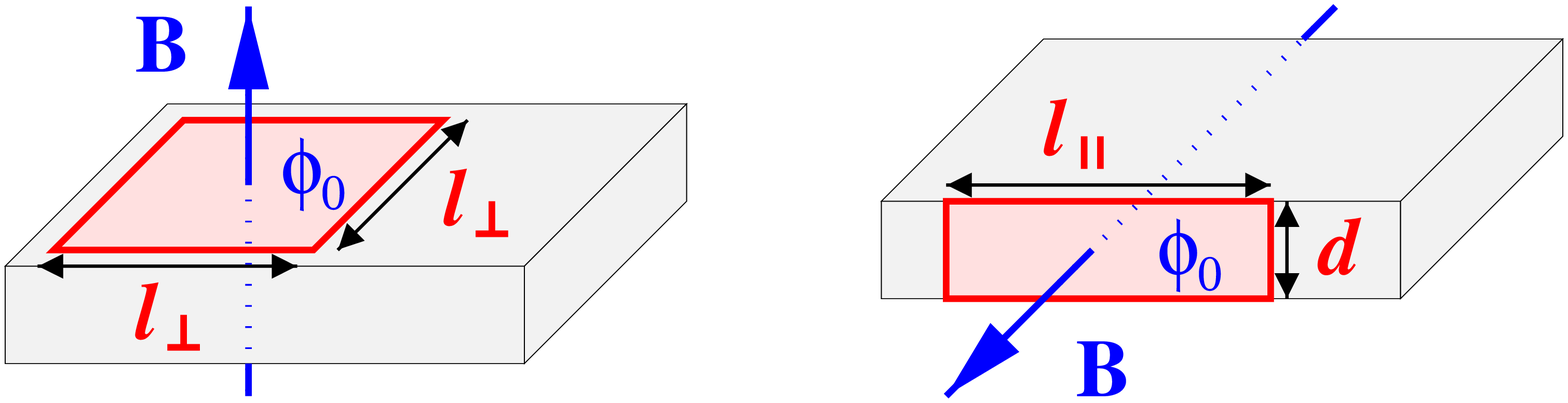}
\caption{Relevant areas for
    perpendicular and parallel magnetic fields. This admits for
    estimating field dependence of the respective phase breaking rates,
    $1/\tau_{B_\perp}\sim 1/l_\perp^2\sim B_\perp$ and
    $1/\tau_{B_\parallel}\sim 1/l_\parallel^2\sim B_\parallel^2$.}
\label{fig-B_pp}
\end{center}
\end{figure}

Quantitatively, $\tau _{B}$ can be found from an
analysis of the lowest eigenvalue of the Cooperon propagator. The Cooperon
propagator is defined by the (diffusion) equation 
\[
\left\{ D(i\nabla_\br -2{\bf A}(\br))^{2}+\frac{1}{\tau _{\phi }}
\right\} {\cal C}({\bf r},{\bf r}^{\prime })=\delta ({\bf r}-{\bf r}^{\prime
})
\]
with the boundary condition 
\[
\left( \nabla _{\bn}+2i{\bf A}\cdot {\bf n}\right) {\cal C}
=0.
\]
Choosing the gauge ${\bf A}=B_{\Vert }z{\bf e}_{y}$, one can separate
variables in the eigenfunction equation for the Cooperon modes, 
\[
-D\left[ \frac{\partial ^{2}}{\partial z^{2}}-Q_{x}^2-\left(Q_{y}-2{B_{\Vert
}z}\right) ^{2}\right] \phi _{k,{\bf Q}}(z)=\Omega _{k,{\bf Q}
}\phi _{k,{\bf Q}}(z),
\]
and find the lowest eigenvalue $\Omega _{0,{\bf Q}}=DQ^{2}+\tau _{B}^{-1}$
within perturbation theory. This yields the magnetic decoherence rate~\cite{AlAr81}  
\[
\tau _{B}^{-1}=\frac{1}{3}D(B_{\Vert }d)^{2}
\]
which can be substituted into the weak localization formula for a diffusive
film. A similar calculation gives 
$\tau _{B}^{-1}=D(B_{\Vert }d)^{2}/8$
for a circular diffusive  wire with diameter $d$.


\subsection{In-plane magnetoresistance in pure metallic films 
and quantum magnetoresistance in quasi-ballistic wires 
with rough edges.}

\label{ssub-Quasiball} 

For a clean metallic film or wire, where the dominant scattering processes
involve surface or edge roughness, the estimate of $\tau_B$ given in Eq.~(\ref{DiffFilmEstimate}) is incorrect due to the exact geometrical phase
cancellation described in~\cite{DuKh84}. Thus, a magnetic field $B_{\Vert }$
cannot affect the interference pattern, i.e.,  neither WL magnetoresistance nor
magneto-fingerprints are to be expected in these systems at any orientation of the
magnetic field.  As a result, magnetic field effects acquire several unusual features. 

Here, we consider a quasi-ballistic wire with width $d$
and length $L$. In such a wire, a diffusive path may contain
ballistic segments  with length $\eta =d/\sin \alpha $  much longer
than the  typical length scale $d$, crossing the wire at the small angle $\alpha$. These segments appear with the
probability 
\[
\rho (\eta )\sim \pi ^{-1}\left| \frac{d\eta (\alpha )}{d\alpha }\right|
^{-1}\sim \eta ^{-2},
\]
for $L>\eta \gg d\gg \lambda _{{\rm F}}$. Thus, the mean free path $\ell$ in such a wire exceeds the width of the wire by a large logarithmic factor, $\ell\sim \langle
\eta \rangle \sim d\,\ln (L/d)$.  

Quantum in-plane magnetoresistance in such a
wire may appear only due to a
curving of the longest ballistic paths by the field itself. The effect of this
curving leads to a switching of the longest segments between the edges of the wire (or surfaces of a film) and, thus, to a sudden suppression of the interference contribution from paths including these longest ballistic flights. However, the
change in the weak localization part of the conductivity in such a situation is
overcome by the change in the classical part of the conductivity: due to the cyclotron curving of
free electron trajectories in combination with the finite width of the
wire, the
longest ballistic flights are suppressed, i.e., $\eta <\sqrt{r_{c}d}$ -- yielding a mean free path
$\ell(B_{\Vert })\sim \frac{1}{2}d\ln (r_{c}/d)$, where $r_{c}=p_{{\rm F}
}/B_{\Vert }$. Thus, the shortening of the longest ballistic
flights happens simultaneously with their involvement into phase accumulation,
immediately bringing in a large time-irreversible phase of 
order of the total flux penetrating through the entire sample. On the other
hand, paths that encircle a large flux can participate in the formation
of a random interference pattern. Therefore, starting from the field value $B_{{\rm defl}}=\phi _{0}d/(L^{2}\lambda _{{\rm F}})$ (implying $\sqrt{r_{c}d}
<L$) --  that coincides with the set in of a classical (positive)
magnetoresistance -- the quantum conductance should acquire a universal random
dependence on the field with a characteristic scale related to
a flux change of order of the flux quantum.  By the same reason, the
magnetic field effect on the localization properties of quasi-ballistic wires is
opposite to the commonly expected crossover between orthogonal and unitary
symmetry classes, i.e., the magnetic field tends to shorten the localization length in
a quasi-ballistic wire. 

These qualitative expectations  have been verified numerically in Ref.~\cite
{numFal}. The results of Ref.~\cite{numFal} are based upon the numerical
solution of a two-dimensional Anderson Hamiltonian on a square lattice, $
H=\sum_{i}|i\rangle \epsilon _{i}\langle i|-V\sum_{\langle ij\rangle
}|i\rangle \langle j|$, where $\langle ij\rangle $ denotes nearest neighbor
sites $i$ and $j$. The structure considered consists of two ideal leads
attached to a scattering region that is $W$ sites wide and $L$ sites long
(all lengths are in units of the lattice constant $a$). For simulating bulk
disorder, the energies $\epsilon _{i}$ in the scattering region are taken uniformly
from the interval $-U/2<\epsilon _{i}-\epsilon _{0}<U/2$, where $U$ is the
disorder strength. For sites on the boundary, $\epsilon _{i}=\epsilon
_{0}+\epsilon _{B}$ with $\epsilon _{B}=10^{4}$. The rough structure of the
boundary was generated by having an equal probability of either 0, 1 or 2
sites at each edge with on-site potential $\epsilon _{0}+\epsilon _{B}$. To
simulate the effect of a magnetic field, a Peierls phase factor has been
incorporated into $V$ in the scattering region.

\begin{figure}[h]
\begin{center}
\includegraphics[scale=0.35]{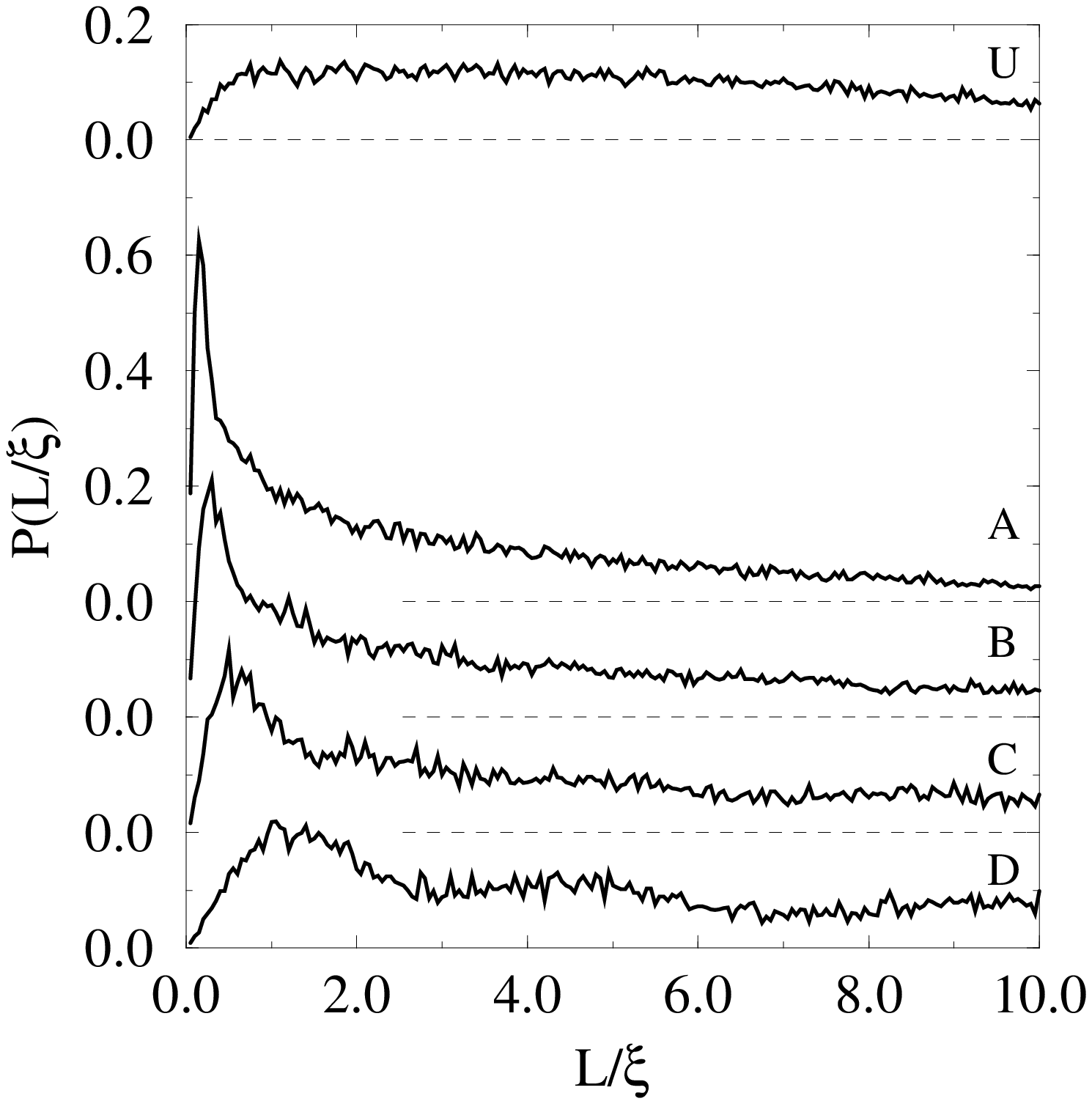}
\hspace*{0.5cm}
\includegraphics[scale=0.35]{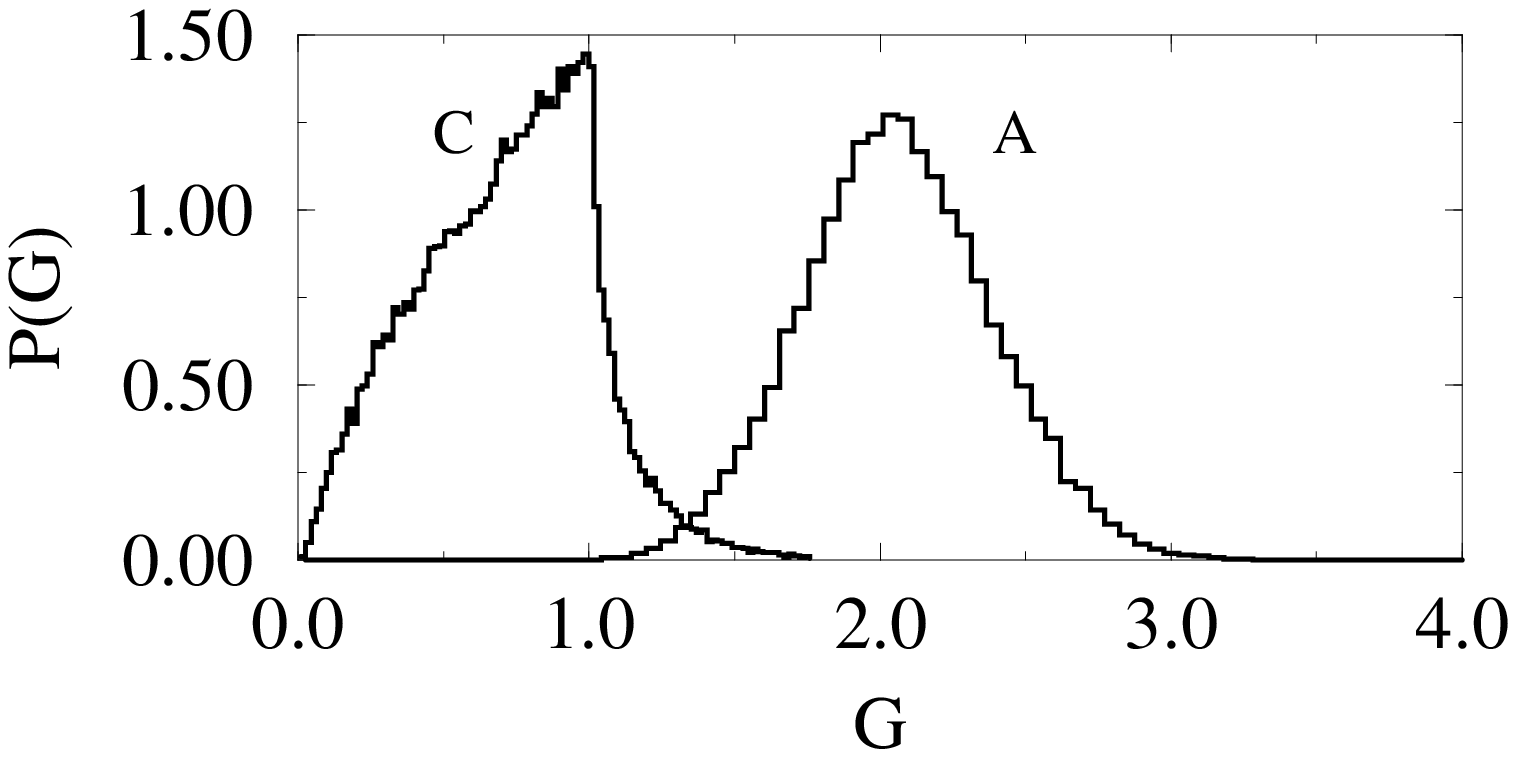}
\hspace*{0.5cm}
\includegraphics[scale=0.35]{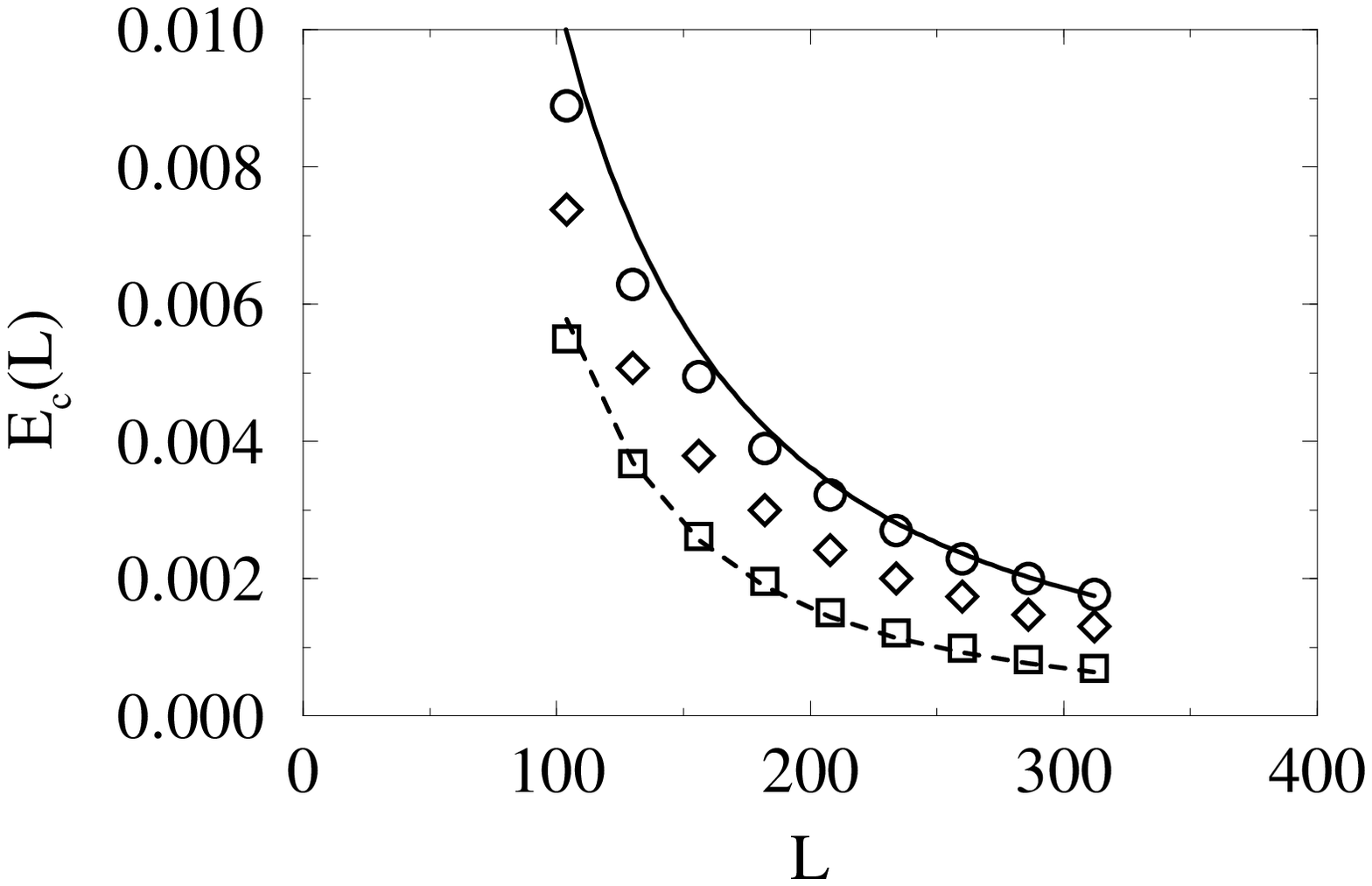}
\caption{(a) Distribution function $P(L/\xi)$ for 4 
 quasi-ballistic
 samples with $W=$15 and $L=$52(A), 104(B), 208(C) and 416(D) and for a
 wire (U) with on-site bulk disorder $U=2.0$ (nominal mean free path $\ell\approx 
 8.5$, $W=15$ and $L= 52$). (b) Distribution function $P(g)$ for the two
 quasi-ballistic structures A and C. (c) $E_{\rm Th}(L)$ for 
 quasi-ballistic
 wires with $\epsilon_0\in [1.5,1.7]$ (circles) and 
 $\epsilon_0\in [1.0,1.2]$ (diamonds) and for a wire from the $U$ 
 series
 (squares). Also shown is the analytical result (cf.~Eq.~(\ref{eq2})
 with $hv_{\rm F}=12.3$ and $W=9.3$ (solid curve). The dashed curve corresponds the ergodic law $E_{\rm 
 Th}\sim hD/L^{2}$ with $hD=62.5$.}
\label{fig-98_1}
\end{center}
 \end{figure}

The effect of L\'{e}vy flights has been identified in
the numerical results in several ways. One was to study the distribution of
transmission coefficients through the wire and the distribution $P(\xi ^{-1})
$ of the corresponding Lyapunov exponents $\xi ^{-1}$. This is shown in Fig.~\ref{fig-98_1}(a) (taken from Ref.~\cite{numFal}) for 4 series of
quasi-ballistic wires (A-D) 
as well as a series of
samples with bulk `defects' (U). As pointed out by Tesanovich {\it et al}~\cite{Tesanovich} and verified numerically in Ref.~\cite{MacKinnon}, the
length of L\'{e}vy flights in quantum systems is limited.
The limitation is due to the fact that the uncertainty in the transverse momentum, $\delta k_{\perp
}\sim 1/d$, in a wire with a finite width sets a quantum limit to the
angles $\alpha\sim d/\eta$ which can be assigned to a classically defined ballistic
segment, namely $\alpha >\delta \alpha \sim \delta k_{\perp }/k_{{\rm F}
}\sim \lambda_{\rm F} /d$. This sets the cut-off $\eta _{{\rm max}
}=d^{2}/\lambda _{{\rm F}}$ and entails a finite localization length $L_{c}\sim \eta _{{\rm max}}$. Samples from the series A and B meet
the criterion $L<L_{c}$ which manifests itself in the distribution of Lyapunov exponents by a pronounced peak at small $\xi ^{-1}$ corresponding to
eigenvalues $T_{n}\approx 1$. This should be compared to the plateau-like distribution~\cite{Chalker} obtained in a sample from the series U using the same
numerical procedure. Note the finite width of
the ballistic peak in $P(\xi ^{-1})$ and that $P(0)=0$. The enhanced density
of small $\xi ^{-1}$ can be identified even in samples from the series C and
D with $L\approx L_{{c}}$. In these samples, $P(\xi ^{-1})$ starts to show a periodic
modulation specific to the localized regime, where the spectrum of Lyapunov
exponents tends to crystallize~\cite{Pichard,Frahm1}.

The distribution of eigenvalues of the transmission matrix results in a
finite-width distribution of conductances and, hence, conductance
fluctuations. The statistics of conductance fluctuations, $P(g)$, for wires
with edge disorder is shown in Fig.~\ref{fig-98_1}(b): the results of numerical simulations~\cite{numFal} are plotted for series A, where $\langle g\rangle \approx 2.0$ and the distribution is almost Gaussian, and series C, where $
\langle g\rangle \approx \,0.71$. (The conductance is expressed in
quantum units $e^2/h$.) The distribution function $P(g)$ is the result of the analysis of various
realizations of disorder. When calculating correlation functions, 
there is an additional averaging over energy, since the energy dependence of conductance
fluctuations is random on the scale of the Thouless energy $E_{
{\rm Th}}$. The latter can be determined from the wire conductance as $
g\sim E_{{\rm Th}}(L)/\Delta (L)$, where $\Delta$ is the mean level
spacing in the wire. Taking into account the logarithmic multiplier in
the dependence of the classical conductance on sample length due to the L\'{e}vy
flights, one obtains
\begin{eqnarray}
g\sim \frac{e^{2}}{h}\frac{d^{2}}{L\lambda _{{\rm F}}}\ln (\frac{L}{1.7d}).
\label{eq2}
\end{eqnarray}
This result for the wire conductance implies an anomalous scaling of the Thouless energy, $E_{
{\rm Th}}=\pi dv_{{\rm F}}L^{-2}\ln (\frac{L}{1.7d})$, with the length 
of the wire. In Fig.~\ref{fig-98_1}(c), it is shown that, indeed, the Levy flights manifest themselves in the
correlation energy of UCF in the quasi-ballistic regime. For comparison, it is 
demonstrated that the results for bulk-disordered samples (U) coincide with the
standard scaling law $E_{{\rm Th}}(L)\sim L^{-2}$. 

\begin{figure}[h]
\begin{center}
 \includegraphics[scale=0.35]{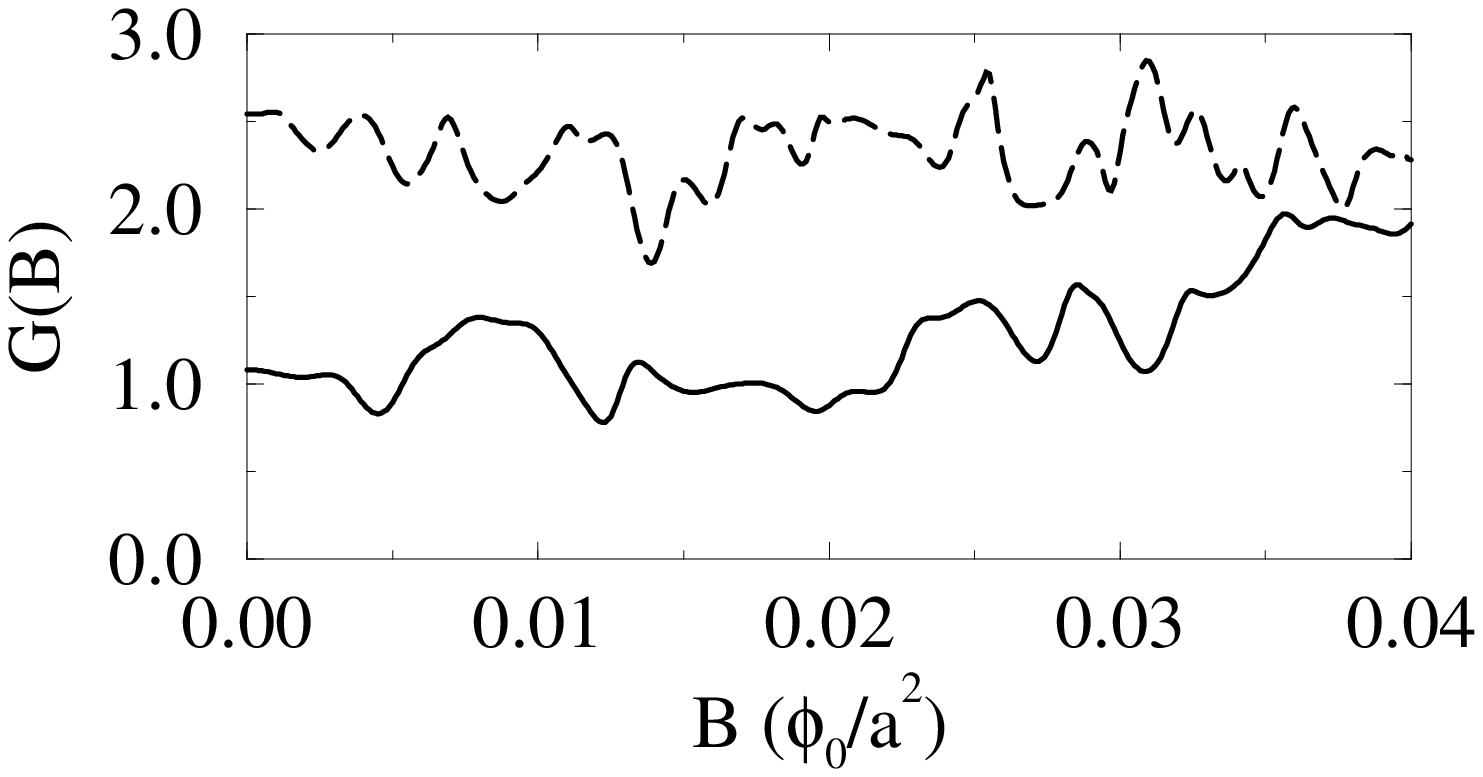}
\hspace*{1cm}
\includegraphics[scale=0.35]{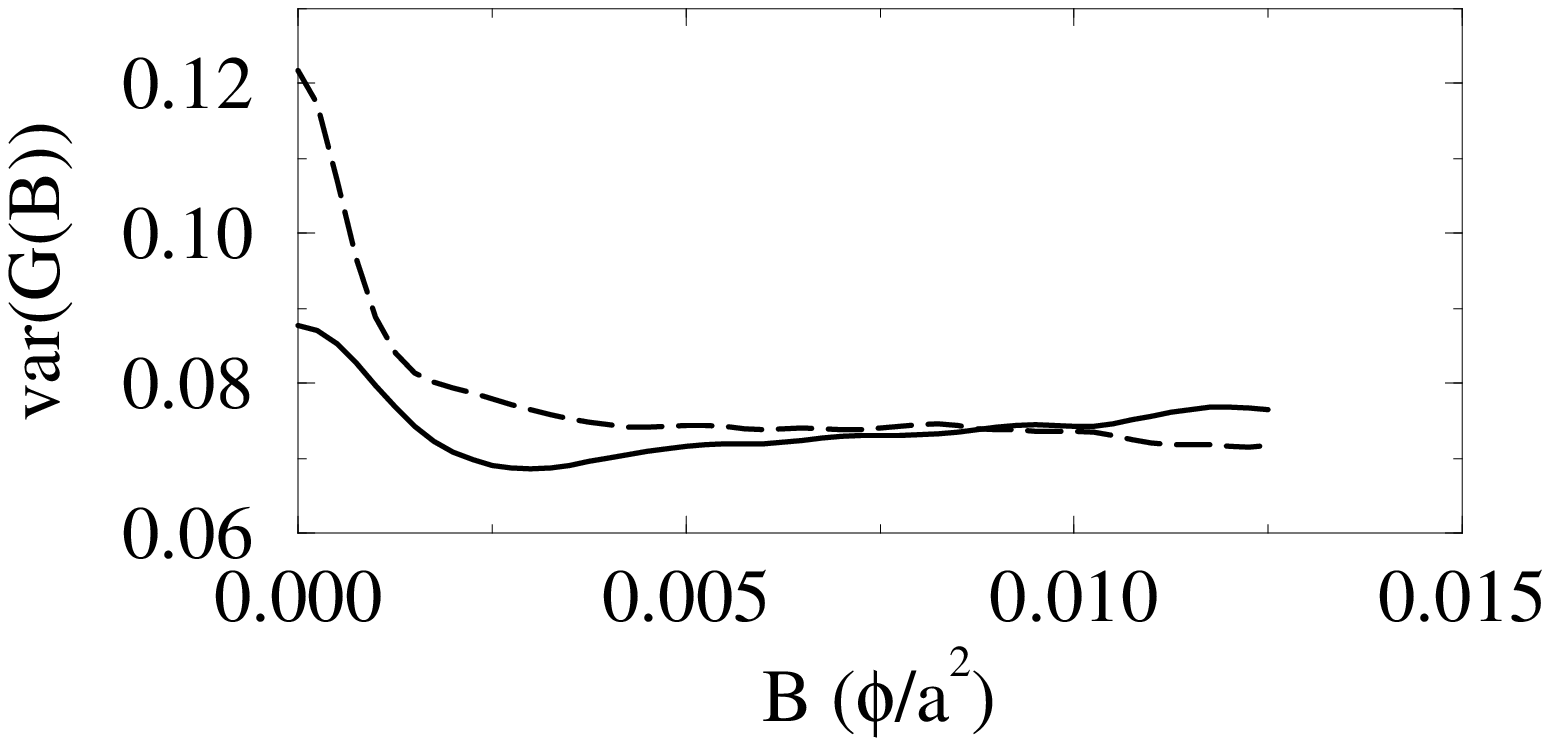}
\caption{(a) Conductance for a quasi-ballistic wire (solid 
 curve) from the B series and a disordered wire (dashed curve) with mean free path $\ell\approx 16$.
 (b) Variance ${\rm var}(g(B))$ for quasi-ballistic wires from the $B$ series
 and disordered wires.}
\label{fig-98_3} 
\end{center}
\end{figure}

Coming now to the magnetic field effects. The expected deflection of L\'{e}vy flights at $B>B_{{\rm defl}}$ entailing a
break of geometrical flux cancellation is observable in Fig.~\ref{fig-98_3}
as pronounced magnetoconductance fluctuations $\delta g(B)=g(B)-\langle
g\rangle $ beginning together with a gradual decrease of the average
conductance value -- which is a purely classical effect. Their variance corresponds to
the usual UCF value, and the auto-correlation function as a function of
magnetic field gives the correlation field $B_{c}$ corresponding to
a change of magnetic flux through the sample area of order $2.5\phi _{0}$ as compared to $1.5\phi _{0}$ in the
bulk-disordered case. This seems to explain an earlier experimental
observation of UCF in quasi-ballistic semiconductor wires~\cite{Houten}.

\begin{figure}[h]
\begin{center}
\includegraphics[scale=0.35]{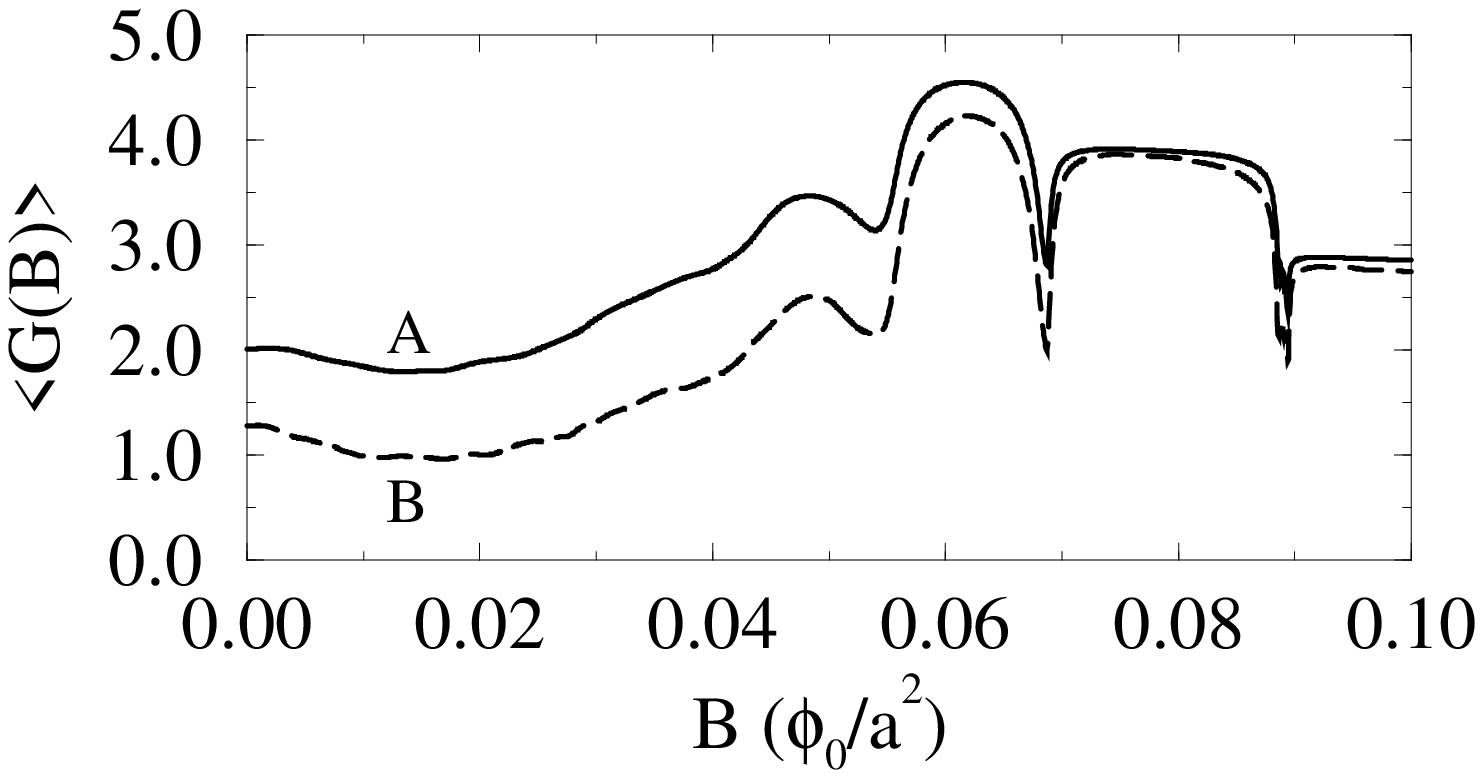}
\hspace*{0.5cm}
\includegraphics[scale=0.35]{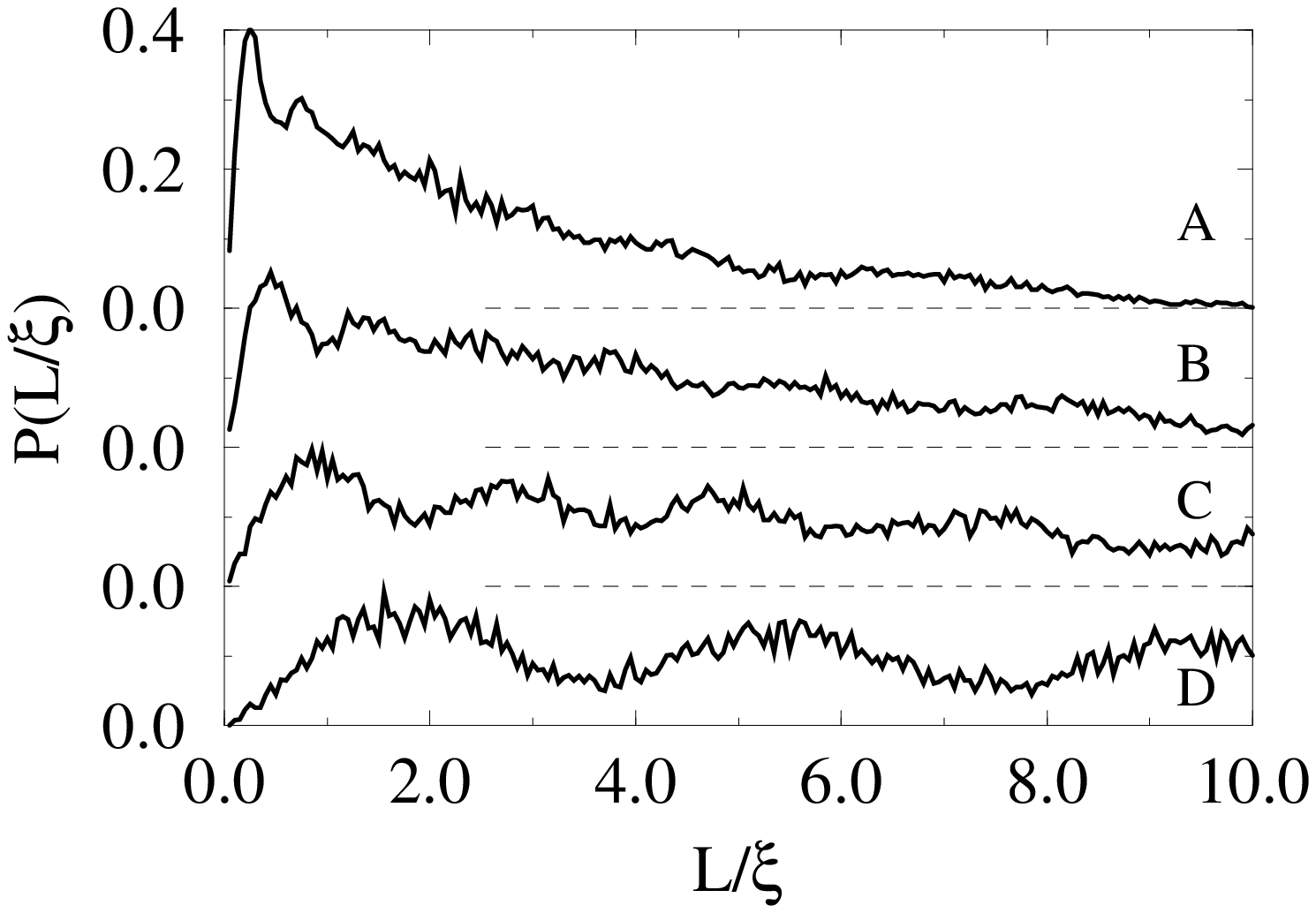}
\hspace*{0.5cm}
\includegraphics[scale=0.35]{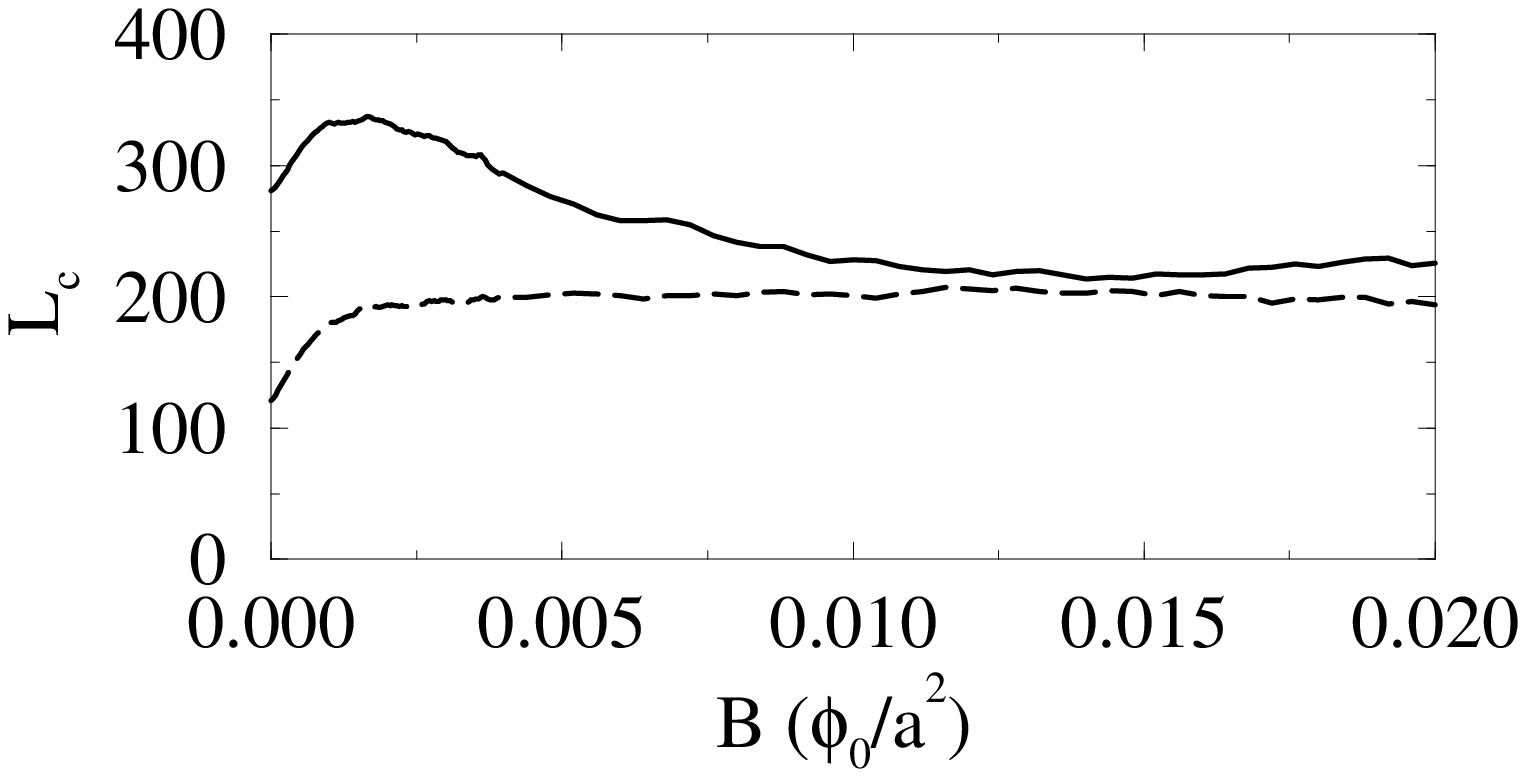}
\caption{(a) The ensemble averaged conductance as a function 
 of magnetic
 field for wires from the A and B series. (b) The same curves (A-D) as in
 Fig.~\ref{fig-98_1}(b), but at finite magnetic field $B=0.02\phi_0/a^2$. (c) The
 localization length for quasi-ballistic wires (solid curve) and disordered wires (dashed
 curve) with mean free path $\ell\approx 8.5$.}
\label{fig-98_2}
\end{center}
 \end{figure}

Furthermore, Fig.~\ref{fig-98_2} shows the tendency of the localization
length in a quasi-ballistic wire to increase with the magnetic field -- as opposed to the
behavior of $L_{{c}}(B)$ in systems with diffusive
transverse motion~\cite{Efetov,Pichard}. The relevant field scale at which
the localization length in quasi-ballistic wires
starts to shorten, $B_c\sim\phi _{0}\lambda _{{\rm F}}/W^{3}$, can be obtained from $B_{{\rm 
defl}}$ by replacing the sample length $L$ with $L_{{c}}$. Formally, this yields
a similar field scale as the one leading to a crossover in $L_{{c}}$ between two symmetry classes due Aharonov-Bohm type phase effects. Thus, the latter
is hindered by the deflection effect and does not appear as an intermediate
crossover regime. 



\section{In-plane magnetoresistance due to spin-orbit coupling}
\label{sec-Spin-orbit}

\subsection{Spin-orbit coupling in 2D heterostructures and quantum dots}

Coupling between the electron spin and its orbital motion is a relativistic
effect inherent to many metallic and semiconductor systems. For example, in
zinc-blend type III-V semiconductors, such as GaAs, which have no inversion
symmetry in the unit cell, the 3D bulk dispersion of conduction band
electrons contains terms that are cubic in the electron momentum and linear in its spin
operator~\cite{Dresselhaus}. This generates a linear spin-orbit coupling for
electrons confined to the 2D plane in a heterostructure or a quantum well~\cite{BychRashba,Rossler}, 
\begin{eqnarray}
H_{0}&=&\frac{{\bf p}^{2}}{2m}+\frac{\alpha _{{\sc br}}}{2m}[{\bf p\times e}_{z}]
\cdot{\underline{\sigma}}+\frac{\alpha _{{\rm cr}}}{2m}(p_{x}{\sigma _{x}}{}-p_{y}{\sigma _{y}}{})=\frac{{\bf p}^{2}}{2m}+\frac{v_{\rm F}}{2\lambda _{{\rm 
SO}}}{\bf n}_{\bf p}\cdot\underline{\sigma }, 
\label{H-SO}
\end{eqnarray}
where ${\bf n}_{\bf p}$ characterizes
the direction and frequency
of spin precession 
(of an electron moving with momentum $\bp$)
measured in units of $v_{\rm F}/\lambda_{\rm SO}$, where
$1/\lambda_{\rm SO}=\sqrt{\alpha_{{\sc br}}^2+\alpha_{{\rm cr}}^2}$.

The form of the spin-orbit (SO) coupling in Eq.~(\ref{H-SO}) is specified for a GaAs
heterostructure grown along the (001) plane which has the symmetry of a
square lattice without inversion symmetry within the unit cell. It includes 
two possible combinations of spin and momentum operators that are
invariant under the symmetry transformation corresponding to such a lattice
symmetry, parameterized by two constants, $\alpha_{\sc br}$ (the so-called Bychkov-Rashba term)
and $\alpha_{\rm cr}$ (the crystalline anisotropy term).

For a particle moving diffusively in an infinite heterostructure, spin-orbit
coupling provides an efficient mechanism for spin relaxation. 
In the frame of the moving electron, its spin undergoes a precession
with angular frequency 
$\underline{\Omega} \sim {\bf n}_{\bf p}v_{\rm F}/\lambda _{{\rm SO}}$ which
randomly changes  direction each time the particle changes its direction of
propagation due to scattering at
impurities. Using the parameterization of the SO coupling by
a spin-orbit length $\lambda _{{\rm SO}}$, as in Eq.~(\ref{H-SO}), the precession
angle acquired between successive scattering events can be estimated as $\delta
\varphi_{\rm so}\sim \tau v_{\rm F}/\lambda _{{\rm SO}}=\ell/\lambda _{{\rm SO}}$. For weak SO
coupling, $\ell/\lambda _{{\rm SO}}\ll 1$, the electron spin precession can be
viewed as a random walk of the electron spin polarization vector around the
unit sphere with a typical step size $\delta\varphi_{\rm so}$. This
causes a loss of spin memory at the (Dyakonov-Perel) rate~\cite{DyakonovPerel} 
\begin{eqnarray}
\tau _{{\rm DP}}^{-1}\sim D/\lambda _{{\rm SO}}^{2}.  
\label{DP}
\end{eqnarray}
In a 2D electron gas, this parameter limits the diffusion time in the triplet Cooperon channel (whereas the singlet Cooperon channel is unaffected). Therefore, it suppresses the triplet contribution to the weak localization correction
formula and controls the crossover between the weak localization (at weak SO coupling) and weak
anti-localization  (at strong SO coupling) regimes in the quantum correction to the conductivity~\cite{HiLa80,LyandaGeller1}.

When the electron motion within the 2D plane is also confined laterally, the use
of the so called Dyakonov-Perel formula, Eq.~(\ref{DP}), dramatically
overestimates the rate of spin relaxation. In particular, if the electron
motion is confined to a single 1D channel (i.e., a quantum wire), no real
spin relaxation takes place -- despite the electron spin precession. In this case, the spin precession is a reversible process that
rotates the initial electron spin to the same definite state at each
point along the wire, no matter how many times the electron moves
forwards and backwards. Mathematically, the above-mentioned property of
the SO coupling becomes transparent in the following way: Consider a 
one-dimensional
electron moving, say, along the $x$-direction. Applying the unitary
transformation, 
\[
U=\exp \left( \frac{ix}{2\lambda _{{\rm SO}}}{\bf n}({\bf 
l}_{x})\cdot\underline{\sigma }\right) ,
\]
where ${\bf l}_x$ is a unit vector along the wire axis,
the SO coupling is eliminated completely from the electron dispersion
represented in a rotated spin-coordinate frame~\cite{Mathur,Aronov}.
Therefore, it does not affect any charge transport properties.
By contrast, the Dyakonov-Perel relaxation of the spin of a 2D electron
(with the rate described in Eq.~(\ref{DP})) is the result of the
diversity of paths the electron may use to move between
two points in the sample. 

A small quantum dot with dimensions $L\ll \lambda _{{\rm SO
}}$ represents an intermediate situation between the 1D and the 2D
system as described above. Here,
we shall consider a dot coupled to metallic leads via two contacts, left ($l$) and right ($
r$), each with $N_{l,r}\gtrsim 1$ open orbital channels. Then the escape
rate~\cite{Beenakker} from the dot into the leads is given as 
\[
\gamma\equiv\tau_{\rm esc}^{-1} =(N_{l}+N_{r})\Delta /(2\pi),
\]
where $\Delta =2\pi/(m{\cal A})$ is the mean level spacing (${\cal A}$ area of the dot). 

Although there is no unitary transformation that eliminates the SO coupling completely by changing to a locally rotated
spin coordinate system, the linear term in the small SO coupling constant can
be cancelled by a proper spin-dependent gauge transformation~\cite{AleinerFalko}, thus, generating only contributions to the free electron
Hamiltonian of order $\lambda _{{\rm SO}}^{-2}$ (or higher).
As a result, the influence of spin-orbit coupling on the transport
characteristics of a small dot is suppressed as compared to
the infinite 2D system with similar material parameters.

However, the application of an in-plane magnetic field changes the situation. The
transformation to the locally rotated  spin-coordinate frame turns an initially
constant external magnetic field, ${\bf B}=B\,{\bf l}$, into an inhomogeneous Zeeman field
that accelerates the loss of spin memory of an electron passing through the
dot. Below, this effect will be described quantitatively.

For the sake of convenience, we choose the coordinate system ($x_{1},x_{2}$)
with axes along the crystallographic directions ${\bf e}_{1}=[110]$ and ${\bf e}
_{2}=[1\bar{1}0]$ such that we can exploit the $C_{2v}$ lattice symmetry of
the (001) plane of GaAs. Then, the single-particle Hamiltonian in the presence of an in-plane magnetic
field  ${\bf B}=B\,{\bf l}_\parallel$ with  ${\bf l}_\parallel=(l_{1},l_{2},0)$ can be rewritten as 
\begin{eqnarray}
H_{0}=\frac{1}{2m}\left[ \left( p_{1}-\frac{\sigma _{2}}{2\lambda _{1}}
\right) ^{2}+\left( p_{2}+\frac{ \sigma _{1}}{2\lambda _{2}}\right) ^{2}
\right] +\frac{\epsilon _{{\rm Z}}}{2}\,{\bf l}_\parallel\cdot\underline{\sigma },
\label{H0}
\end{eqnarray}
where $\lambda _{1,2}=1/(\alpha_{\sc br} \pm \alpha_{\rm cr})$ characterize the
length scales associated with the strength of the SO coupling for
electrons moving along the principal crystallographic directions. Here,
$\sigma_{1,2,3}$ 
are Pauli matrices with  $\sigma _{2}=-\sigma _{2}^{T}$ and 
$\sigma_{1,3}=\sigma _{1,3}^{T}$. $\epsilon_Z=g\mu_B B_\parallel$ 
is the Zeeman energy associated with the parallel magnetic field. 
Furthermore, ${\bf p}={\bf P}-{\bf A}$ is the kinetic
momentum with ${\bf P}=-i \nabla$ and 
the vector potential ${\bf A}=B_{z}[{\bf r\times e}
_{z}]/2$ corresponding to an additional perpendicular magnetic field. 

In the following, we assume the dot to be sufficiently small to fulfill the conditions $L_{1,2},\ll \lambda _{1,2}$ and $
\gamma ,\epsilon _{{\rm Z}}\ll E_{{\rm Th}}$, where $E_{{\rm Th}}=h D/L^{2}$ is the Thouless energy.

\subsection{Effect of spin-orbit coupling on weak localization and conductance fluctuations}

For a spin-$\frac{1}{2}$ particle in a quantum dot connected to adiabatic ballistic
contacts, the WL corrections to the conductance can be related to the
lowest-lying modes of Cooperons in the singlet and triplet channels. The expressions for the different Cooperon modes can summarized in the equation
\begin{eqnarray}
{\cal C}^{LM}=\frac{1}{2}\left\langle {\rm tr}\left[\sigma _{L}\sigma _{2}\hat{G}
_{R}^{T}(\varepsilon )\sigma _{2}\sigma _{M}\hat{G}_{A}(\varepsilon -\omega
)\right]\right\rangle, 
\end{eqnarray}
where $L,M=0,\dots,3$. Note that the use of the (perturbative) diagrammatic technique in the description
of quantum dots is justified if the number of channels in the leads is
large, $N_{l,r}\gg 1$.

In the absence of SO coupling and Zeeman splitting, the Cooperon modes ${\cal C}^{LM}$ can be separated into four completely independent channels, i.e., one singlet channel ${\cal C}_0={\cal C}^{00}$ and three
triplet channels ${\cal C}_1^{M}={\cal C}^{MM}$ ($M=1,2,3$). Or, in a matrix representation ,
\[
\hat{{\cal C}}={\cal C}\hat{\delta},\qquad{\rm where}\enspace\hat{\delta}\equiv \delta _{LM},
\]
and ${\cal C}$ obeys the conventional diffusion equation. This leads to the familiar result for the WL corrections~\cite{HiLa80},
\[
g^{{\rm WL}}\propto \frac{e^{2}}{h}\left( {\cal C}_{0}-\sum_{M=1,2,3}{\cal C}_{1}^{M}\right).
\]
However, the SO coupling and Zeeman splitting
mix the various components~\cite{LyandaGeller1} and, thus, split their spectra. 
This changes the conventional diffusion equation into the matrix equation
\begin{eqnarray}
\hat{\Pi}\,\hat{{\cal C}}({\bf x},{\bf x}^{\prime })= \delta ({\bf x}
\!-\!{\bf x}^{\prime })\,\hat{\delta},
\end{eqnarray}
where
\begin{eqnarray}
\hat{\Pi} &=&-D \left((\partial _{x_{1}}+2i
A_{1})\hat{\delta}-\frac i{\lambda _{1}}\hat{S}_{2}\right)^{2}
-D \left((\partial _{x_{2}}+2iA_{2})\hat{
\delta}+\frac i{\lambda _{2}}\hat{S}_{1}\right)^{2}+\gamma \,\hat{\delta}+i\epsilon _{{\rm Z}}\,\hat{\eta}.
\label{Coop}
\end{eqnarray}
Here, $\hat{S}_{K}^{LM}=-i\varepsilon ^{KLM}$ are spin-1 operators ($K,L,M=1,2,3$
), and $\varepsilon ^{KLM}$ is the antisymmetric tensor. As a $
4\times 4$ matrix, $\hat{S}$ also has zero elements when $L=0$ or $M=0$. The
other relevant matrix $\hat{\eta}$ is defined as 
\[
\eta ^{LM}=l_{L}\,\delta _{0M}+l_{M}\,\delta _{0L}
\]
(where $l_K$ is the $K$th component of ${\bf l}_\parallel$) indicating that coherence between electrons with opposite polarization is lost on
the time scale $\epsilon _{{\rm Z}}^{-1}$. 

Eq.~(\ref{Coop}) is supplemented with a boundary
condition at the edge of the dot characterized by the normal direction ${\bf
n}_{\parallel }=(n_{1},n_{2})$, 
\begin{eqnarray}
\left[ {\bf n}_{\parallel }\cdot (\nabla +2i{\bf A}
)\hat{\delta}-in_{1}\lambda _{1}^{-1}\hat{S}_{2}+in_{2}\lambda _{2}^{-1}\hat{S}_{1}
\right] \hat{{\cal C}}=0.  \label{boundary}
\end{eqnarray}
Because of the boundary condition, the lowest mode of the Cooperon cannot be 
a mere constant solution, such as ${\bf q}=0$. To find the lowest Cooperon mode 
in this problem, one should make such a gauge transformation (that is, a 
unitary rotation of the Cooperon spin components)
\[
\hat{{\cal C}}=\hat{U}\hat{\cal C}'\hat{U}^{-1},
\] 
that would transform the original boundary condition into 
${\bf n}_{\parallel }\cdot \nabla \hat{\cal C}'=0$. In the new spin 
coordinate system, 
one can now approximate the lowest Cooperon mode 
for $\hat{\cal C}'$ by a ${\bf q}=0$ 
solution, i.e., $\hat{\cal C}'= {\rm const.}$, and evaluate its eigenvalue 
using the standard methods of the 
Hamiltonian perturbation theory with respect to the terms generated 
by such a rotation in the initial differential operator $\hat{\Pi}$.
This program can be realized by applying the transformation 
\[
\hat{U}=\exp \left[i\left(\frac{x_2}{\lambda_2}\hat{S}_{1}-\frac{x_1}{\lambda_1}\hat{S}_{2}\right)\right]\exp [-i\varphi _{s}({\bf x})\hat{S}_{3}]\exp [-i\varphi _{a}(
{\bf x})],
\]
where function $\varphi =\varphi _{s}+\varphi _{a}$ 
transforms from the symmetric 
gauge to a gauge, 
where the normal component of the vector potential at the boundary of the dot vanishes. This eliminates the lowest orders SO coupling terms from
the boundary condition, and, thus, (in a small dot, 
$L_{1,2}\ll
\lambda _{1,2}$) can be followed by a perturbative analysis of 
extra terms in Eq.~(\ref{Coop})
generated by the rotation $\hat{U}$. This results in
the 0D matrix equation for the Cooperon, 
\begin{eqnarray}
\hat{{\cal C}}=\!\left[ \gamma \hat{\delta}+i\epsilon _{{\rm Z}}\hat{\eta}+\!\left( \frac1{\sqrt{\tau _{B}}}\hat{\delta}-\!\sqrt{\epsilon _{\bot }^{{\rm so}}}\hat{S}
_{3}\right) ^{2}\!\!+\epsilon _{\bot }^{{\rm Z}}(\hat{\delta}\!-\!\hat{S}
_{3}^{2})+\epsilon _{\parallel }^{{\rm so}}(\hat{\bf S}^{2}\!-\!\hat{S}
_{3}^{2})\right] ^{-1}\!\!\!\!\!\!,  \label{CoopMatrix}
\end{eqnarray}
which depends on six different energy scales to be discussed in the following.
($\gamma$ and $\epsilon_Z$ have been introduced earlier.)

In Eq.~(\ref{CoopMatrix}), the two parameters 
\begin{eqnarray}
\tau _{B}^{-1} =\frac{4\pi B_{z}^{2}}{\Delta }\left\langle |M_{\alpha \beta
}|^{2}\right\rangle =\kappa E_{{\rm Th}}( 2eB_{z}{\cal A})^{2}, \qquad\qquad\qquad \epsilon _{\bot }^{{\rm so}} =\kappa E_{{\rm Th}}\left( {\cal A}/\lambda
_{1}\lambda _{2}\right) ^{2},\nonumber
\end{eqnarray}
possess the same dependence on the shape of the dot and 
the disorder in the sample. Here, 
$\kappa $ is a geometry-dependent coefficient. 
Furthermore, the random quantities 
$M_{\alpha \beta }$ are the non-diagonal
matrix elements of the magnetic moment of the electron in the dot.  
Note that the difference in the third term in brackets of 
Eq.~(\ref{CoopMatrix}), containing $\tau_B^{-1}$ and 
$\epsilon _{\bot }^{{\rm so}}$, reflects the
addition or subtraction of the Berry and Aharonov-Bohm phases, as was
pointed out in Ref.~\cite{Mathur,Aronov}. 

The
parameter $\epsilon _{\bot }^{{\rm Z}}$ 
in Eq.~(\ref{CoopMatrix}) is the result of a parallel field induced
additional Zeeman splitting, 
\begin{eqnarray}
\epsilon _{\bot }^{{\rm Z}}=\frac{\epsilon _{Z}^{2}}{2\Delta }\sum_{i,j=1,2}
\frac{l_{i}l_{j}}{\lambda _{i}\lambda _{j}}L_iL_j\;\Xi _{ij},\qquad {\rm where}\enspace\Xi
_{ij}=\frac{\pi}{L_iL_j} \langle X_{i}^{\alpha \beta }X_{j}^{\beta \alpha }\rangle ,\nonumber
\end{eqnarray}
and $x_{1,2}^{\alpha \beta }$ are the non-diagonal matrix elements of the
dipole moment of the electron in the dot. The quantities $ \Xi _{ij}$ depend on
geometry and the disorder and may be estimated as $\Xi
\simeq \Delta/E_{{\rm Th}}=g^{-1}$. 
This yields $\epsilon _{\bot }^{{\rm Z}}\ll
\epsilon _{{\rm Z}}$. 
A similar energy scale has been found in other recent
publications~\cite{Halperin}.

Finally, the parameter
\begin{eqnarray}
\epsilon _{\parallel }^{{\rm so}}\sim \left[ \left( L_{1}/\lambda _{1}\right)
^{2}+\left( L_{2}/\lambda _{2}\right) ^{2}\right] \epsilon _{\bot }^{{\rm so}
}\ll \epsilon _{\bot }^{{\rm so}}\nonumber
\end{eqnarray}
introduces the smallest energy scale through 
which the SO coupling affects the Cooperon propagator. 

In principle, the form of Eq.~(\ref{CoopMatrix}) is applicable beyond the
diffusive approximation as it follows purely form symmetry
considerations. 
Now, the weak localization
corrections can be found from Eq.~(\ref{CoopMatrix}) as
\begin{eqnarray}
g^{{\rm WL}}\propto 
{\rm tr}\left[\,\hat{{\cal C}}\left(\hat{\delta}-{\hat {\bf S}^{2}}\right)\right].
\end{eqnarray}
 In the absence of a perpendicular magnetic field 
 ($\tau _{B}^{-1}=0$) and using the assumptions made above, namely 
$\gamma ,\epsilon _{{\rm Z}}, D/\lambda
_{1,2}^{2}\ll E_{{\rm Th}}$, this yields 
\begin{eqnarray}
g^{{\rm WL}}\approx \frac{e^{2}}{h}\frac{a_{lr} }{4}\left( -\frac{\gamma }{
\gamma +\epsilon _{\bot }^{{\rm so}}}-\frac{\gamma }{\gamma +\epsilon _{\bot
}^{{\rm Z}}+2\epsilon _{\parallel }^{{\rm so}}}+\frac{\epsilon _{\bot }^{{\rm so}
}}{\gamma +\epsilon _{\bot }^{{\rm so}}+\frac{\epsilon _{{\rm Z}}^{2}}{
\gamma }}\right) ,  
\label{crossoverWL}
\end{eqnarray}
where the fact that $\epsilon _{\bot }^{{\rm Z}}\ll \epsilon _{{\rm Z
}}$ has been used. Furthermore, we introduced the notation
\[
a_{lr} =\frac{4N_{l}N_{r}}{(N_{l}+N_{r})^{2}}.
\]
The formula in Eq.~(\ref{crossoverWL}) describes the average tendency of the
in-plane magnetoresistance of a dot due to the interplay between SO coupling
and Zeeman splitting.

It is interesting to note that the application of a Zeeman field (i.e., an in-plane magnetic field~\cite{MarcusSpin}) alone
 does not suppress the weak
localization corrections completely as long as $\epsilon _{{\rm Z}}\ll E_{{\rm Th}}$ -- whereas in
the opposite case, $\epsilon _{{\rm Z}}\gtrsim E_{{\rm Th}}$, it does.
However, the orbital effect of such a strong in-plane field is already sufficient to suppress weak localization as
discussed in the previous section.

In the regime 
$\epsilon _{\parallel }^{{\rm so}}/\gamma,\epsilon _{\bot }^{{\rm so}}/\gamma \rightarrow 0$  
relevant for the experiments on small dots~\cite{MarcusSpin}, 
the form of Eq.~(\ref{crossoverWL}) is dominated by the
first two terms, 
\begin{eqnarray}
g^{{\rm WL}}(B)\approx -\frac{e^{2}}{h}\frac{a_{lr} }{4}\left( 1+\frac{1}{
1+\epsilon _{\bot }^{{\rm Z}}(B)/\gamma }\right).
\end{eqnarray}
This suggests a possible procedure for measuring the ratio $\lambda
_{1}/\lambda _{2}$. By fitting the experimental magnetoresistance data to 
$g^{{\rm WL}}(B)$, one can determine the characteristic in-plane magnetic field 
${\cal B}$ at which the weak localization part of the dot
conductance gets suppressed by the factor of two. 
For a dot with a strongly anisotropic shape, this parameter
would depend on the orientation of the in-plane magnetic field. In
particular, ${\cal B}$ should be measured for two orientations of the in-plane
field: namely ${\cal B}_{[110]}$ for ${\bf l}=[110]$ and ${\cal B}_{[1\bar{1
}0]}$ for ${\bf l}=[1\bar{1}0]$. Furthermore, one should perform a simultaneous
measurement of the two characteristic fields ${\cal B}_{[110]}^{\prime }$ and $
{\cal B}_{[1\bar{1}0]}^{\prime }$ in a dot produced on the same chip by
rotating the same lithographic mask by 90${{}^{\circ }}$. The anisotropy of
the SO coupling is then obtained directly from the ratio 
\[
\left( {\cal B}_{[110]}{\cal B}_{[110]}^{\prime }/{\cal B}_{[1\bar{1}0]}
{\cal B}_{[1\bar{1}0]}^{\prime }\right) =\left( \lambda _{1}/\lambda
_{2}\right) ^{4},
\]
that is, independently of details of the sample geometry.

The results for weak localization part of a two-terminal conductance of a
dot and the variance of its universal fluctuations are summarized for the
limiting regimes in the following equation, 
\begin{eqnarray}
g^{{\rm WL}}=-\,\frac{2-\beta }{2\beta \Sigma }\,a_{lr} \,\frac{e^{2}}{h}
,\qquad\langle \delta g^{2}\rangle =\frac{s}{4\beta \Sigma }\,a_{lr} ^{2}\left( 
\frac{e^{2}}{h}\right) ^{2}.
\end{eqnarray}
Here, the conventional parameter $\beta $ describes time-reversal symmetry of the
orbital motion, $s$ is the Kramers' degeneracy parameter, and $\Sigma $ is an
additional parameter characterizing the mixing of states with different
spins for strong Zeeman splitting. In a small dot~\cite{MarcusSpin}, where $
\epsilon _{\parallel }^{{\rm so}}/\gamma \rightarrow 0$ as well as $\epsilon _{\bot }^{
{\rm so}}/\gamma \rightarrow 0$, one obtains the following values for the different parameters: $\beta =1$ indicates that for  weak SO
coupling the electron spin splitting cannot violate the time-reversal
symmetry of the orbital motion;  Kramers' degeneracy is preserved ($s=2$) for $\epsilon _{{\rm Z}}<\gamma $, but lifted
($s=1$) for $\epsilon _{{\rm Z}}>\gamma $; finally, $\Sigma =1$ for 
$\epsilon _{\bot }^{{\rm Z}}<\gamma $ and $\Sigma =2$ for $(\epsilon_{{\rm Z}}\gg)\,\epsilon
_{\bot }^{{\rm Z}}>\gamma $ at very strong Zeeman splitting.


\section{In-plane magnetoresistance in systems with magnetic impurities}
\label{sec-Spin}

In addition to a purely potential disorder, metals and semiconductors may
contain magnetic impurities~\cite{Geim,Benoit}. 
In fact, most of real materials certainly do have them to some
extent~\cite{Birge,Pothier}. In this section, we discuss the influence
of a dilute
magnetic contamination  on the quantum transport
characteristics of disordered conductors.  In particular, we describe
the suppression of weak
localization, and its restoration by an in-plane magnetic field due to 
a polarization of the localized magnetic moments -- 
which slows down the decoherence of conducting electrons and produces
an observable magnetoresistance, 
$g(B_{\Vert })=g_{{\rm class}}+\delta g^{\rm WL}(B_{\Vert })$.

\subsection{Electron spin relaxation and Korringa time for magnetic
impurities}

\label{sub-magimp}

The coupling between a spin-$\frac{1}{2}$ electron and the magnetic impurity
spin originates from the exchange interaction and 
can be described by the Hamiltonian
\[
\hat{{\cal H}}_{{\rm s}}=J\,\,{\bf S} ({\bf r})\cdot \underline{\sigma}, 
\]
where ${\bf S}$ is the impurity spin and $J$ is the exchange
coupling. Furthermore, $\underline{\sigma}$ is a vector of Pauli matrices. When
scattering on a magnetic impurity, the electron can change its spin state.
Thus, electrons diffusing in a disordered conductor containing
magnetic impurities loose their spin memory after
the time scale $\tau _{{\rm s}}$ determined by the spin-flip relaxation rate 
\[
\tau _{{\rm s}}^{-1}\sim \nu \,J^{2}\,S(S+1)\sim \nu \,n_{{\rm s}
}\,j^{2}\,S(S+1). 
\]
Here, $j$ the
exchange coupling due to a single impurity
and  $n_{\rm s}$ is the density of magnetic impurities. 
For dilute magnetic impurities, 
$\tau _{{\rm s}}\gg \tau $.

To incorporate spin-flip scattering into the quantitative
Cooperon/diffuson analysis of weak localization and UCFs, 
one has to make assumptions concerning
correlation properties of this additional source of disorder. Similarly to
the non-magnetic impurities, we assume the  magnetic disorder to be Gaussian 
$\delta $-correlated with zero mean and variance 
\[
\langle JS_{\alpha }({\bf r},t)JS_{\beta }({\bf r}^{\prime },t)\rangle =
\frac{1}{6\pi \nu \tau _{{\rm s}}}\delta _{\alpha \beta }\,\delta ({\bf r}-
{\bf r}^{\prime })
\]
at coinciding moments of time.

There is, however, an important difference between potential disorder, 
which is static
for a fixed configuration of impurities in the sample, and magnetic
scatterers, which change their spin state after each electron-impurity
spin-flip event. As a result, magnetic disorder has its own dynamics, and a
magnetic scatterer forgets about its initial spin state with the
so-called Korringa relaxation rate~\cite{Korringa} 
\begin{eqnarray}
\tau _{{\rm K}}^{-1}\sim \nu \,n_{{\rm e}}\,j^{2}\frac{T}{\epsilon _{{\rm F}}
},  \label{KorringaRate}
\end{eqnarray}
where $n_{\rm e}$ is the electron density.

Korringa relaxation, as it is known, takes place due to spin-flip
scattering  at the impurity of any electron with energy 
$|\epsilon-\epsilon _{{\rm F}}|\sim T$ close to the Fermi level. 
Therefore, the corresponding relaxation rate is
temperature-dependent, whereas the single-electron spin relaxation rate 
$1/\tau _{{\rm s}}$ is temperature independent. As a result, for
temperatures $T>T_{{\rm s}}$ higher than a certain temperature 
\[
T_{{\rm s}}\sim \frac{n_{{\rm s}}}{n_{{\rm e}}}\epsilon _{{\rm F}}, 
\]
an impurity changes its spin state faster than any free electron whose
individual propagation we may follow, i.e., 
$1/\tau _{{\rm K}}>1/\tau _{{\rm s}}$.
By contrast, for $T<T_{{\rm s}}$,  the single-electron spin relaxation
is faster, $1/\tau _{{\rm s}}>1/\tau _{{\rm K}}$, and the
exchange field of the impurities can be treated as static 
when analyzing weak localization effects.

The relaxation of the impurity spin sub-system should be taken into
account in the correlation properties of this source of
randomness. This can be achieved by saying that, for the same magnetic scatterer, 
\[
\langle S_{\alpha }({\bf r},t)S_{\beta }({\bf r},0)\rangle =\delta _{\alpha
\beta }\,S(S+1)\,e^{-|t|/\tau _{{\rm K}}}. 
\]

\subsection{Spin-flip scattering, impurity spin dynamics and polarization of
impurities in weak localization}

After an electron flips its spin a few times at encountered magnetic impurities,
the coherence between waves propagating in clockwise and anti-clockwise
directions along the same geometrical path is destroyed. In the regime of
fast Korringa relaxation, $1/\tau _{{\rm K}}>1/\tau _{{\rm s}}$, this happens since electron waves traveling
towards each other along a loop encounter the same magnetic
scatterer at different moments of time and, therefore, in uncorrelated
initial states. In the regime of slow Korringa relaxation, $1/\tau _{{\rm K}}<1/\tau _{{\rm s}}$, decoherence
between these two waves accumulates due to a non-commutativity of 
electron spin operators, whose product would
appear due to the coupling to spin-flip transitions between
the same initial and final states of the same set of magnetic impurities
visited in the opposite order.

\begin{figure}[h]
 \begin{center}\includegraphics[scale=0.3]{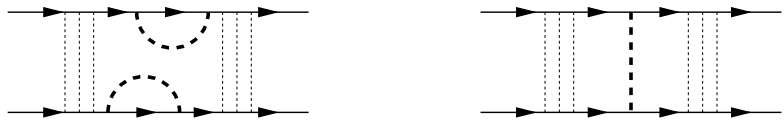}
 \caption{Cooperon for $|t|<\tau_K$ (both contributions) and $|t|>\tau_K$ (only left diagram). 
 (The thin dashed line represent scattering from the non-magnetic 
 impurities while the thick dashed lines stand for (spin-flip) scattering from 
 the magnetic impurities.)}
\label{coop-tau_k}
 \end{center}
 \end{figure}

As a result, in a magnetically contaminated conductor, the time of flight
along trajectories forming the enhancement of backscattering is limited by
the spin-relaxation time, $\tau _{{\rm s}}$. As in the previous section, a
quantitative description of weak localization corrections to the conductivity, 
\begin{eqnarray}
g^{\rm WL} =\frac{e^{2}}{2\pi h}\int_{\tau }^{\infty }dt\;\left[ {\cal C}
_{0}(t;0)-3\,{\cal C}_{1}(t;0)\right] ,  \label{Spin-10}
\end{eqnarray}
requires a diagrammatic evaluation of the singlet and triplet Cooperon
correlation functions. In Eq.~(\ref{Spin-10}), the variable of integration 
$t$ is the difference between moments of time when clockwise and anti-clockwise
propagating waves pass through the same point in space. As shown in
Fig.~\ref{coop-tau_k}, when deriving the
equation of motion for Cooperons, one should include magnetic impurity
scattering into both, (a) the self-energy of the
impurity-averaged electron Green functions, i.e., as a correction to
the mean free time, and (b) the two-particle correlation functions as additional impurity lines in the ladder
diagram. Taking into account the Clebsh-Gordan coefficients that
appear upon splitting the Cooperon into singlet and triplet channels, the equation for the
Cooperon takes the form 
\begin{eqnarray}
\left\{ \partial _{t}-D\nabla ^{2}+\frac{1}{\tau _{{\rm s}}}\left(
1+c_{J}e^{-|t|/\tau _{{\rm K}}}\right) \right\} {\cal C}_{J}(t,t^{\prime };
{\bf r},{\bf r}^{\prime })=\delta (t-t^{\prime })\delta ({\bf r}-{\bf r}
^{\prime }),  \label{Spin-CooperonEq}
\end{eqnarray}
where $J=0$ (spin-singlet) or $1$ (spin-triplet), and $c_{0}=1$,
$c_{1}=-1/3$. The time-dependence accounts for the fact that, for clockwise and anti-clockwise trajectories, electron waves
test the same scatterer at different moments of time. If the relevant time
scale determining the value of the integral in Eq.~(\ref{Spin-10}) is
large such
that $|t|>\tau _{{\rm K}}$, the states of the same impurity seen by waves
propagating in opposite directions are uncorrelated. In this regime,
the solution of Eq.~(\ref{Spin-CooperonEq}) is
\begin{eqnarray}
{\cal C}({\bf q},\omega )=\frac{1}{D{\bf q}^{2}-i\omega +\tau _{{\rm s}}^{-1}
}\;,  \label{Spin-CooperonSol1}
\end{eqnarray}
with the same relaxational pole $\tau _{{\rm s}}^{-1}$, i.e., the single-particle spin relaxation rate, in all spin
channels. This
regime is realized when $\tau _{{\rm K}}<\tau _{{\rm s}}$, that is, at high
temperatures $T>T_{{\rm s}}$.

By contrast, for $\tau _{{\rm K}}>\tau _{{\rm s}}$, the integral in Eq.~(\ref{Spin-10}) is determined by the behavior of Cooperons at $|t|<\tau _{{\rm 
K}}$, where the decoherence in the triplet channel takes about
thrice longer than in the singlet channel, 
\[
{\cal C}_{0}({\bf q},\omega )=\frac{1}{D{\bf q}^{2}-i\omega +2\tau _{{\rm s}
}^{-1}}\;,\;\;{\cal C}_{1}^{\alpha\beta}({\bf q},\omega )=\frac{\delta _{\alpha\beta}}{D{\bf q}
^{2}-i\omega +\frac{2}{3}\tau _{{\rm s}}^{-1}}\;. 
\]
The latter feature indicates that a single spin-flip does not completely
destroy coherence between clock and anti-clockwise paths, given that in both
propagation scenarios the magnetic impurity state undergoes exactly the same
transition.

Eq.~(\ref{Spin-CooperonEq}) can be solved using the following
substitution, 
\[
{\cal C}_{J}(t,t^{\prime };{\bf r},{\bf r}^{\prime })=e^{-f(t)}\,\tilde{
{\cal C}}_{J}(t,t^{\prime };{\bf r},{\bf r}^{\prime })\,e^{\,f(t^{\prime
})}\,, 
\]
with 
\[
f(t)=\frac{1}{\tau _{{\rm s}}}\left( t+c_{J}\tau _{{\rm K}}\,{\rm sign}
\,t\,(1-e^{-|t|/\tau _{{\rm K}}})\right) \,. 
\]
Then the `modified' Cooperon $\tilde{{\cal C}}_{J}(t,t^{\prime };{\bf r},{\bf r}
^{\prime })$ obeys the conventional diffusion equation $(\partial
_{t}-D\nabla ^{2})\tilde{{\cal C}}_{J}(t,t^{\prime };{\bf r},{\bf r}^{\prime
})=\delta (t-t^{\prime })\delta ({\bf r}-{\bf r}^{\prime })$. After
substituting the resulting expression for the Cooperon ${{\cal
    C}}_{J}$ back into Eq.~(\ref{Spin-10}), we arrive at 
\begin{eqnarray}
g^{{\rm WL}} =\frac{e^{2}}{2\pi h}(D\tau _{{\rm s}})^{1-\frac{d}{2}
}\int\limits_{\tau /\tau _{{\rm s}}}^{\infty }\frac{d\theta }{\theta ^{d/2}}
\left\{ \exp \left[ -2\theta -\frac{2\tau _{{\rm K}}}{\tau _{{\rm s}}}
(1-e^{-\theta \tau _{{\rm s}}/\tau _{{\rm K}}})\right] -3\exp \left[ -2\theta +\frac{2\tau _{{\rm K}}
}{3\tau _{{\rm s}}}(1-e^{-\theta \tau _{{\rm s}}/\tau _{{\rm K}}})\right]
\right\} .&&  \label{Spin-WLtime}
\end{eqnarray}

In the `low-temperature' limit, where $\tau _{{\rm K}}^{-1}\rightarrow 0$,
\begin{eqnarray}
g^{{\rm WL}}(\tau _{{\rm K}}^{-1})\rightarrow 0=-\frac{e^{2}}{\pi h}(D\tau _{{\rm s}})^{1-\frac{d}{2}}\left\{ 
\begin{array}{cc}
{\frac{\pi }{\sqrt{2}}(3^{3/2}-1)} &  \qquad d=1, \\ 
{\ln \frac{3^{3/2}\tau _{{\rm s}}}{4\tau }} &  \qquad d=2.
\end{array}
\right. \label{Spin-noKorr}
\end{eqnarray}
For comparison, in the limit $\tau _{{\rm K}}^{-1}\rightarrow \infty $, the
above equation changes into 
\begin{eqnarray}
\label{Spin-Korr}
g^{{\rm WL}}(\tau _{{\rm K}}^{-1}\rightarrow \infty) =-\frac{e^{2}}{\pi h}(D\tau _{{\rm s}})^{1-\frac{d}{2}}\left\{ 
\begin{array}{cc}
\sqrt{\frac{\pi }{2}} & \qquad d=1, \\ 
{\ln \frac{\tau _{{\rm s}}}{\tau }} & \qquad d=2.
\end{array}
\right. \nn
\end{eqnarray}
The difference between the dynamic and static results in Eqs.~(\ref{Spin-Korr}) and (\ref{Spin-noKorr})  manifests itself in a weak
temperature dependence of the conductivity, 
\begin{eqnarray}
\sigma (T>T_{{\rm s}})-\sigma (T\ll T_{{\rm s}})\approx \frac{e^{2}}{2\pi h}
(D\tau _{{\rm s}})^{1-\frac{d}{2}}\left\{ 
\begin{array}{cc}
{1.2} & \qquad d=1, \\ 
1.9 & \qquad d=2.
\end{array}
\right.  \label{Spin-diff}
\end{eqnarray}

So far in this section, we addressed only the suppression of weak localization
by spin-flip scattering, whereas our main goal was to identify
possible effects of an external magnetic field leading to an observable
in-plane magnetoresistance. An in-plane magnetic fields tends to polarize
the paramagnetic impurities, thus, opening a gap $\epsilon _{{\rm Z}}^{{\rm (imp)}
}=g^{{\rm (imp)}}\mu _{{\rm B}}B_{\parallel }$ for the electron spin-flip
relaxation process. At high magnetic fields, $\epsilon _{{\rm Z}}^{{\rm (imp)}
}\gtrsim T$, the allowed electron energies are insufficient to
flip the impurity spin, i.e., spin-flip scattering is suppressed. Thus, at $
B_{\parallel }>B_{{\rm s}}$, where 
\begin{eqnarray}
B_{{\rm s}}(T)=T/(g^{{\rm (imp)}}\mu _{{\rm B}}),  
\label{Spin-Bs}
\end{eqnarray}
magnetic impurities act as non-magnetic `potential' scatterers, with the
only difference that their potential profiles are different for
spin-$\uparrow $ and spin-$\downarrow $ 
electrons. Treating the spin-$\uparrow $ 
and spin-$\downarrow $ electron subsystems separately, one may
conclude that, for $T\ll \epsilon _{{\rm Z}}^{{\rm (imp)}}$, each of them
contributes a `restored' value towards the weak localization correction to
the conductivity (conductance), which results in its overall drop across a
broad range of in-plane magnetic fields,
\begin{eqnarray}
g(B_{\parallel }\gg B_{{\rm s}})-g(B_{\parallel }=0)\sim
-\,\frac{e^{2}}{\pi h}\times \left\{ 
\begin{array}{cl}
\ln 
{\displaystyle \frac{\tau _{\varphi }}{\tau _{s}}}
, & \qquad {\rm 2D;} \\ 
\sqrt{
{\displaystyle \frac 1{D\tau _{s}}}
} & \qquad {\rm Q1D\enspace wire;} \\ 
1 & \qquad {\rm 0D\enspace dot.}
\end{array}
\right.  
\label{Spin-MR}
\end{eqnarray}
In Eq.~(\ref{Spin-MR}), $\tau _{\varphi }>\tau _{{\rm s}}$ is a `true'
inelastic decoherence rate due to the electron-electron interaction or
external electromagnetic noise.

The quantitative analysis of a system with fully polarized static magnetic
impurities implies the evaluation of all Cooperon channels, ${\cal C}_{0}$, $
{\cal C}_{10}$,  and ${\cal C}_{1\pm 1}$. The first two happen to have a
gap in the spectrum, both due to i) the splitting between up/down-spin
bands caused by the combination of $B_{\parallel }$ and the mean exchange
splitting due to impurities, and ii) the difference between the
scattering conditions for up/down-spin
electron. By contrast, ${\cal C}_{11}$ and ${\cal C}
_{1-1}$ remain massless. When $\tau _{\varphi }\gg \tau _{{\rm s}}$, the
crossover between the two limits in Eq.~(\ref{Spin-MR}) can be roughly
described by substituting the electron spin-flip rate at intermediate 
fields, 
\begin{eqnarray}
\tau _{{\rm s}}^{-1}(B_{\parallel })=\frac{u}{e^{u}-e^{-u}}\times \tau _{
{\rm s}}^{-1},\quad {\rm where} \enspace u=B_{\parallel }g^{{\rm (imp)}}\mu
_{{\rm B}}/T=B_\parallel/B_{\rm s},
\label{Spin-TauS1}
\end{eqnarray}
into Eq.~(\ref{Spin-noKorr}). The field-dependent electron spin-flip rate 
in the latter expression manifests the activational character of the impurity 
spin flip process and was calculated after taking into account 
thermal occupancy of spin-split impurity states.
For a2D electron gas or a thin metallic film, this yields a
magnetoconductivity with the characteristic form, 
\begin{eqnarray}
g(B_{\parallel })-g(0)=\frac{e^{2}}{h}\ln \frac{u}{e^{u}-e^{-u}}\approx -\,
\frac{e^{2}}{h}\times \left\{ 
\begin{array}{cc}
\frac{1}{6}\left( \frac{B_{\parallel }}{B_{{\rm s}}}\right) ^{2} & 
\qquad B_{\parallel }<B_{{\rm s}}, \\ 
\frac{B_{\parallel }}{B_{{\rm s}}} & \qquad B_{\parallel }>B_{{\rm s}}.
\end{array}
\right.   \label{MR2}
\end{eqnarray}

\subsection{Effect of impurity spin dynamics on mesoscopic conductance
fluctuations}

\label{sub-ucfspin}

At $B_{\parallel }=0$, contamination by magnetic impurities also suppresses mesoscopic conductance
fluctuations -- as a  result of the combination of
electron spin-flip scattering with the impurity spin dynamics. In a small
sample, the mesoscopic part of the interference correction to the conductance, 
$\delta g(t)$, is finite for any instant configuration of magnetic
scatterers, but it is specific for each particular snapshot of the magnetic
subsystem. As the localized magnetic moments undergo a random temporal evolution,
this interference correction to the conductance also fluctuates in time,
thus, leading to a self-averaging of the UCFs to much smaller
values. By contrast, in the regime of  strong magnetic fields, $B_{\parallel }\gg B_{
{\rm s}}$, the impurity spins stay polarized, such that the random
potentials for spin-$\uparrow $ and spin-$\downarrow $ electrons are
static (though different), and
UCFs are fully restored.

To describe the reproducible part of conductance fluctuations in a
DC-current measurement over the entire range of fields, one has to analyze
the time-averaged conductance, 
\[
\bar{g}=\lim_{\Upsilon \rightarrow \infty }\frac
1\Upsilon\int_{0}^{\Upsilon }dt \; g(t)=\langle g\rangle +\delta \bar{g}, 
\]
and its variance,
\[
\langle \delta \bar{g}^{2}\rangle =\lim_{\Upsilon \rightarrow \infty
}\frac 1\Upsilon\int_{0}^{\Upsilon }dt\; \langle \delta g(t)\delta
g(0)\rangle =\lim_{t\rightarrow \infty }\langle \delta g(t)\delta
g(0)\rangle . 
\]
The correlation function $\langle \delta g(t)\delta
g(0)\rangle $ determining $\langle \delta \bar{g}^{2}\rangle$ can be
expressed in terms of diffusons ${\cal D}_{J}(\eta
,\eta ^{\prime };t;{\bf r},{\bf r}^{\prime })$ with an additional time
variable $t$ and Cooperons as in Eq.~(\ref{Spin-CooperonEq}) in the usual way
-- via the standard set of perturbation theory diagrams.

The diffusons ${\cal D}_{J}(\eta ,\eta ^{\prime };t;{\bf r},{\bf r}^{\prime })$
describe the quantum diffusion of electrons across two distinctive time
intervals separated by $t$. They obey the following equation, 
\[
\left\{ \partial _{\eta }-D\nabla ^{2}+\frac{1}{\tau _{{\rm s}}}\left(
1-c_{J}e^{-|t|/\tau _{{\rm K}}}\right) \right\} {\cal D}_{J}(\eta ,\eta
^{\prime };t;{\bf r},{\bf r}^{\prime })=\delta (\eta -\eta ^{\prime })\delta
({\bf r}-{\bf r}^{\prime }), 
\]
where the loss of spin memory of the  magnetic impurities due to the Korringa
relaxation is taken into account. In the relevant limit of $|t|/\tau _{{\rm K}
}\rightarrow \infty $, where $e^{-|t|/\tau _{{\rm K}}}\rightarrow 0$, the
same simplification occur as the ones leading to Eq.~(\ref{Spin-CooperonSol1}) for the Cooperon: the poles of all (singlet and triplet) diffuson components acquire the
same relaxational gap equal to the single-electron spin-relaxation
rate, i.e., 
\[
{\cal D}(\bq,\omega)=\frac{1}{D{\bf q}^{2}-i\omega +\tau _{{\rm s}}^{-1}}. 
\]
As a result, one arrives at strongly suppressed fluctuations. For a wire or
dot geometry, their variance for a given value of the magnetic field follows
the field-dependence of the spin-flip scattering rate for an electron in Eq.~(\ref{Spin-TauS1}), as 
\[
\langle \delta \bar{g}^{2}\rangle \sim \left( \frac{e^{2}}{h}\right)
^{2}\times \left\{ 
\begin{array}{ll}
\left( \tau _{{\rm s}}/\tau _{{\rm fl}}\right) ^{3/2} & \qquad {\rm wire \enspace geometry,}
\\ 
\left( \tau _{{\rm s}}/\tau _{{\rm esc}}\right) ^{2} & \qquad {\rm dot \enspace geometry,}
\end{array}
\right. 
\]
where $\tau_{\rm s}=\tau_{\rm s}(B_\parallel)$, $\tau _{{\rm esc}}$
is the electron escape rate from a dot defined in the
previous chapter, and $\tau_{{\rm fl}}=L^2/D$.

The suppression of mesoscopic fluctuation in magnetically contaminated
conductors as well as their restoration by a magnetic field have been
observed, both in metallic~\cite{Benoit} and semiconductor~\cite{Geim}
microstructures. It is worth mentioning that this characteristic behavior
of weak localization and mesoscopic fluctuations may be used for testing
the nature of decoherence~\cite{Mohanty1,Mohanty2} even in purified
materials -- in particular, when its suspected origin is spin-flip
scattering at residual magnetic impurities~\cite{Birge,Pothier}.


\section{2D layer symmetry, in-plane magnetoresistance and the non-linear
sigma-model}
\label{sec-SUSY}
In the previous chapters, we have shown how to analyze ``simple'' problems using
the diagrammatic perturbation theory technique. In this section, we give an
example of a problem that is more convenient to tackle with the supersymmetric (SUSY)
sigma-model method:
the analysis of the interplay between spatial symmetry of a quasi-2D
electron gas with few occupied subbands and an in-plane magnetic field.\footnote{
In fact, the present, perturbative problem is accessible by
diagrammatic methods as well. However, to get the interplay between inter-band
correlations and disorder scattering reliably under control, the formalism
of field integration has the advantage that the fully microscopic aspects of
the problem are processed in the early stages of the derivation~\cite{b-Efe}.
} The main results have been discussed in chapter \ref{sec-Orbital}. Here, we first give a brief overview of the SUSY technique and, then,
present the derivation of the SUSY sigma-model action for a multi-subband 2D system subject
to an in-plane magnetic field. Finally, we discuss the properties of the Cooperon matrix and the consequent implications for the magnetoresistance.

\subsection{The field theoretic technique}
\label{app-Techniques} 

In Sec.~\ref{sec-Techniques} the diagrammatic approach to calculating
correlation functions in the perturbative regime has been introduced. An
alternative approach is provided by the coherent state path integral. The
(retarded) Green function can be represented as a field integral, 
\begin{eqnarray}
G^+(\br,\br^{\prime};\epsilon)\eq\langle\br|(\epsilon+i0-\hat\cH)^{-1}|\br
^{\prime}\rangle=\frac i\cZ\int Ds\, Ds^* \;s^*(\br)s(\br^{\prime})\; e^{\,i\int d\br
\,s^*(\epsilon^+-\hat\cH)s},\nn
\end{eqnarray}
where 
$\cZ=\int Ds\, Ds^* \, \exp[{i\int d\br\,s^*(\epsilon^+\!-\!\hat\cH)s}]$,
and $\hat\cH=\hat\bp^2/(2m)+V(\br)$. Here $V$ represents the impurity
potential which is assumed to be drawn from a Gaussian white noise
distribution, cf.~(\ref{wn}).

Unfortunately, in this form it is not possible to carry out the disorder
averaging: Due to the presence of the partition function as a normalization
factor, $\cZ^{-1}[V]$, the random potential appears in the numerator as well
as in the denominator.
One method to circumvent this problem is {supersymmetry}~\cite{b-Efe}.
When one is considering only {single-particle} properties of a
system, there are two equivalent formulations of the path integral, namely
by using bosonic or fermionic fields. Supersymmetry now exploits the
following property of commuting ($s$) versus anti-commuting or Grassmann ($
\chi$) variables: 
\begin{eqnarray}
\int ds^*\,ds\,e^{-s^* Ms}={\det}^{-1} M, \qquad \int
d\bar\chi\,d\chi\,e^{-\bar\chi M\chi}=\det M.\nn
\end{eqnarray}
Thus, combining both variables into a `supervector', $\psi^T=(s,\chi)$, yields 
the result $\int d\psi^\dagger\,d\psi\,\exp[{-\psi^\dagger \,M\otimes{{\bf 1}}^{\rm{bf}
}\,\psi}]=1$,
where the superscript `$\rm{bf}$' stands for `boson-fermion'.
Applying this to the partition sum, it is automatically normalized to unity, 
${\mathcal{Z}}=1$, and the impurity
averaging is straightforward. The evaluation of a two-particle correlation
function requires the introduction of two sets of fields, covering the
advanced and retarded sector. With $\psi^T=(s_1,\chi_1,s_2,\chi_2)$ the
correlator can be written as 
\begin{eqnarray}
\langle G^+(\epsilon\!+\!\frac\omega2)G^-(\epsilon\!-\!\frac\omega2)\rangle=
-\left\langle\int D[\psi,\bar\psi]\,s_1^*s_1s_2^*s_2\, e^{-i\int d\br
\,\bar\psi(\epsilon-\frac{\omega^+}2\sar_3-\hat\cH)\psi}\right\rangle,\nn
\end{eqnarray}
where $\omega^+=\omega+i0$ and $\sar_3$ is a Pauli matrix in
advanced/retarded space. Furthermore, $\bar\psi=\psi^\dagger L$ with $L=\sar
_3\otimes E_{\rm{bb}}+{{\bf 1}}^{\rm{ar}}\otimes E_{\rm{ff}}$
, where $E_{\rm{bb}}$ and $E_{\rm{ff}}$ are projectors onto the
boson-boson and  fermion-fermion block, respectively. 
By introducing a source term, 
$S_J=-\int d\br\,(J^\dagger\psi+\bar\psi J)$,
different correlators of Green functions can be obtained from the generating
functional $\cZ[J]$ by taking derivatives with respect to the source field $J$.
In the following, we will suppress the sources and consider only $\cZ[0]$.

Now the impurity averaging of the partition function leads to a quartic term
in the fields $\psi$, 
\begin{eqnarray}
\langle e^{\,i\int d\br \,\bar\psi V \psi} \rangle = \exp[{
-\frac1{4\pi\nu\tau}\int d\br \,(\bar\psi\psi)^2}].\nn
\end{eqnarray}
By Fourier transformation to momentum representation, one can identify the
slow modes, 
\begin{eqnarray}
\int d\br\, (\bar\psi\psi)^2&=&\sum_{\sum\bp_i=0}(\bar\psi_{\bp_1}\psi_{\bp
_2})(\bar\psi_{\bp _3}\psi_{\bp_4})\nn\\
&\approx&\sum_{\bp,\bp^{\prime};\bq}\Big((\bar\psi_{\bp}\psi_{-\bp+\bq
})(\bar\psi_{-\bp ^{\prime}}\psi_{\bp^{\prime}-\bq})+(\bar\psi_{\bp}\psi_{-\bp^{\prime}})(\bar\psi_{\bp^{\prime}-\bq}\psi_{-\bp
+\bq})+(\bar\psi_{\bp}\psi_{\bp^{\prime}-\bq})(\bar\psi_{-\bp+\bq}\psi_{-\bp
^{\prime}})\Big),\nn
\end{eqnarray}
where $|\bq|\ll \ell^{-1}$.

The first term corresponds to slow fluctuations of the energy which can be
absorbed by a local redefinition of the chemical potential. Thus, we
concentrate on the remaining terms: The second term generates the diffuson
contribution while the third term yields the Cooperon contribution.
Enlarging the field space\footnote{Generally, each discrete symmetry leads to a doubling of the low-lying modes
and, thus, should be incorporated by doubling the field space~\cite{SiAg97}.}
by defining $\Psi^T=(\psi^T,\psi^\dagger)/\sqrt2$, the last two terms can be
rewritten into a single contribution, 
$\Tr[\sum_\bq\zeta(\bq)\zeta(-\bq)],$
where $\zeta(\bq)=\sum_\bp\Psi(\bp-\bq)\bar\Psi(-\bp)$. The components of
the newly defined vector $\Psi$ fulfill the symmetry relation $
\Psi^\dagger=(C\Psi)^T$, where $C=\str_1\otimes E_{\rm{bb}}+i\str
_2\otimes E_{\rm{ff}}$. This symmetry corresponds to time-reversal ($
\psi\to\psi^*$, $\hat\cH\to\hat\cH^T$).
In the absence of the symmetry-breaking energy difference, $\omega=0$, the
action is invariant under rotations $\Psi\to U\Psi$, where $ULU^\dagger=L$
and $U^T=CU^\dagger C^T$. Thus, $U\in$~Osp(4$|$4).

As a next step the quartic interaction is decoupled by a
Hubbard-Stratonovich transformation, introducing the new (supermatrix-)
fields $Q$: 
\begin{eqnarray}
\exp\left[-\frac1{4\pi\nu\tau}\int\! d\br\,(\bar\Psi\Psi)^2\right] \eq\int
DQ \, \exp\left[\frac{\pi\nu}{8\tau}\int\! d\br\,\Tr Q^2\!-\!\frac1{2\tau}
\int\! d\br \,\bar\Psi Q\Psi\right],\nn
\end{eqnarray}
where $\Tr M=\tr M_{\rm{bb}}-\tr M_{\rm{ff}}$. The symmetries of $Q
$ reflect the symmetries of the dyadic product $\Psi\otimes\bar\Psi$, namely 
$Q=CLQ^T(CL)^T$.
Now the resulting exponent is only quadratic in the original $\Psi$-fields.
Therefore, the Gaussian integral can be readily evaluated, yielding the
action 
\begin{eqnarray}
S[Q]=-\frac{\pi\nu}{8\tau}\int d\br \,\Tr Q^2+\frac12\int d\br \,\Tr\ln \cG
^{-1},  \label{sq}
\end{eqnarray}
where $\langle\cZ\rangle=\int \!DQ\,\exp(-S[Q])$ and 
$\cG^{-1}=\frac1{2m}\hat\bp^2-\epsilon_{\mathrm{F}}+\frac{\omega^+}2\sar
_3+\frac i{2\tau}Q$.

To extract an effective low-energy, long-wavelength field theory from this
action, a saddle point analysis has to be performed. Variation of (\ref{sq})
with respect to $Q$ yields 
$Q_{\rm{sp}}(\br)=i\cG(\br,\br)/({\pi\nu})$.
Neglecting the small energy $\omega$, the Ansatz $Q_{\rm{sp}}$ constant
and diagonal leads to 
\begin{eqnarray}
Q_{\rm{sp}}=-\frac i\pi \int d\xi\,\frac1{\xi-\frac i{2\tau}Q_{\rm{
sp}}}=\mathrm{sgn}(Q_{\rm{sp}}).
\end{eqnarray}
Thus, the saddle point $Q_{\rm{sp}}$ has the meaning of a self-energy.
Analytic properties of the Green function single out the solution $Q_{
\rm{sp}}=\sar_3$.

In fact, the action is invariant under transformations $Q\to T Q T^{-1}$,
where $T$ constant: instead of one saddle point one obtains -- at $\omega=0$
and in the absence of symmetry breaking sources -- a degenerate saddle point
manifold $Q^2={{\bf 1}}$. Fluctuations around the saddle point can be
subdivided into longitudinal modes, $[\delta Q_l,Q_{\rm{sp}}]=0$, and
transverse modes, $\{\delta Q_t,Q_{\rm{sp}}\}=0$. The longitudinal
modes $\delta Q_l$ leave the saddle point manifold $Q^2={{\bf 1}}$.
Therefore, they are massive and do not contribute to the low-energy physics
of the system. In the following, we concentrate on the transverse modes $
\delta Q_t$. The parameter which stabilizes this distinction is $k_{\mathrm{F
}}\ell\sim\epsilon_{\mathrm{F}}\tau$, i.e., the following considerations are
valid in the {quasi-classical} limit.
We proceed by expanding the action around the saddle point in the slowly
varying fields $Q(x)=T(x)Q_{\rm{sp}}T^{-1}(x)$. Separating the fast and
slow degrees of freedom with $\hat\bp\to\bp+\hat\bq$, this expansion yields 
\begin{eqnarray}
S\simeq\frac12\int\! d\br\int\! d\bp\,\Tr\left[\frac{\omega^+}2\cG_0T^{-1}
\sar_3T-\frac1{2m^2}\left(\cG_0\,T^{- 1}\,\bp\hat\bq\, T\right)^2\right].\nonumber
\end{eqnarray}
The integral over fast momenta, $\bp$, can be performed using the following
representation for the Green function, 
\begin{eqnarray}
\cG_0(\bp)=\frac12\sum_{s=\pm}\frac{1+s\sar_3}{-\xi_p+s\frac i{2\tau}}
\equiv\frac12\sum_{s=\pm}(1+s\sar_3)G_0^s(\bp).\nn
\end{eqnarray}
Then, $\int\! d\bp \; \cG_0(\bp)=-i\pi\nu\sar_3$, and 
\begin{eqnarray}
\int d\bp \, \Tr\left[\cG_0(\bp)\,T^{-1}\,\bp\hat\bq\, T\,\cG_0(\bp
)\,T^{-1}\,\bp\hat\bq \,T\right]
\eq\underbrace{\frac{\nu p_{\mathrm{F}}^2}{4d}\int d\xi\;G_0^+G_0^-}_{
\displaystyle={m^2\pi\nu D}/{2}}\Tr\left[(1\!+\!s\sar_3)T^{-1}\,\hat\bq \,
T(1\!-\!s\sar_3)T^{-1}\,\hat\bq \,T\right].\nn
\end{eqnarray}
Finally, summing over $s$ and using the cyclic invariance under the trace,
the effective action takes the form of a non-linear $\sigma$ model, 
\begin{eqnarray}
S[Q]=-\frac{\pi\nu}8\int d\br \,\Tr\left[D(\partial Q)^2+2i\omega^+\sar_3Q
\right].
\end{eqnarray}
Note that the effect of a weak magnetic field is to generalize the
derivatives to $\tilde\partial=\partial-i\bA[\str_3,\,.\,]$. For the
quasi-twodimensional system in an in-plane magnetic field this will be
discussed in more detail in the following.

In the perturbative regime, an expansion of the effective action around the
saddle point in the generators $W$, where $Q=e^{-W/2}\sar_3e^{W/2}$ and $\{W,
\sar_3\}=0$, reproduces the diagrammatic results. By contrast, in the {
non}-perturbative regime, the action is dominated by zero-modes which
require an integration over the whole saddle point manifold 
Osp(4$|$4)/(Osp(2$|$2)$\otimes$Osp(2$|$2)). For our purposes, a perturbative expansion will be sufficient.

\subsection{Magnetoresistance in a multi-subband electron layer with
a possible Berry-Robnik symmetry}

As pointed out earlier, the magnetoresistance in quasi-twodimensional electron systems subject to parallel fields is very sensitive to the presence or absence of $\cP_z$-symmetry. While in chapter \ref{sec-Orbital} we restricted ourselves to a qualitative discussion of the effect, here the quantitative results are presented.
To explore such type of
phenomena one needs to construct an approach which on the one hand is
sensitive to microscopic details in $z$-direction while on the other hand
should be capable of efficiently describing large scale in-plane properties.
This task can efficiently be addressed within a field integral formalism.
The starting point of the derivation is a supersymmetric field integral with
action 
\begin{eqnarray}
S[\psi] \eq i \int\! d^3r \,\bar\psi \Big(\epsilon_{\mathrm{F}} -\omega^+ 
\sar_3 +\frac1{2m} ( \partial_x^2 + (\partial_y \!-\!iBz)^2 + \partial_z^2 )
- W(z) - V(x,y)\Big)\psi,\label{S-orig}
\end{eqnarray}
where $W$ is the confining potential of the 2DEG, and $V$ a disorder
potential. To simplify the analysis, 
it is assumed that the disorder potential does not depend on the $z$
-coordinate. Given the typical architecture of 2DEGs, this is  certainly a
justified zeroth order assumption. In, e.g., high mobility 2DEGs (or 2DHGs)
in GaAs/AlGaAs heterostructures, the mobility is limited by a long-range
random potential, created by charged impurities located far from the plane.
Later on, this condition will be  relaxed by generalizing $V$ according to $
V(x,y) \to V(x,y) +  U(x,y,z)$, where $U$ is weak and can be treated
perturbatively in a  sense to be specified below. 

The confinement in $z$-direction is responsible for the size quantization
which entails a subband structure of the system. To make progress with the
action, Eq.~(\ref{S-orig}), an orthonormalized set of wavefunctions $\{\phi_k\}$, diagonalizing the $z$-dependent part of the problem, is
introduced: 
\begin{eqnarray}
\left(-\frac1{2m} \partial_z^2 +W(z) -\epsilon_k\right)\phi_k = 0.
\end{eqnarray}
Expanding the original fields $\psi$ in the complete set of eigenfunctions $\phi_k
$, that is $\psi(x,y,z) = \sum_k \psi_k(x,y) \phi_k(z)$, the action takes
the form 
\begin{eqnarray}
S[\psi] \eq i \int\! d^2r \, \bar\psi_k \Big( \big[\epsilon_{\mathrm{F}
}-\omega^+ \sar_3 -\epsilon_k + \frac{\partial_x^2}{2m} - V(x,y)\big]
\delta_{kk^{\prime}}+ \frac1{2m} \left((\partial_y -i\hat A)^2\right)_{kk^{\prime}} 
\Big) \psi_{k^{\prime}},\nn
\end{eqnarray}
where the integration extends over the $x$-$y$--plane, summation over $
k,k^{\prime}$ is implied, and 
\begin{eqnarray}
A_{kk^{\prime}} \equiv B \int\! dz \, \phi_k(z) z \phi_{k^{\prime}}(z)
\end{eqnarray}
is the vector potential. I.e.~the magnetic field couples to the dipole matrix
elements $d_{kk^{\prime}}=\int\! dz \, \phi_k(z) z \phi_{k^{\prime}}(z)$
that contain detailed information about the microscopic symmetry properties
of the system.

Using the same steps as explained in the previous section , the action in terms of the slow supermatrix fields $Q_{kk^{\prime}}$ reads 
\begin{eqnarray}
S[Q] \eq - \frac{\pi \nu}{8\tau} \int\! d^2r \, \Tr Q^2+   \frac12\int\! d^2r \,\Tr\ln \left(\epsilon_{\mathrm{F}} \!-\! \omega^+ 
\sar_3 \!-\!\hat \epsilon \!+\! \frac1{2m} (\partial_x^2 \!+\! (\partial_y
\!-\!i\hat A\str_3)^2) \!-\!\frac i{2\tau}Q \right),\label{S-Q1}
\end{eqnarray}
where a compact $k$-index free notation has been introduced. Here the energy matrix $\hat
\epsilon \equiv \mathrm{diag}(\epsilon_0,\epsilon_1,\,\dots )$ contains the
subband energies. The next step in the construction of the effective theory
is the saddle point analysis. Functional differentiation of the action with
respect to $Q$ obtains the equation 
\begin{eqnarray}
\Lambda_k = \frac i{\pi\nu} \int\! d^2p \,\frac1{i\delta \sar_3 - \epsilon_{
\mathrm{F}} - \epsilon_k + \frac{p^2}{2m} + \frac i{2\tau} \Lambda_k}\nn
\end{eqnarray}
for the diagonal elements $\Lambda_k$ of the saddle point matrix $
Q_{kk^{\prime}}$. At this stage one has to specify the relative position of
the Fermi energy $E_{\mathrm{F}}$ and the subband energies $\epsilon_k$.
Below, we will explore the case where $M$ bands with energy $\epsilon_k < E_{
\mathrm{F}}$ ($k=0,\dots,M-1$) exist. This leads to 
$\Lambda_k = \sar_3$ for $k<M$, 
and $\Lambda_k =0$ otherwise, 
where it has been assumed that the highest occupied subband $\epsilon_{M-1}$
lies well below (farther than $\tau^{-1}$) the Fermi level. Based on this
solution, the low-lying fields of the theory can be represented as $
Q=T\Lambda T^{-1}$, where $\Lambda = \{\Lambda_k \delta_{k k^{\prime}}\}$,
and the final expression for the general slow action reads ~\cite{MAA01,Thesis}
\begin{eqnarray}
S[Q] \eq-\frac{\pi\nu}8\int\! dS\,\;\tilde {\sum_k}\;\Tr\Big(4i\omega \sar
_3Q_k+D_k(\tilde\partial_kQ_k)^2\Big)+\frac{\pi\nu}4\int\! dS\,\;\tilde {\sum_{k,k^{\prime}}}\;\Tr\Big(\cX
_{kk^{\prime}}\str_3Q_k\str_3Q_{k^{\prime}}\Big),  \label{S-slow}
\end{eqnarray}
where $\tilde\partial_k=\partial-i{\mathbf{e}}_yA_{kk}[\str_3,\,.\,]$. The sum 
$\tilde {\sum}_k$ involves only the occupied subbands $k=0,\dots,M-1$.
Furthermore, 
\begin{eqnarray}
\cX_{kk^{\prime}}=\frac12(D_k+D_{k^{\prime}})\frac{1}{(\epsilon_{kk^{
\prime}}\tau)^2+1}A_{kk^{\prime}}A_{k^{\prime}k}(1 -\delta_{kk^{\prime}}).
\label{x-kk}
\end{eqnarray}
Here $D_k$ is the diffusion constant of subband $k$, and $
\epsilon_{kk^{\prime}}=\epsilon_k-\epsilon_{k^{\prime}}$.
The first line of (\ref{S-slow}) gives the conventional result of a $2d$
system while the second line describes the coupling of the subbands induced
by the magnetic field.

To prepare the 
one Cooperon approximation to the conductivity, one has to expand the fields $Q$ to
lowest non-trivial order in some generators. It is convenient to decompose
the generators into `diffuson' ($d$) and `Cooperon' ($c$) blocks. It is
clear from the structure of the second order action, that it does not couple
between the `$d$' and the `$c$' sector, i.e., $S=S^d + S^c$. The diffuson action $S^d$ is generated by those fields that commute with $\str_3$ and, thus, do not couple to the magnetic field, while the Cooperon action $S^c$ is generated by those fields that anticommute with $\str_3$ and, thus, are field-sensitive.

Here we are only interested in the Cooperon action being responsible for WL
corrections. The kernel appearing in $S^c$ is the `inverse of the Cooperon'.
More explicitly, the Cooperon $\cC$, which in our formulation is a matrix in
the discrete space of $k$-indices and diagonal in $\mathbf{q}$-space, is
obtained by inverting the matrix 
\begin{eqnarray}
(\cC^{-1}_{\mathbf{q}})_{kk^{\prime}}\eq \left(-\frac{2i\omega}{D_k}+ (\bq 
\!-\! 2 \bA_{kk})^2 +\frac2{D_k} \sum_{k^{\prime\prime}}\cX
_{kk^{\prime\prime}} \right)\delta_{k k^{\prime}} +\frac{2}{\sqrt{
D_kD_{k^{\prime}}}}\,\cX_{kk^{\prime}}.\nn
\end{eqnarray}
The magnetoconductance is determined by the specific form of this matrix.
The conductivity is given as $\sigma=\sigma^0+\Delta\sigma$ with $\sigma^0=
\sum_{k=0}^{M-1}\sigma_k^0,$ where $\sigma_k^0=\nu D_k$ is the Drude
conductivity of subband $k$, and 
\begin{eqnarray}
\Delta\sigma=- \frac2\pi\, \sum_{k=0}^{M-1}\sum_\bq\;(\cC_{\bq
,\omega=0})_{kk}.  \label{res-cond2}
\end{eqnarray}
In general, the field-dependent terms will render $\cC$ massive, i.e., the
weak localization corrections will suffer from a field induced suppression.
However, there is the situation mentioned above, where $z$-inversion, $\cP
_z:z \mapsto -z$, is an (approximate) symmetry of the Hamiltonian, $[\cH,\cP
_z]\approx 0$. 

For systems with an exact $\cP_z$-symmetry, the eigenfunctions $\phi_k$ obey 
$\cP_z \phi_k = (-)^k \phi_k$. The definition of the vector potential matrix $\hat
A$ then implies 
\begin{eqnarray}  \label{z-sym1}
A_{kk^{\prime}}= \left\{ 
\begin{array}{ll}
A_{kk^{\prime}} \qquad & k+k^{\prime}\enspace {\rm odd}, \\ 
0 & k+k^{\prime}\enspace {\rm even},
\end{array}
\right.\nn
\end{eqnarray}
and, thus, the same holds true for $\cX_{kk^{\prime}}$.

This structure bears consequences on the Cooperon mass. To analyze this
point, consider the spatial Cooperon zero-mode, 
\begin{eqnarray}
(\cC_{\mathbf{0},\omega =0}^{-1})_{kk^{\prime }}=2\left( \frac{1}{D_{k}}
\sum_{k^{\prime \prime }}\cX_{kk^{\prime \prime }}\delta _{kk^{\prime }}+
\frac{\cX_{kk^{\prime }}}{\sqrt{D_{k}D_{k^{\prime }}}}\right) .
\label{coop-zero}
\end{eqnarray}
This matrix has determinant zero implying that there is a Cooperon mode
which is not affected by the field. In fact, it is straightforward to verify
that the $M$-dimensional vector 
\begin{eqnarray}
\bX\equiv {\mathcal{N}}^{-1/2}\sum_{k}(-)^{k}\sqrt{D_{k}}\,{\mathbf{e}}_{k},
\label{EV}
\end{eqnarray}
where ${\mathcal{N}}=\sum_{k}D_{k}$, is annihilated by $\cC_{{\mathbf{0}},\omega =0}^{-1}$. As the zero-mode matrix is symmetric, one can, in principle,
construct a complete set of orthonormal eigenvectors, $\{\bX_{0}\equiv \bX,
{\mathbf{X}}_{1},\dots \bX_{M-1}\}$, with eigenvalues $\{0,\lambda _{1},\dots
,\lambda _{M-1}\}$.

Furthermore, due to $A_{kk}=0$, the full Cooperon kernel is
separable (i.e., it is the sum of a spatial and an `internal' operator).
Inserting the result into Eq.~(\ref{res-cond2}) yields 
\begin{eqnarray}
\Delta\sigma(B)= - \frac2\pi \,\tilde{\sum_{k;\bq}}\, \frac1{q^2+\lambda_k}.
\label{res-cond}
\end{eqnarray}
 This is our final result for the conductivity. Notice
that, due to $\lambda_0=0$, the weak localization corrections do survive the
magnetic field; carrying out the $\bq$-summation leads to the usual
logarithmic correction to the Drude conductance. Thus, even at high magnetic
fields, a logarithmic temperature dependence -- see Chap.~\ref{sec-Orbital} --
of the conductance should be observable. All other eigenvalues are
proportional $B^2$ and, thus, display the usual field dependence.

In the following, let us concentrate on the behavior of the lowest eigenvalue if the system is not exactly inversion
symmetric. An asymmetry can be caused either by the confining potential or
by a $z$-dependence of the random impurity potential.

In the case of an {asymmetric confining potential}, Eq.~(\ref{z-sym1}) generalizes to $A_{kk^{\prime}}\to A_{kk^{\prime}} + \delta A_{kk^{\prime}}$,
where $\delta A_{kk^{\prime}}$ is assumed to be much weaker than the
symmetry allowed elements $A_{kk^{\prime}}$, $k+k^{\prime}$ odd. Similarly,
there are non-vanishing but small matrix elements $\delta\cX_{kk^{\prime}}$
for both $k+k^{\prime}$ even and odd.
To lowest order in perturbation theory, the presence of these matrix elements shifts the zero-mode eigenvalue 
$\lambda_0(\bq)$ of the unperturbed Cooperon mode at momentum $\bq$ by the amount 
$\delta \lambda_0^{\mathrm{(as)}}(\bq) = \bX^T \delta \cC^{-1}_{\mathbf{q}} 
\bX$,
where $\delta \cC^{-1}_{\mathbf{q}}$ is the perturbation contribution to the
Cooperon operator. Explicitly, 
\begin{eqnarray}
\left(\delta \cC^{-1}_{\mathbf{q}}\right)_{kk^{\prime}}=\Big[ -4 q_y \delta
A_{kk} + 4 \delta A_{kk}^2 + 2 \sum_{k^{\prime\prime}}\delta \cX
_{kk^{\prime\prime}} \Big] \delta_{kk^{\prime}} + 2 \delta \cX_{kk^{\prime}}.
\end{eqnarray}
Combining these equations and making use of the definition of the zero-mode
eigenvectors (\ref{EV}) yields 
\begin{eqnarray}
\delta\lambda_0^{\mathrm{(as)}}(q_{\mathrm{min}})\eq\frac2{\cN
^2}\sum_{k,k^{\prime}}D_kD_{k^{\prime}}(\delta A_{kk}-\delta
A_{k^{\prime}k^{\prime}})^2 + \frac4\cN \sum_{k+k^{\prime}\,\mathrm{even}
}\delta \cX_{kk^{\prime}}.\nn
\end{eqnarray}
Thus, the Cooperon acquires a mass term $\sim B^2$.

The influence of {$z$-dependent impurities} has a similar effect. A potential with a generic $z$-dependence will not be inversion symmetric,
implying that, somehow, the Cooperon must pick up a mass. Assuming a
Gaussian distributed potential $U(\br,z)$, 
\begin{eqnarray}
\langle U(\br,z)\rangle = 0,\qquad \langle U(\br,z) U(\br^{\prime},z^{
\prime})\rangle = \gamma^2 \delta(\br -\br^{\prime}) \delta(z-z^{\prime}),\nn
\end{eqnarray}
for small $\gamma$, it is sufficient to consider the lowest order
non-vanishing contribution in $U$ to the action. We arrive at 
\begin{eqnarray}
S[Q]= S_0[Q] + \left(\frac{\pi \gamma \nu}2 \right)^2 \int\! dS \, \tilde{
\sum_{kk^{\prime}}}\, \Gamma_{kk^{\prime}} \Tr(Q_k(\br) Q_{k^{\prime}}(\br)),
\end{eqnarray}
where the coefficient 
$\Gamma_{kk^{\prime}k^{\prime\prime}k^{\prime\prime\prime}} = \int\! dz
\,\phi_k(z) \phi_{k^{\prime}}(z)\phi_{k^{\prime\prime}}(z)
\phi_{k^{\prime\prime\prime}}(z)$, 
and $\Gamma_{kk^{\prime}}\equiv\Gamma_{kkk^{\prime}k^{\prime}}$ is positive.
This expression tells that the $z$-dependent scattering tends to lock the
fields $Q_k$. For $\gamma$ large, only field configurations $\{Q_k \equiv Q\}
$ with no $k$-dependence survive. The physical mechanism is the following:
Scattering in $z$-direction leads to a coupling between the different $k$
-bands. Thus, the formerly independent diffusons and Cooperons are coupled,
too.
The formerly massless Cooperon channel does not survive
this coupling as the $k$-space eigenvector $\bX_{0}$
associated with the eigenvalue $\lambda _{0}=0$ is
staggered in $k$, cf.~Eq.~(\ref{EV}), i.e., it stands orthogonal on the
field configurations that are compatible with the locking.

If the coupling due to the impurity scattering is smaller than the field
induced subband coupling, the shift of the lowest eigenvalue is again
obtained by first order perturbation theory. Then, 
\begin{eqnarray}
\delta \lambda _{0}^{\mathrm{(imp)}}=\frac{1}{\mathcal{N}}\pi \nu \gamma
^{2}\sum_{k+k^{\prime }\;\mathrm{odd}}\Gamma _{kk^{\prime }}.
\label{lam0i}
\end{eqnarray}
Or, $\delta \lambda _{0}^{\mathrm{(imp)}}\sim 1/(\cN\tau ^{\prime })$, where 
$\tau ^{\prime }$ has the meaning of a scattering time perpendicular to the
plane, i.e., between the subbands.

This result which does not depend on the magnetic field holds true only for
sufficiently large fields. For smaller fields, the disorder induced mass
term fixes the preferred eigenvector. To compute the mass of the completely
locked Cooperon, consider 
$\lambda _{l}\equiv \bX_{l}^{T}\cC_{{\mathbf{0}},\omega =0}^{-1}\bX_{l}$, 
where the Cooperon operator is given by Eq.~(\ref{coop-zero}), and the
`locked' vector $\bX_{l}$ reads 
$\bX_{l}\equiv {\mathcal{N}}^{-1/2}\sum_{k}\sqrt{D_{k}}\,{\mathbf{e}}_{k}$. 
Explicitly computing the matrix element leads to 
\begin{eqnarray}
\lambda _{l}=\frac{2}{\cN^{2}}\sum_{k,k^{\prime }}D_{k}D_{k^{\prime
}}(A_{kk}-A_{k^{\prime }k^{\prime }})^{2}+\frac{4}{\cN}\sum_{k,k^{\prime }}
\cX_{kk^{\prime }}.
\label{laml}
\end{eqnarray}
At low fields, the mass of the Cooperon increases quadratically with $B$
according to Eq.~(\ref{laml}), but then, due to Eq.~(\ref{lam0i}), it levels
off at large fields. The characteristic field $B_{c}$ can be estimated by
comparing Eqs.~(\ref{lam0i}) and (\ref{laml}) which yields $B_{c}\sim E/v_{
\mathrm{F}}\sqrt{\tau /\tau ^{\prime }}/d$, where $E$ stands for the typical
energy separation between subbands and $d$ sets the scale for the width of
the quantum well.

As the expressions obtained above are rather lengthy, it is helpful to
consider some specific examples. We concentrate on the experimentally most
relevant case $M=2$ and, for simplicity, choose\footnote{
Admitting for different diffusion  constants $D_0\neq D_1$ does not change
the results qualitatively.} $D_0=D_1\equiv D$. Diagonalization of the $
2\times2$ Cooperon matrix, 
\begin{eqnarray}
{\cC}^{-1}=\pmatrix{(\bq-\bA)^2+\frac2{D}\cX_{01}& \frac2{D}
\cX_{01} \cr
\frac2{D}\cX_{01}&(\bq+\bA)^2+\frac2{D}\cX_{01}},\nn
\end{eqnarray}
yields 
\begin{eqnarray}
\lambda=q^2+A^2+\frac2{D}\cX_{01}\pm2\sqrt{(\bA\bq)^2+\frac1{D^2}
\cX_{01}^2},
\end{eqnarray}
where $\bA=\bA_{00}-\bA_{11}$, and ${\cX}_{01}=DA_{01}^2/(1+(
\epsilon_{10}\tau)^2)$ obtains from (\ref{x-kk}). The corresponding magnetic
decoherence times read $1/\tau_B=D\lambda$.
Note that at small magnetic fields, $2\cX_{01}\ll1/\tau_\phi$, the symmetry
mechanism is ineffective. Irrespective of $A$, the magnetoconductance yields 
\begin{eqnarray}
\sigma(B)-\sigma(0)\simeq2\frac{e^2}{\pi h}\cX_{01}\tau_\phi,\nn
\end{eqnarray}
which shows the usual low-field quadratic dependence on $B$. However, the
coefficient is diminished by the factor $1/(1+(\epsilon_{10}\tau)^2)$.

At large magnetic fields, $2\cX_{01}\gg1/\tau_\phi$, if the confining
potential is fully symmetric ($\bA=0$), the result reduces to 
$1/\tau_B=2\cX_{01}(1\pm1)$.
While $1/\tau_B=4\cX_{01}$ leads to a logarithmic field dependence
(see Eq.~(\ref{ManySubbands})), due to the field-insensitive $1/\tau_B=0$, the
conductance maintains its temperature dependence through $\tau_\phi$ even at
large fields. A slight asymmetry of the confining potential entails a finite 
$\bA$, which leads to 
$1/\tau_\phi(B)\simeq DA^2+2{\cX}_{01}(1\pm1)$. 
Thus, the temperature dependence remains as long as $DA^2<1/\tau_\phi$.

\begin{figure}[h]
\begin{center}
\includegraphics[scale=0.35]{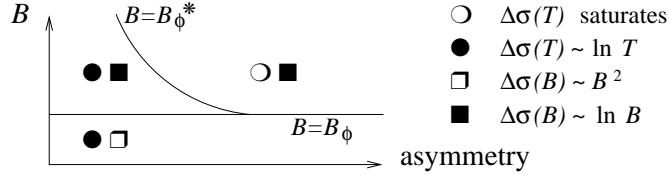}
\caption{Different regimes of $B$- and $T$-dependence of the weak
localization corrections.}
\label{fig-phases}
\end{center}
\end{figure}

The regimes with different field and temperature dependences of the
conductance are shown schematically in Fig.~\ref{fig-phases}~\cite{MAA01,Thesis}. The fields $
B_{\phi }$ and $B_{\phi }^{\ast }$ are defined through $2\cX_{01}\tau _{\phi
}=1$ and $DA^{2}\tau _{\phi }=1$, respectively. I.e. 
\[
B_{\phi }\equiv \frac{1}{\sqrt{D\tau _{\phi }}\,d_{01}\sqrt{\frac{2}{
1+(\epsilon _{10}\tau )^{2}}}},\qquad \qquad B_{\phi }^{\ast }\equiv \frac{1}{
\sqrt{D\tau _{\phi }}\,(d_{00}\!-\!d_{11})},\nn
\]
where $d_{kk^{\prime }}$ are the dipole matrix elements defined above.



\section{Summary}
\label{sec-Summary}
In summary, we presented three mechanisms of 
quantum in-plane magnetoresistance in two-dimensional 
electron systems, such as quantum wells, heterostructures,
and inversion layers in field transistors,
in lateral quantum dots prepared of these materials, and in thin 
metallic films. These mechanisms include:
the purely orbital effects possible due to 
subband mixing in a 2D structure by the magnetic field, 
the effect of an interplay between the spin-orbit coupling 
in the 2D electron dispersion and Zeeman splitting 
by a magnetic field, and the field effect on the
efficiency of spin-flip processes in materials 
contaminated by paramagnetic impurities.

{\textsc{Acknowledgments:}} We would like to thank I. Aleiner, A. Altland, 
A.K. Geim, C. Marcus, and B.D. Simons for useful discussions. 
Some of the recent work described in this article has been supported by EPSRC and 
NATO CLG Programme. We also thank the Max-Planck-Institut f{\"u}r
Physik komplexer Systeme in Dresden for their hospitality
at an early stage of preparation of this review during one of the
MPI-PKS research workshops.


\bibliography{rev-cm}

\begin{thebibliography}{10}

\bibitem{2D}
T. Ando, A. Fowler, and F. Stern, Rev. Mod. Phys. {\bf 54},  437  (1982).

\bibitem{PepperWires}
K.~J. Thomas {\it et~al.}, Phys. Rev. Lett. {\bf 77},  135  (1996).

\bibitem{MarcusSpin}
J.~A. Folk {\it et~al.}, Phys. Rev. Lett. {\bf 86},  2102  (2001).

\bibitem{b-AlLe}
B.~L. Altshuler, P.~A. Lee, and R.~A. Webb, {\em Mesoscopic Phenomena in
  Solids} (North-Holland, Amsterdam, Oxford, New York, Tokio, 1991).

\bibitem{b-leshouches95}
{\em Mesoscopic Quantum Physics (Les Houches Session LXI)}, edited by E.
  Akkermans, G. Montambaux, J.-L. Pichard, and J. Zinn-Justin (Elsevier,
  North-Holland, Amsterdam, 1995).

\bibitem{b-meso}
{\em Mesoscopic Electron Transport}, Vol.~345 of {\em NATO Advanced Study
  Institute, Series E: Applied Sciences}, edited by L.~L. Sohn, L.~P.
  Kouwenhoven, and G. Sch{\"o}n (Kluwer, Dordrecht, 1997).

\bibitem{ib-SiAl}
B.~D. Simons and A. Altland,  in {\em Theoretical Physics at the End of the
  XXth Century, Banff}, {\em CRM Series in Mathematical Physics} (Springer, New
  York, 2001), Chap.~Mesoscopic Physics.

\bibitem{GoLa79}
L.~P. Gor'kov, A.~I. Larkin, and D.~E. Khmel'nitskii, JETP Lett. {\bf 30},  228
   (1979).

\bibitem{AltshulerKhm1}
B.~L. Altshuler and D.~E. Khmel'nitskii, JETP Lett. {\bf 42},  291  (1985).

\bibitem{AltshulerKhm2}
P.~A. Lee and A.~D. Stone, Phys. Rev. Lett. {\bf 55},  1622  (1985).

\bibitem{AltshulerKhm3}
P.~A. Lee, A.~D. Stone, and H. Fukuyama, Phys. Rev. B {\bf 35},  1039  (1987).

\bibitem{Washburn}
R.~A. Webb, S. Washburn, C.~P. Umbach, and R.~B. Laibovitz, Phys. Rev. Lett.
  {\bf 54},  2696  (1985).

\bibitem{Falko89}
V.~I. Fal'ko, J. Phys. Cond. Matt. {\bf 2},  3797  (1990).

\bibitem{FalkoRMF}
V.~I. Fal'ko, Phys. Rev. B {\bf 50},  17406  (1994).

\bibitem{MAA01}
J.~S. Meyer, A. Altland, and B.~L. Altshuler, preprint cond-mat/0105623.

\bibitem{Thesis}
J.~S. Meyer, Mesoscopic phenomena driven by parallel magnetic fields, PhD
  thesis, 2001.

\bibitem{FaJu01}
V.~I. Fal'ko and T. Jungwirth, Phys. Rev. B {\bf 65},  81306  (2002).

\bibitem{AnisMass1}
L. Smrcka {\it et~al.}, Phys. Rev. B {\bf 51},  18011  (1995).

\bibitem{AnisMass2}
J.~M. Heisz and E. Zaremba, Phys. Rev. B {\bf 53},  13594  (1996).

\bibitem{Spectroscopy1}
I. Kukushkin {\it et~al.}, JETP Lett. {\bf 53},  334  (1991).

\bibitem{Spectroscopy2}
V. Kirpichev, I. Kukushkin, V. Timofeev, and V.~I. Fal'ko, JETP Lett. {\bf 51},
   436  (1990).

\bibitem{AlAr81}
B.~L. Altshuler and A.~G. Aronov, JETP Lett. {\bf 33},  499  (1981).

\bibitem{DuKh84}
V.~K. Dugaev and D.~E. Khmel'nitskii, Sov. Phys. JETP {\bf 59},  1038  (1984).

\bibitem{HoutenBeenakker}
C.~W.~J. Beenakker and H. van Houten, Phys. Rev. B {\bf 37},  6544  (1988).

\bibitem{numFal}
M. Leadbeater, V.~I. Fal'ko, and C.~J. Lambert, Phys. Rev. Lett. {\bf 81},
  1274  (1998).

\bibitem{DyakonovPerel}
M. Dyakonov and V. Perel, Sov. Phys. JETP {\bf 33},  1053  (1971).

\bibitem{HiLa80}
S. Hikami, A.~I. Larkin, and Y. Nagaoka, Prog. Theor. Phys. {\bf 63},  707
  (1980).

\bibitem{LyandaGeller1}
S. Iordanskii, Y. Lyanda-Geller, and G.~E. Pikus, JETP Lett. {\bf 60},  207
  (1994).

\bibitem{LyandaGeller2}
Y. Lyanda-Geller and A.~D. Mirlin, Phys. Rev. Lett. {\bf 72},  1894  (1994).

\bibitem{Beenakker}
C.~W.~J. Beenakker, Rev. Mod. Phys. {\bf 69},  731  (1997).

\bibitem{AleinerFalko}
I.~L. Aleiner and V.~I. Fal'ko, Phys. Rev. Lett. {\bf 87},  256801  (2001).

\bibitem{MarcusDephasing1}
A.~G. Huibers {\it et~al.}, Phys. Rev. Lett. {\bf 81},  200  (1998).

\bibitem{MarcusDephasing2}
A.~G. Huibers {\it et~al.}, Phys. Rev. Lett. {\bf 83},  5090  (1999).

\bibitem{WeakLoc}
G. Bergmann, Phys. Rep. {\bf 101},  1  (1982).

\bibitem{FalkoJETP}
V.~I. Fal'ko, JETP Lett. {\bf 53},  342  (1991).

\bibitem{BobkovFalkoKhm}
A.~A. Bobkov, V.~I. Fal'ko, and D.~E. Khmel'nitskii, Sov. Phys. JETP {\bf 71},
  393  (1990).

\bibitem{Chandrasekhar}
V. Chandrasekhar {\it et~al.}, Phys. Rev. B {\bf 42},  6823  (1990).

\bibitem{FalkoJP}
V.~I. Fal'ko, J. Phys. Cond. Matt. {\bf 4},  3943  (1992).

\bibitem{Benoit}
A. Benoit {\it et~al.}, Superlatt. Microstr. {\bf 11},  313  (1992).

\bibitem{Geim}
A.~K. Geim, S.~V. Dubonos, and I.~Y. Antonova, JETP Lett. {\bf 52},  247
  (1990).

\bibitem{Efetov}
K.~B. Efetov, Phys. Rev. Lett. {\bf 74},  2299  (1995).

\bibitem{LaNe66}
J.~S. Langer and T. Neal, Phys. Rev. Lett. {\bf 16},  984  (1966).

\bibitem{Ber84}
G. Bergmann, Phys. Rep. {\bf 107},  2  (1984).

\bibitem{RoBe86}
M. Robnik and M.~V. Berry, J. Phys. A {\bf 19},  669  (1986).

\bibitem{Georges}
J.-P. Bouchaud and A. Georges, Phys. Rep. {\bf 195},  127  (1990).

\bibitem{Pippard}
A.~B. Pippard, {\em Magnetoresistance in Metals} (Cambridge University Press,
  New York, 1989).

\bibitem{MathurBaranger}
H. Mathur and H.~U. Baranger, Phys. Rev. B {\bf 64},  235325  (2001).

\bibitem{UCF-ends}
R. Serota {\it et~al.}, Phys. Rev. B {\bf 36},  5031  (1987).

\bibitem{Marcus-private}
C.~M. Marcus, private communication.

\bibitem{Tesanovich}
Z. Tesanovich {\it et~al.}, Phys. Rev. Lett. {\bf 57},  2760  (1986).

\bibitem{MacKinnon}
K. Nikolic and A. MacKinnon, Phys. Rev. B {\bf 50},  11008  (1994).

\bibitem{Chalker}
J.~T. Chalker and A. Macedo, Phys. Rev. Lett. {\bf 71},  3693  (1993).

\bibitem{Pichard}
J.-L. Pichard {\it et~al.}, J. Phys. (Paris) {\bf 51},  587  (1990).

\bibitem{Frahm1}
K. Frahm, Phys. Rev. Lett. {\bf 74},  4706  (1995).

\bibitem{Houten}
H. van Houten {\it et~al.}, Superlatt. Microstruct. {\bf 3},  497  (1988).

\bibitem{Dresselhaus}
G. Dresselhaus, Phys. Rev. {\bf 100},  580  (1958).

\bibitem{BychRashba}
Y. Bychkov and E. Rashba, JETP Lett. {\bf 39},  78  (1984).

\bibitem{Rossler}
G. Lommer, F. Malcher, and U. Rossler, Phys. Rev. Lett. {\bf 60},  728  (1988).

\bibitem{Mathur}
H. Mathur and A.~D. Stone, Phys. Rev. Lett. {\bf 68},  2964  (1992).

\bibitem{Aronov}
A.~G. Aronov and Y.~B. Lyanda-Geller, Phys. Rev. Lett. {\bf 70},  343  (1993).

\bibitem{Halperin}
B.~I. Halperin {\it et~al.}, Phys. Rev. Lett. {\bf 86},  2106  (2001).

\bibitem{Birge}
A.~B. Gougam {\it et~al.}, J. Low Temp. Phys. {\bf 118},  447  (2000).

\bibitem{Pothier}
A. Anthore {\it et~al.}, preprint cond-mat/0109297.

\bibitem{Korringa}
J. Korringa, Physica {\bf 16},  601  (1959).

\bibitem{Mohanty1}
P. Mohanty and R.~A. Webb, Phys. Rev. Lett. {\bf 84},  4481  (2000).

\bibitem{Mohanty2}
P. Mohanty, E.~M.~Q. Jariwala, and R.~A. Webb, Phys. Rev. Lett. {\bf 78},  3366
   (1997).

\bibitem{b-Efe}
K.~B. Efetov, {\em Supersymmetry in Disorder and Chaos} (Cambridge University
  Press, New York, 1997).

\bibitem{SiAg97}
B.~D. Simons, O. Agam, and A.~V. Andreev, J. Math. Phys. {\bf 38},  1982
  (1997).

\end{thebibliography}
\bibliographystyle{prsty}
\end{document}